\documentclass[fleqn,usenatbib]{mnras}
\usepackage{newtxtext,newtxmath}
\usepackage[T1]{fontenc}
\DeclareRobustCommand{\VAN}[3]{#2}
\let\VANthebibliography\thebibliography
\def\thebibliography{\DeclareRobustCommand{\VAN}[3]{##3}\VANthebibliography}

\usepackage{graphicx}	
\usepackage{amsmath}	
\usepackage{verbatim}   
\usepackage{balance}   

\usepackage[dvipsnames]{xcolor}	

\usepackage{geometry}
\newcommand{\secref}[1]{\hyperref[#1]{Section~\ref*{#1}}}
\newcommand{\appref}[1]{\hyperref[#1]{Appendix~\ref*{#1}}}

\usepackage{color}

\newcommand{\deleted}[1]{}
\title[MeerKAT multi-dish autocorrelation calibration]{\textsc{Hi} intensity mapping with MeerKAT: Calibration pipeline for multi-dish autocorrelation observations}

\author[Wang et al.]{Jingying Wang\thanks{astro.jywang@gmail.com}$^{1}$,
Mario G. Santos$^{1,2}$,
Philip Bull$^{3,1}$,
Keith Grainge$^{4}$,
Steven Cunnington$^{3}$,
\and
Jos\'e Fonseca$^{5,6,3,1}$,
Melis O. Irfan$^{1,3}$,
Yichao Li$^{1}$,
Alkistis Pourtsidou$^{3,1}$,
Paula S. Soares$^{3}$,
\and
Marta Spinelli$^{7,8}$,
Gianni Bernardi$^{9,10,2}$, 
Brandon Engelbrecht$^{1}$\\
$^{1}$Department of Physics and Astronomy, University of the Western Cape, Cape Town 7535, South Africa\\
$^{2}$South African Radio Astronomy Observatory (SARAO), 2 Fir Street, Observatory, Cape Town, 7925, South Africa\\
$^{3}$Astronomy Unit, Queen Mary University of London, Mile End Road, London E1 4NS, United Kingdom\\
$^{4}$Jodrell Bank Centre for Astrophysics, Department of Physics and Astronomy,
The University of Manchester, Manchester M13 9PL, UK\\
$^{5}$Dipartimento di Fisica ``G. Galilei'', Universit\`a degli Studi di Padova, Via Marzolo 8, 35131 Padova, Italy\\
$^{6}$INFN -- Istituto Nazionale di Fisica Nucleare, Sezione di Padova, Via Marzolo 8, 35131 Padova, Italy\\
$^{7}$INAF-Osservatorio Astronomico di Trieste, Via G.B. Tiepolo 11, 34143 Trieste, Italy\\
$^{8}$IFPU - Institute for Fundamental Physics of the Universe, Via Beirut 2, 34014 Trieste, Italy\\
$^{9}$INAF - Istituto di Radioastronomia, via Gobetti 101, 40129 Bologna, Italy\\
$^{10}$Department of Physics and Electronics, Rhodes University, PO Box 94, Grahamstown, 6140, South Africa\\
}
\date{Accepted XXX. Received YYY; in original form ZZZ\\}
\pubyear{2020}

\begin{document}
\label{firstpage}
\pagerange{\pageref{firstpage}--\pageref{lastpage}}
\maketitle
\begin{abstract}
While most purpose-built 21cm intensity mapping experiments are close-packed interferometer arrays, general-purpose dish arrays should also be capable of measuring the cosmological 21cm signal.
This can be achieved most efficiently if the array is used as a collection of scanning autocorrelation dishes rather than as an interferometer. As a first step towards demonstrating the feasibility of this observing strategy, we show that we are able to successfully calibrate dual-polarisation autocorrelation data from 64 MeerKAT dishes in the L-band (856 -- 1712 MHz, 4096 channels), with 10.5 hours of data retained from six nights of observing. We describe our calibration pipeline, which is based on multi-level RFI flagging, periodic noise diode injection to stabilise gain drifts and an absolute calibration based on a multi-component sky model. We show that it is sufficiently accurate to recover maps of diffuse celestial emission and point sources over a $10^\circ \times 30^\circ$ patch of the sky overlapping with the WiggleZ 11hr field.
The reconstructed maps have a good level of consistency between per-dish maps and external datasets, with the estimated thermal noise limited to $1.4~\times$ the theoretical noise level ($\sim 2$ mK). The residual maps have rms amplitudes below 0.1 K, corresponding to $<1\%$ of the model temperature. 
The reconstructed Galactic \textsc{Hi} intensity map shows excellent agreement with the Effelsberg-Bonn \textsc{Hi} Survey, and the flux of the radio galaxy 4C+03.18 is recovered to within 3.6\%, which demonstrates that the autocorrelation can be successfully calibrated to give the zero-spacing flux and potentially help in the imaging of MeerKAT interferometric data. Our results provide a positive indication towards the feasibility of using MeerKAT and the future SKA to measure the \textsc{Hi} intensity mapping signal and probe cosmology on degree scales and above.
\end{abstract}


\section{Introduction}

Intensity mapping provides a way of measuring the cosmological
clustering signal from large numbers of galaxies without having to
resolve them individually. 
Using a spectral line, such as the 21cm
line of neutral hydrogen, it is possible to accurately recover the spatial distribution of brightness temperature fluctuations of the line in three dimensions (as a function of angle and redshift), and therefore -- assuming that the neutral hydrogen is a biased tracer of the dark matter distribution -- recover familiar clustering signals such as the baryon acoustic oscillations (BAO) and redshift space distortions (RSDs) in the late Universe, or the evolving ionisation state of the intergalactic medium during Cosmic Dawn and the Epoch of Reionisation.
This method, known as \textsc{Hi} intensity mapping (\textsc{Hi} IM hereafter), is a promising technique for
efficiently mapping the large-scale structure of the
Universe and thus delivering precision constraints on cosmological models (\citealt{2008PhRvL.100i1303C};
\citealt{2008PhRvL.100p1301L}; \citealt{2008PhRvD..78b3529M}; \citealt{2008PhRvD..78j3511P};
\citealt{2008MNRAS.383..606W}; \citealt{2008MNRAS.383.1195W}; \citealt{2009astro2010S.234P};
\citealt{2010MNRAS.407..567B}; \citealt{2010ApJ...721..164S}; \citealt{2011ApJ...741...70L};
\citealt{2012A&A...540A.129A}; \citealt{2013MNRAS.434.1239B}; 
\citealt{Bull:2014rha}; \citealt{Kovetz:2017agg}).

\textsc{Hi} IM experiments differ from traditional spectroscopic galaxy surveys in their ability to recover 3D clustering information on large scales with a high survey speed over a very wide redshift range, potentially allowing us to survey extremely large fractions of the total available comoving volume of the cosmos along our past lightcone using this method. While the 21cm signal itself is much fainter than many foreground sources -- by a factor of around $10^4 - 10^5$ compared to Galactic synchrotron emission for example -- the advent of large radio telescope arrays, consisting of many elements with low-noise, high-bandwidth receivers, has now made it practical to perform large cosmological surveys with sufficient depth for detection within a reasonable observing time.

The majority of radio telescopes currently targeting the cosmological 21cm signal are large interferometric arrays observing at low frequencies, i.e., at $z > 6$. These include the Low-Frequency Array (LOFAR)\footnote{\url{http://www.lofar.org/}}, 
the Murchison Widefield Array (MWA)\footnote{\url{http://www.mwatelescope.org/}},
the Precision Array for Probing the Epoch of Reionization (PAPER)\footnote{\url{http://eor.berkeley.edu/}}, the Hydrogen Epoch of Reionization Array (HERA)\footnote{\url{http://reionization.org}} and 
the 21 Centimetre Array (21CMA)\footnote{\url{http://21cma.bao.ac.cn}}. 
Several arrays targeting intermediate redshifts are also either operating or under construction, including the Canadian Hydrogen Intensity Mapping Experiment (CHIME)\footnote{\url{https://chime-experiment.ca/en}} and Tianlai\footnote{\url{http://tianlai.bao.ac.cn/}}, and Hydrogen Intensity and Real-Time Analysis experiment (HIRAX)\footnote{\url{https://hirax.ukzn.ac.za/}}.

While these interferometric experiments have tremendous sensitivity in principle, the only positive detections of the cosmological 21cm signal to date have come from observations by ``single-dish'' (or autocorrelation) telescopes. In particular, the Green Bank Telescope (GBT)\footnote{\url{https://greenbankobservatory.org/}} and
Parkes telescope\footnote{\url{https://www.parkes.atnf.csiro.au/}} have pioneered the detection of the \textsc{Hi} signal in cross-correlation with optical galaxy surveys (\citealt{2010Natur.466..463C}; \citealt{2013ApJ...763L..20M}; \citealt{2017MNRAS.464.4938W}; \citealt{2018MNRAS.476.3382A}; \citealt{2021RAA....21...30L}; \citealt{2021arXiv210204946W}).
While neither Parkes nor GBT were able to definitively detect the auto-power spectrum of the \textsc{Hi} signal (but see \citealt{Switzer:2013ewa} for upper limits), other single-dish IM projects are planned, including 
the purpose-built BINGO\footnote{\url{http://www.bingotelescope.org/en/}} instrument, and surveys on existing or planned telescopes such as 
the Five-hundred-meter Aperture Spherical Telescope (FAST)\footnote{\url{https://fast.bao.ac.cn/}},
and the Square Kilometre Array (SKA)\footnote{\url{https://www.skatelescope.org/}}.

In the case of the SKA, an \textsc{Hi} IM survey has been proposed that would use SKA1-MID in autocorrelation mode. SKA1-MID is a mid-frequency array of 197 parabolic-dish antennas that will be built in South Africa, and a large 21cm autocorrelation survey has been ranked as one of its science priority cases \citep{2015aska.confE..19S}. While an interferometric intensity mapping survey would also be possible with SKA1-MID, this is expected to be insufficiently sensitive to large-scale cosmological features such as the BAO \citep{Bull:2014rha}.

MeerKAT\footnote{\url{https://www.sarao.ac.za/}}, as a pathfinder (and, ultimately, a component) of SKA1-MID, will play a crucial role in the development of a successful multi-dish autocorrelation survey method. In this work, we address the most critical calibration and systematics issues that are expected to affect dual-polarisation autocorrelation observations with MeerKAT. With SKA1 scheduled to begin full operations in 2028, there is a good window of opportunity for MeerKAT to robustly demonstrate the single-dish survey method and make a significant first detection before the SKA comes online \citep{2017arXiv170906099S}.

\noindent This paper is organized as follows:

\noindent {\bf Sect.~\ref{sec:survey}} describes the basic setup of our MeerKAT pilot survey.
  
\noindent {\bf Sect.~\ref{sec:pipeline}} gives a detailed overview of our calibration pipeline. 
  
\noindent {\bf Sect.~\ref{sec:mapping}} explains how the calibrated data is combined into multi-frequency maps. 
  
\noindent {\bf Sect.~\ref{sec:characteristics}} studies the characteristics of the calibrated autocorrelation data, and presents a variety of quality and consistency checks on the calibrated data.
  
\noindent {\bf Sect.~\ref{sec:comparison}} presents a variety of comparisons with previous observations of diffuse Galactic emission and point sources.
  
\noindent {\bf Sect.~\ref{sec:conclusions}} presents our conclusions.
  
\noindent Appendix~\ref{sec:flux_eqs} presents expressions for weighting the flux from point sources.

\begin{figure*} 
\centering
\includegraphics[width=1.6\columnwidth]{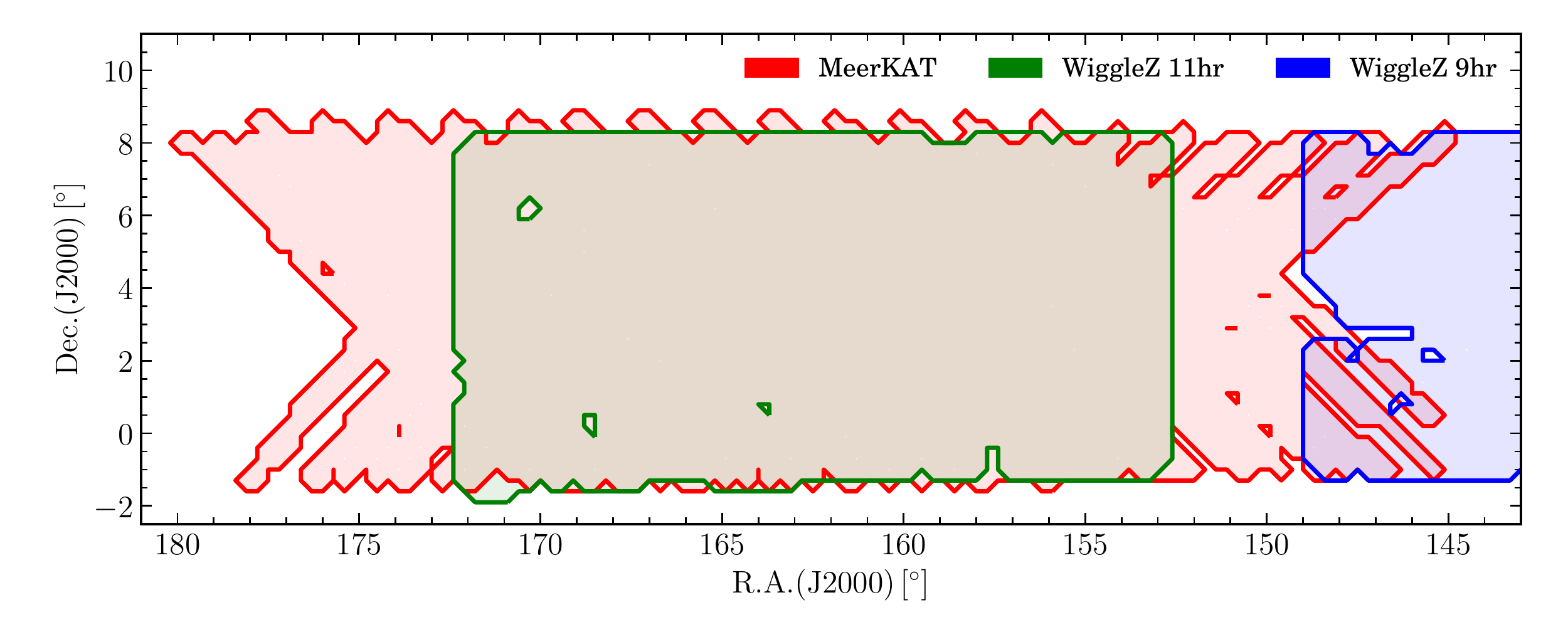}
\caption{The footprint of the MeerKAT survey field and the WiggleZ survey fields.}
\label{fig:footprint}
\end{figure*}

\section{Description of the survey data}\label{sec:survey}
In this Section, we review the basic parameters of our survey.

\subsection{Target field}
To check our observation and calibration strategy, we proposed a pilot survey through the MeerKAT open time call. The observation was done in the L-band and targeted a single patch of about 200 deg$^2$. 
The main factors affecting the choice of the sky region were to avoid the strong galactic emission and to pick an area with good spectroscopic coverage in the redshifts we are observing ($z < 0.5$). This spectroscopic coverage will ultimately allow us to attempt a detection of the cross-correlation power spectrum between the \textsc{Hi} IM signal and optical galaxy surveys, which is easier to achieve than a direct autocorrelation detection of the cosmological \textsc{Hi} signal because residual foregrounds and other systematics that bias the autocorrelation drop out in cross-correlation with optical surveys as they are uncorrelated, (e.g., see \citealt{2010Natur.466..463C, 2013ApJ...763L..20M, 2017MNRAS.464.4938W, 2018MNRAS.476.3382A}). 

As in \cite{2013ApJ...763L..20M}, we set our scans to cover the 11hr field of the WiggleZ Dark Energy Survey \citep{2010MNRAS.401.1429D}, which is a large-scale spectroscopic survey of emission-line galaxies selected from UV and optical imaging. The footprint of our observation and the WiggleZ coverage is shown in \autoref{fig:footprint}. Note that this area is also covered by Baryon Oscillation Spectroscopic Survey (BOSS) \citep{2016MNRAS.455.1553R}. Our observations took place between February and July 2019, when the WiggleZ 11h field appeared with high elevation ($>40 ^{\circ}$) during the night. The total usable observation time was about 10.5 hours, which corresponds to around 630 hours in total (before any flagging) when combined over 60 dishes on average. This can be compared to \citet{2013ApJ...763L..20M}, which covered the same region with a total area of $\sim 41 \, \text{deg}^2$ and 190 hrs of total integration time, although at lower frequencies (700-900 MHz) and higher angular resolution due to the large ($\sim$100m) diameter of the GBT dish.

\subsection{Scanning Strategy}
\begin{figure}
\centering
\includegraphics[width=\columnwidth]{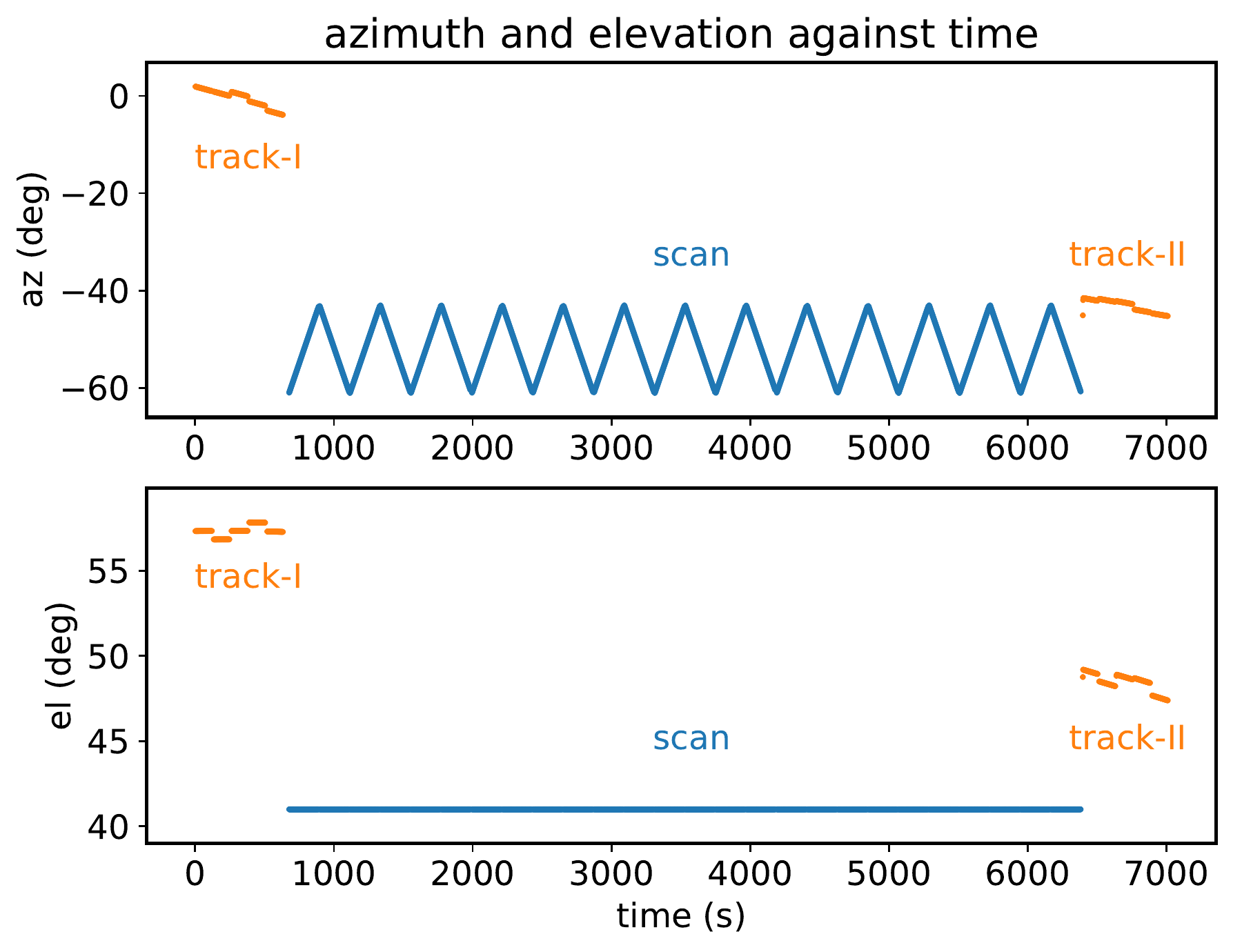}
\caption{An example of the azimuth and elevation against time during a single scan for MeerKAT dish {\tt m000} in {\tt obs190225}. Blue denotes the scanning of the target field, orange denotes the tracking of the calibrator source 3C 273.}
\label{fig:azel}
\end{figure}

Observations with single dishes typically require a scanning strategy where the dishes are rapidly moved across the sky. The aim is to cover the relevant angular scales within the stability time scale of the instrument, when the gains are approximately constant (set by the so-called 1/f noise knee, e.g., \citealt{2018MNRAS.478.2416H}). This required the development of a new observing mode not previously available with MeerKAT. 
The MeerKAT antennas were set to scan in azimuth at constant elevation (see an example in \autoref{fig:azel}) to minimise fluctuations of ground spill and airmass. 

The telescope scan speed was set to 5 arcmin/s along azimuth, corresponding to a projected speed on the sky of $5\cos(el)$ arcmin/s (where $el$ is the elevation angle). This ensures that the telescope pointing does not move significantly compared to the width of the primary beam ($\sim1$ deg) during a single time dump. With a time resolution of 2 sec, this gives a scan speed of no more than 10 arcmin per time sample, which is well within the beam size. The system-induced $1/f$-type variations for the MeerKAT receiver are well under the thermal noise fluctuations over $\sim 100$ second time scales after appropriate frequency filtering is performed (see \citealt{2020arXiv200701767L} for details). For a scan speed of 5 arcmin/s, we can scan about 10 deg in 100 seconds while retaining gain stability. In order to maintain gain stability over longer time scales, noise diodes attached to each receiver were fired for 1.8 s once every 20 s during the observation 
to provide a relative time-ordered data (TOD) calibration reference. The 1.8 s duration was chosen so that we can clearly see when the noise diode overlaps two time dumps, since we are currently unable to synchronise the noise diode fires with the correlator dumps (see \secref{sec:diode}). 

\begin{figure}
\centering
\includegraphics[width=\columnwidth]{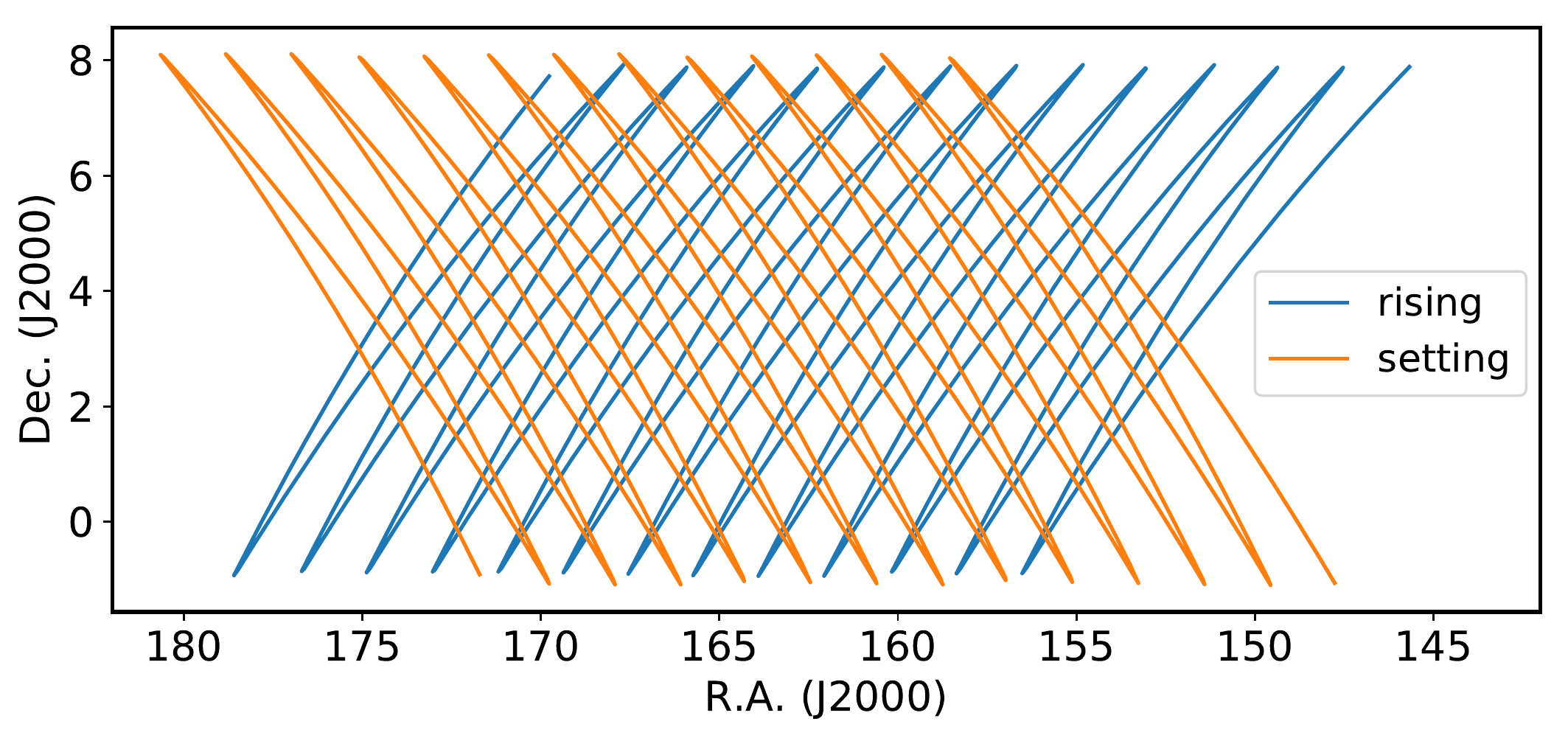}
\caption{Coverage map of the cross routes for the field rising (blue) and setting (orange) scans in the same night. }
\label{fig:cross}
\end{figure}

The dishes are moved back and forth with a slew of 18 deg in each direction, corresponding to an observing time of about 200\,s per stripe.
At fixed elevation, two scans can be performed of each stripe per night, corresponding to when the field is rising and setting respectively. The two scans will cross each other as shown in \autoref{fig:cross}, to achieve good sky coverage in the region of overlap. The duration of each set of scans is about 1.5 hours, and before and after each scan we spent about 15 minutes tracking a nearby celestial point source to use as a bandpass calibrator and absolute flux calibrator. 
The technical details of the observations are listed in \autoref{tab:basic}.
Not all planned observations were successful due to equipment issues. After basic data quality checks, we were able to retrieve 
seven blocks of data that were suitable for further data analysis.
These blocks are labelled by their observation start Unix time, as listed in \autoref{tab:block}.

\begin{table}
\centering
 \caption{Specifications of the MeerKAT Observations}
 \label{tab:basic}
 \begin{tabular}{lcc}
  \hline\hline
 Antennas & All 64 MeerKAT dishes\\
 Observation mode & Single-dish \\
 Polarisation & Linear (horiz. + vert.) feeds \\
 System temperature & $\sim$16 K \\
 Target field & WiggleZ 11hr field ($10^{\circ} \times 30 ^{\circ}$)\\
 Frequency range & 856-1712 MHz \\
 Frequency resolution & 0.2 MHz \\
 Number of channels & 4096 \\
 Time resolution & 2s \\
 Exposure time & 1.5hr x 7 scans \\
 Extent of azimuth scan & 18 deg \\
 Scan speed (along azimuth) & 5 arcmin/s\\
 Diode injection & Pattern mode (1.8s per 20s)\\
\hline
\end{tabular}
\end{table}

\begin{table*}
\centering
 \caption{Basic information of observation blocks used in this work.} 
 \label{tab:block}
 \begin{tabular}{cccccccc}
  \hline
  \hline
Block ID  & Short name & Observation start time & Sunset & az range & el & Calibrator source & Motion of field \\
(Unix Timestamp) &(in this paper) &(UTC time)&&($^{\circ})$&($^{\circ})$&&\\
\hline
obs1551037708  & {\tt obs190224} & 2019-02-24 19:48:28  & 02-24 17:09:57 & [41.6, 59.6] & 42.0 & 3C 237 & Rising\\
obs1551055211  & {\tt obs190225} & 2019-02-25 00:40:11  & 02-24 17:09:57 &[-61.0, -43.0] & 41.0 & 3C 273 & Setting \\
obs1553966342  & {\tt obs190330} & 2019-03-30 17:19:02 & 03-30 16:29:35 &[43.7, 61.7] & 40.5 & Pictor A & Rising \\ 
obs1554156377 & {\tt obs190401} & 2019-04-01 22:06:17 & 03-31 16:28:22  &[-61.0, -43.0] & 41.0 & 3C 273 & Setting \\
obs1556052116 & {\tt obs190423} & 2019-04-23 20:41:56 & 04-23 16:02:05 & [-60.3, -42.3] & 41.5 & 3C 273 & Setting  \\
obs1556138397 & {\tt obs190424} & 2019-04-24 20:39:57 & 04-24 16:01:04 &[-60.3, -42.3] & 41.5 & 3C 273 & Setting  \\
obs1562857793 & {\tt obs190711} & 2019-07-11 15:09:53 & 07-11 15:43:57 & [-55.3, -37.3] & 43.4 & 3C 273 & Setting \\
\hline
\end{tabular}
\end{table*}

\section{Data reduction pipeline}\label{sec:pipeline}
\begin{figure}
\includegraphics[width=\columnwidth]{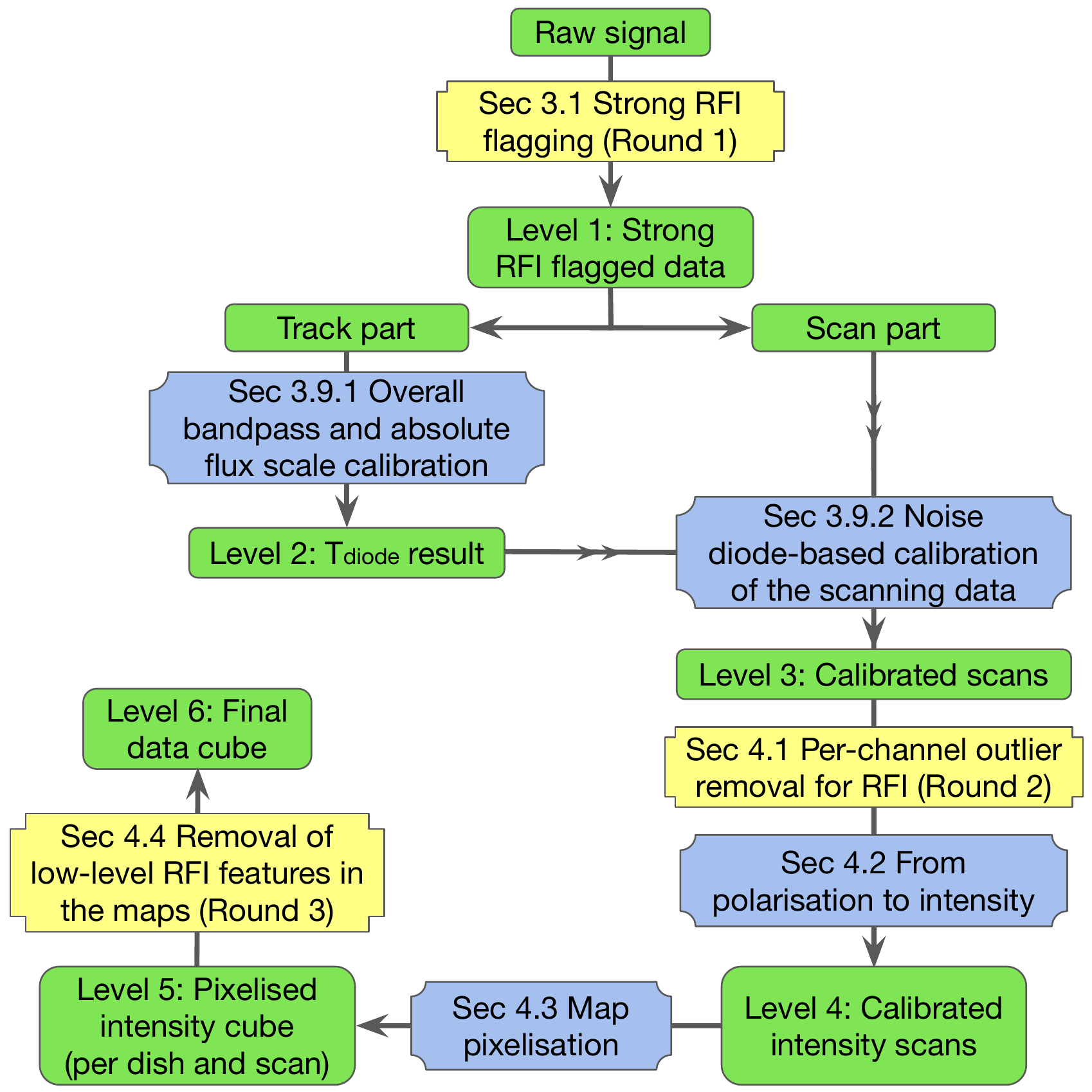}
\caption{Flowchart showing each step in the calibration and map-making pipeline. The numbers of the sections from this paper where each step is described are also shown.}
\label{fig:pipe}
\end{figure}

The data reduction pipeline for the time-ordered data includes steps for flagging of human-made radio frequency interference (RFI), bandpass and absolute calibration using a known point source, and calibration of receiver gain fluctuations based on interleaved signal injection from a noise diode (map making is described in section \secref{sec:mapping}). \autoref{fig:pipe} shows the flowchart of the pipeline, with an indication of the different stages of calibrated data and a reference to the sections in the paper where they are described.

\subsection{Strong RFI flagging (Round 1)} \label{sec:rfi}
\begin{figure}
\centering
\includegraphics[width=\columnwidth]{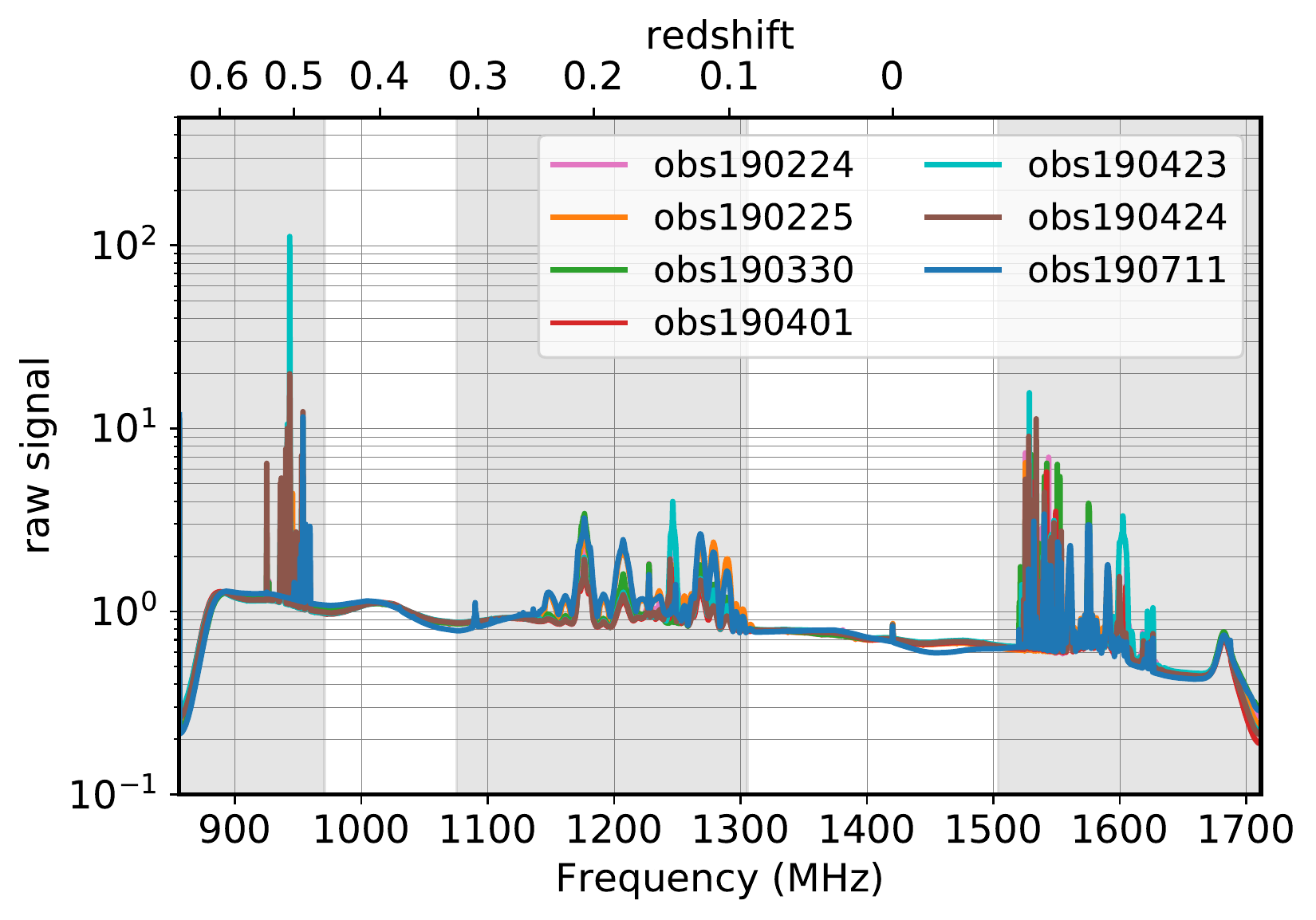}
\caption{The normalised time-averaged raw signal for {\tt m010h}, all seven observations. The clusters of bright spike features correspond to RFI-dominated bands. Unshaded regions are the frequency bands for further data analysis.}
\label{fig:rfi_band}
\end{figure}

The RFI from modern telecommunications and satellites is an important  contaminant of radio observations \citep{2018MNRAS.479.2024H}. In \autoref{fig:rfi_band} we show the time-averaged raw signal in a typical
observation. Several RFI groups can be seen clearly.
MeerKAT has RFI mitigation systems to prevent RFI before and during the observation \citep{2016mks..confE...1J}. 
However, as even the best RFI mitigation methods cannot completely prevent all RFI \citep{2010rfim.workE...1B}, we must employ methods to reduce the effect of RFI after observation. 
One method is to flag those parts of the spectrum which are dominated by RFI and not use those frequencies/times in scientific data analysis. In this work we use the RFI package of Signal Extraction and Emission Kartographer (SEEK; \citealt{2017A&C....18....8A}), which follows the SumThreshold algorithm \citep{2010MNRAS.405..155O}.

We denote the raw input data as $\hat T_{\rm raw}(t, \nu)$, where the hat indicates data in arbitrary correlator units.
We group the data in time according to the diode status ($t_{\rm nd}^0$, $t_{\rm nd}^{1a}$, and $t_{\rm nd}^{1b}$, corresponding to diode off, diode on/starting, and diode on/stopping; see \secref{sec:diode} for more details), as well as its scan state (whether tracking a calibrator source or scanning across the survey field). The SEEK flagging algorithm is then applied to these six groups of raw data separately, followed by an additional filter that discards any frequency channels or time dumps where $>70\%$ of the data are flagged.

This strong RFI flagging progress is applied to the data in two steps: First, on all of the raw data directly, to flag the strongly RFI-contaminated channels that are occupied by satellite communications; and second, focusing only on the target frequency bands, 971--1075 MHz (channels 550--1050) and 1305--1504 MHz (channels 2150-3100). In the second step, the value of $\hat T_{\rm raw}$ at each frequency is normalised by the mean value along the time direction to account for the shape of the bandpass. At the same time, the locations of bright point sources are protected by temporary masks, as otherwise they would show as transient peaks in the time stream as the dishes scan across the sky, potentially leading to them being erroneously flagged. 

This two-step SEEK flagging is applied to the per-dish HH and VV raw polarisation signals separately. The union of the flags from the two polarisations is then adopted as the primary set of flags. 
Typically, about 35\% of the raw data are flagged in this step. 
Note that additional rounds of RFI flagging are performed during and after the calibration to prevent contamination from weaker RFI signals that can only be detected after further processing (see \secref{sec:weakRFI}).

In the following calibration process, we focus on the 971-1075 MHz (channels 550-1050) and 1305-1504 MHz (channels 2150-3100) bands, considering 
the RFI distribution and the bandpass flatness. In these frequency ranges, most of the satellite communications are avoided, with only a few strong RFIs and some weak RFIs appearing intermittently.  

\subsection{Calibration model and strategy}
\label{sec:calibmodel}
Our autocorrelation calibration strategy is based on propagating an absolute flux and bandpass calibration obtained from tracking observations of bright point celestial point sources through to scanning observations of the survey field by using periodic noise diode fires as a relative calibration reference. We use a Bayesian approach to jointly fit the free parameters in our calibration and sky models, as described in the next section.

Bright celestial sources (e.g., AGN, supernova remnants) are commonly taken as bandpass calibrators and absolute flux calibrators for single-dish observations. Despite being resolved in interferometric observations, well-known bright sources such as 3C 273 and Pictor A are sufficiently small compared to the single-dish beam size that we can treat them as point sources.
The noise diode injections are taken as stable-in-time calibrators to remove receiver gain drifts, which are otherwise known to limit the sensitivity of single-dish observations. We examine the limitations of some of these assumptions in \secref{sec:characteristics}.

We construct a model for all of the components that will contribute to the total signal, and fit the free parameters by comparing the model to the time ordered data using a prescribed likelihood with priors.
Our model, in temperature units, is expressed as 
\begin{gather}\label{equ:model}
  T_{\rm model}(t, \nu)= T_{\rm ps}(t,\nu)  + T_{\rm diffuse}(t, \nu) + T_{\rm el}(t, \nu)     +  T_{\rm diode}(t, \nu) \\ \notag 
  ~~~~~~~~~~~~~~~~~~  + T_{\rm rec} (t, \nu),
\end{gather}
where $T_{\rm ps}$, $T_{\rm diffuse}$ and $T_{\rm el}$  are the antenna temperature models of the point source contributions, celestial diffuse component, and elevation-dependent terrestrial emission (atmosphere and groundspill) respectively, while $T_{\rm diode}$ is the noise diode contribution and  $T_{\rm rec}$ is the receiver temperature, i.e., the equivalent temperature of the noise power injected by the receiver. All quantities external to the dish are considered to be already convolved by the primary beam e.g., $T_{\rm ps}(t,\nu)$, $T_{\rm diffuse}(t, \nu)$ and $T_{\rm el}(t, \nu)$ should already have the beam applied. As will be described later, the models for $T_{\rm diffuse}$ and $T_{\rm el}$ are fixed while $T_{\rm rec}$ is assumed a smooth function of time (at each frequency). Therefore, any unknown offset on the fixed models (or indeed any unknown, slowly time changing, contamination) will be absorbed into $T_{\rm rec}$.

In order to compare with the raw data from the correlator, we need to multiply the signal model by the gain, $g(t, \nu)$, so that
\begin{gather}\label{equ:model2}
  \hat T_{\rm model}(t, \nu)= g(t, \nu)\, T_{\rm model}(t, \nu), 
\end{gather}
where again the hat indicates a temperature in the (arbitrary) correlator units. 
The signal and gain models are then fitted to the time ordered data for each polarisation, frequency channel, dish, and observation scan, all of which are treated independently (i.e., we do not consider correlations in parameters across these dimensions). Using the gain solution, we can then obtain the calibrated temperature, 
\begin{gather} \label{eq:T}
T_{\rm cal}(t,\nu)\equiv \hat T_{\rm raw}(t, \nu)/g(t, \nu).
\end{gather}
The residual temperature difference between the signal model and calibrated data is then
\begin{gather} \label{eq:T_resi}
T_{\rm res}(t,\nu)\equiv T_{\rm cal}(t,\nu) - T_{\rm model}(t, \nu).
\end{gather}
Note that throughout most of the paper we consider each polarisation separately, so temperature values should be taken to refer to a single polarisation. We will make it explicit when the two polarisations are combined.

\subsection{Bayesian model fitting framework}
The fitting process is done within a Bayesian framework, where the posterior probability is 
\begin{gather}
    p_{\rm post}\propto p_{\rm prior}\cdot \mathcal{L}(\mathbf{\hat T}_{\rm raw}|\mathbf{\hat T}_{\rm model}),
\end{gather}
where $p_{\rm prior}$ is the prior probability for all the parameters and $\mathcal{L}(\mathbf{\hat T}_{\rm raw}|\mathbf{\hat T}_{\rm model})$ is the likelihood for the data given the model. $\mathbf{\hat T}_{\rm raw}$ and $\mathbf{\hat T}_{\rm model}$ are vectors representing all the time stamps.

We assume that the difference between the data and our model is noise dominated and that this noise is uncorrelated in time. In that case, for fitting at a certain frequency in a single scan, the joint likelihood can be written as  
\begin{gather}
\mathcal{L}(\mathbf{\hat T}_{\rm raw}|\mathbf{\hat T}_{\rm model}) =  \prod_{t_i=0}^{t_{\rm max}} \frac{1}{\sqrt{2\pi \sigma^2(t_i)}} 
 \exp\Large\left\{-\frac{[\hat T_{\rm raw}(t_i)-\hat T_{\rm model}(t_i)]^2}{2 g^2(t_i,\nu) \sigma^2(t_i)}\Large\right\},
\end{gather}
where $\sigma({t_i})$ is the expected noise level for which we take the standard radiometer equation \citep{2009tra..book.....W}: 
\begin{gather}
\sigma^2({t_i}) = \frac{T^2_{\rm model}}{\delta t \delta\nu},
\end{gather}
where $\delta t$ and $\delta\nu$ account for the time and frequency resolution.

We assume independent prior probabilities, such that the total prior for the vector of all parameters $\mathbf{p}$, denoted by $p_{\rm prior}(\mathbf{p})$,
is the product of normalised 1D Gaussian distributions with mean $\bar{p}_j$ and standard deviation $\sigma_j$ for each parameter $p_j$.

In principle, by simply maximizing the posterior probability $p_{\rm post}$, we can obtain the optimal $\mathbf{\hat{p}}$. Usually we take the log of this probability and, ignoring any constant terms (which are irrelevant for the fitting process), we get 
\begin{gather}\label{eq:fit}
  \ln p_{\rm post} = -\sum_{t_i=0}^{t_{\rm max}} w(t_i) \Big[\ln \sigma(t_i) + 
  \frac{[\hat T_{\rm raw}(t_i)-\hat T_{\rm model}(t_i)]^2}{2 g^2(t_i,\nu) \sigma^2(t_i)}\Big]\\ \notag 
  ~~~~~~~~~~~~~~~~~~~~~~~~~~~~~~~~-\sum_j \frac{(p_j-\bar{p}_j)^2}{2\sigma_j^2},
\end{gather}
where $w(t_i)$ is an extra weight we apply if we want to change the contribution for a given time $t_i$.
As explained later, in the tracking part of the timestream, we set $w({t_i})=1$ for the entire integration time, while in the scan part we use $w({t_i})=1$ for diode-off and $w({t_i})=4$ for diode-on time samples. 
For every 10 time samples, two have noise injection, therefore to equal the relevance of diode-on/off calibration we up-weight the diode-on by a factor of 4. This ensures that the noise diode contribution acts as the main calibrator of the gain drift in the scan part.
Note that we cannot take $\sigma(t_i)$ as a constant since it also depends on the model.
Finally we can use a gradient descent (specifically, the modified Powell's method)
to find 
\begin{gather}
  \mathbf{\hat{p}} = \arg \min_{\mathbf{p}} \left[-\ln p_{\rm post}(\mathbf{p})\right].
\end{gather}

\subsection{Beam model}\label{sec:beam}

\begin{figure}
\centering
\includegraphics[width=\columnwidth]{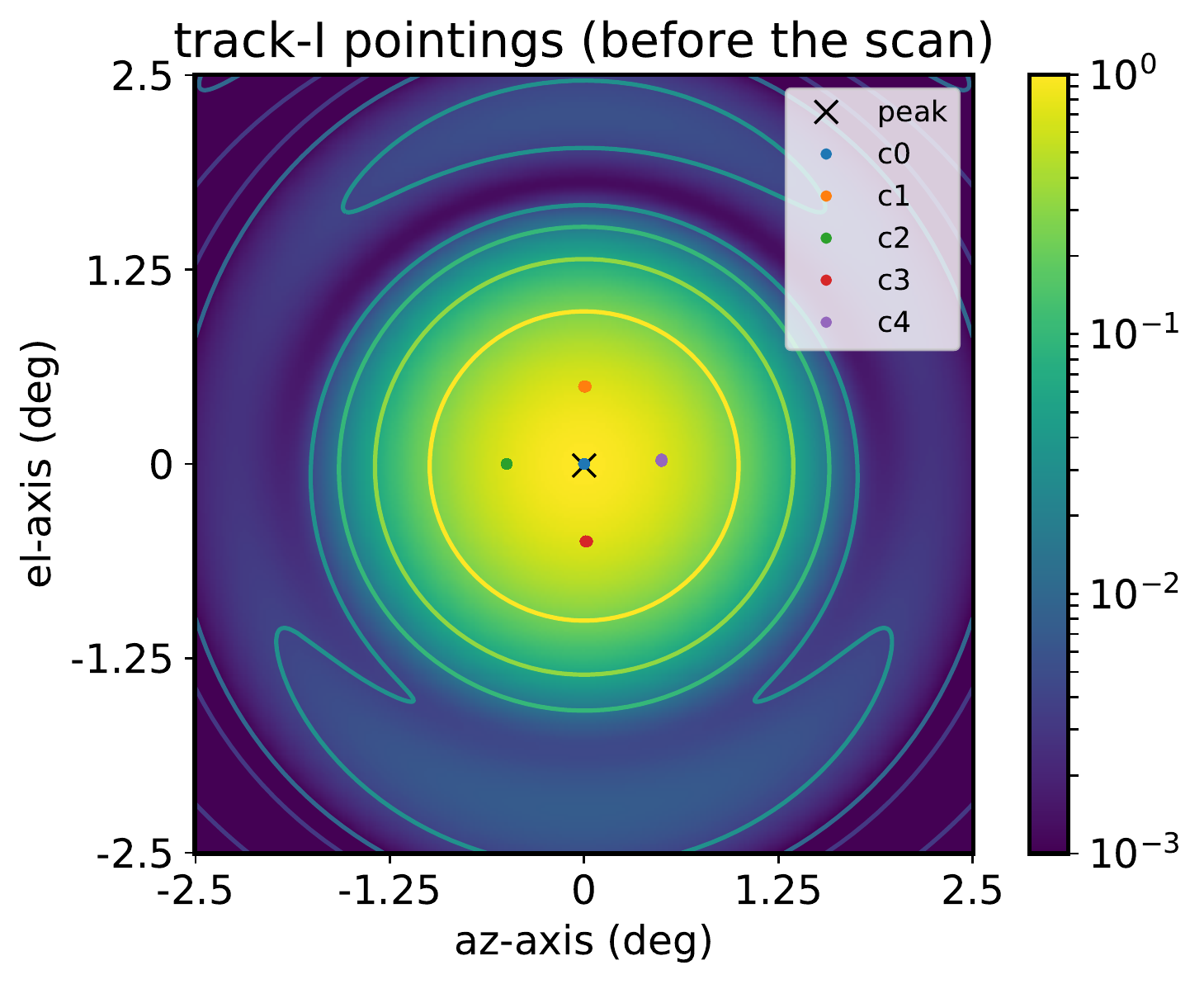}
\caption{Normalised beam pattern for polarisation HH at 1023 MHz, taken from the \citet{2019arXiv190407155A} model. Note the asymmetry of the beam pattern beyond the mainlobe. Also shown are the track-I pointing locations of the calibrator for observation {\tt obs190225}.}
\label{fig:beam}
\end{figure}

The large dynamic range in brightness between foreground emission and the 21cm fluctuations makes it particularly important to understand the beam pattern of the dishes. This is particularly relevant for calibrator point sources, since, as the beam are pointing on the source peak and outskirts (see an example in \autoref{fig:beam}), we will measure the calibrator source contribution at different positions of the beam.

We adopt a beam model based on astro-holographic observations and electromagnetic simulations for MeerKAT, as described in \citet{2019arXiv190407155A}. This model exhibits a degree of asymmetry (see \autoref{fig:beam}), which is necessary to match multi-pointing observations of the calibrator sources, and the beam pattern has a non-trivial frequency and polarisation dependence.
While this level of complexity in the beam model comes at the cost of requiring point source coordinates to be projected to the telescope coordinate system for every time sample,
the computational overhead involved is not significant compared with the overall runtime of the calibration pipeline. Nevertheless, in \secref{sec:comparison} we will also make use of a simple frequency-dependent 2D Gaussian fit to the beam when performing certain comparisons with external data.

\subsection{Calibrator point sources}
We use a small number of bright point sources as flux and bandpass calibrators, selected for their low polarisation fractions, and visibility at high elevations before and after each scan. For the WiggleZ 11hr field, our main calibrators were 3C273, 3C237, and Pictor A. We modelled the sources as strictly point-like, with frequency spectra
\begin{gather}\label{equ:spectrum}
  S_{\rm ps}(\nu)=S_{\rm ps}^{\rm 1.41}\left(\frac {\nu}{1.41~{\rm GHz}}\right)^{\alpha},
\end{gather}
where $S_{\rm ps}^{\rm 1.41}$ is the flux density at 1.41 GHz derived from the Parkes catalogue \citep{1990PKS...C......0W}, and $\alpha$ is the spectral index, derived using the observed fluxes at 1.41 GHz and 408 MHz. All three sources are bright in the L-band (3C237: 6.6~Jy; 3C273: 42~Jy; and Pic A: 66~Jy at 1.41 GHz), and have polarisation fractions of a few percent or less \citep[e.g.][]{1997A&A...328...12P, 1998A&ARv...9....1C}. 3C273 in particular has a rather flat spectrum ($\alpha=-0.219$), which makes it an ideal bandpass calibrator (whereas $\alpha=-0.744$ and $-0.683$ for Pic A and 3C237 respectively).

At a given frequency and time, the antenna temperature contribution from a point source to each polarisation is
\begin{gather}\label{equ:Tptr2}
  T_{\rm ps}(t,\nu)=  P_{\rm B}(t,\nu) \frac{1}{2k_B}S_{\rm ps}(\nu) A_{\rm eff}(\nu) \, ,
\end{gather}
where ${P_{\rm B}(t,\nu)}$ is the normalised beam power pattern that depends on the relative displacement of the point source from the beam centre, $k_B$ is the Boltzmann constant, $A_{\rm eff}(\nu)$ the effective beam area that computed by integrating over the 2D beam pattern within 10 deg width.

\subsection{Noise diode model} \label{sec:diode}

The temperatures of the noise diodes have previously been measured in the lab at several frequencies. We interpolate them to other frequency channels using a radial basis function (RBF) spline and produce a reference value $T_{\rm diode}^{\rm ref} (\nu)$ to be used as a starting point in our fitting. The noise diode signals can be modeled as  
\begin{gather}
  T_{\rm diode}(t, \nu) = f_{\rm diode}(t)\, \overline T_{\rm diode}(\nu) \, ,
   \end{gather}
where $\overline{T}_{\rm diode}(\nu)$ is the noise diode injected temperature and $f_{\rm diode}(t)$ is the fraction of time 
the noise diode is on in each 2-second time sample. The noise diodes are supposed to be very stable and in our calibration process we assume them to be constant in time.

\begin{figure} 
\centering
\includegraphics[width=\columnwidth]{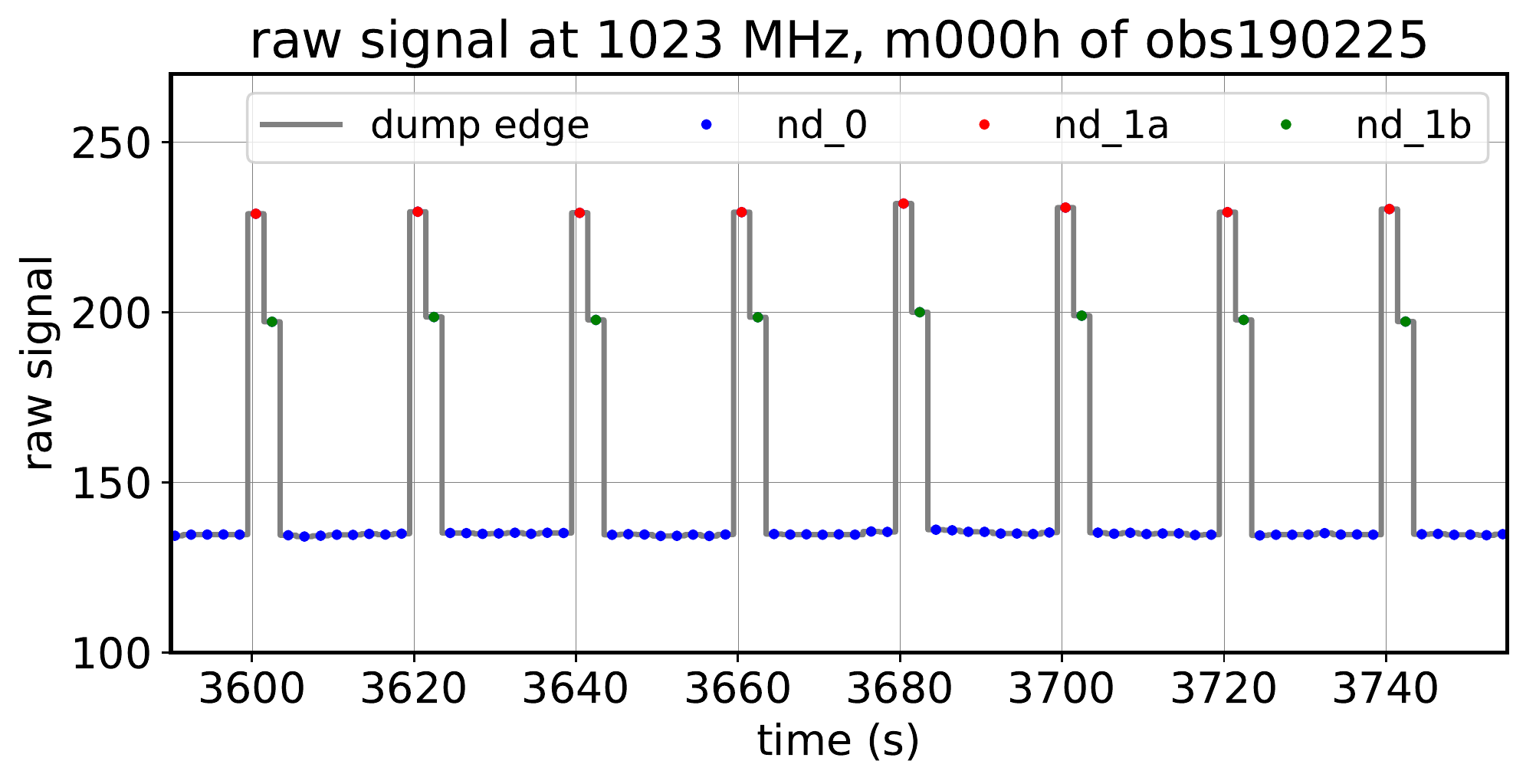}
\caption{Enlarged plot for the raw signal along time in one frequency channel to show the distribution of nd\_0 (time dumps without diode injection), nd\_1a (the first dump with diode injection) and nd\_1b (the second dump with diode injection).}
\label{fig:his_part}
\end{figure} 

\begin{figure*}
\centering
\includegraphics[width=2\columnwidth]{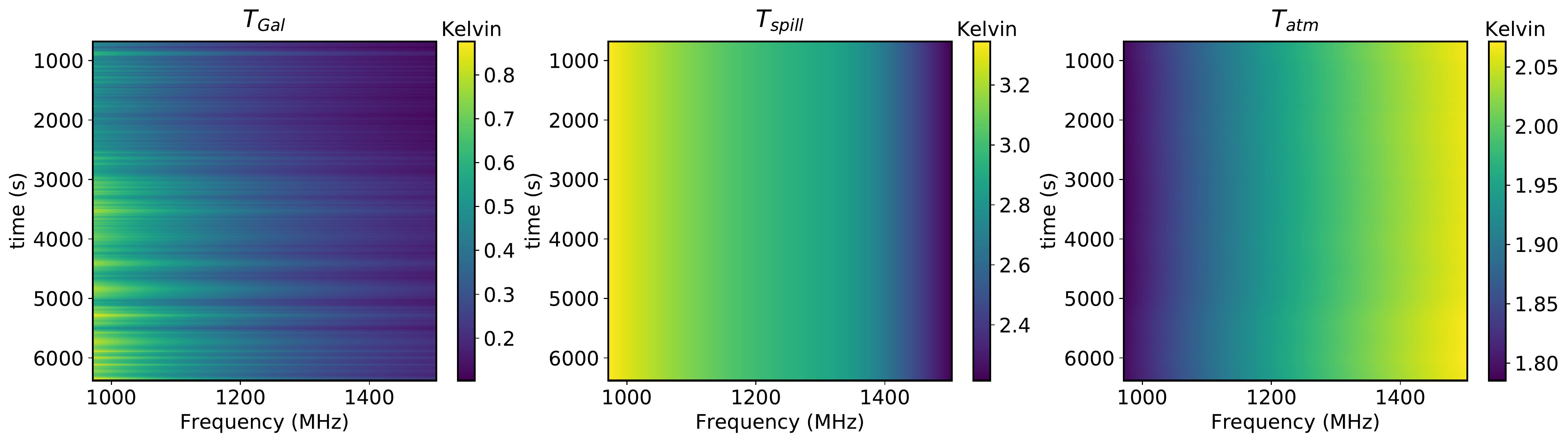}
\caption{The models of Galactic emission,  spillover from the ground and the atmospheric emission, {\tt m000h} of {\tt obs190225} as an example.}
\label{fig:models}
\end{figure*} 

Since at present the noise diode is not synchronised with the data sampling we cannot guarantee that the injection only happens in a single time dump. This could be problematic if only a small fraction falls in one of the time samples making it difficult to know when the noise diode was fired. To avoid this, we made the choice of injecting noise with a duration of 0.9 time samples ($1.8~ {\rm s}$). This way there is always a high chance that a single instance of the injection falls into two neighbouring samples ($t_{\rm nd}^{1a}$ and $t_{\rm nd}^{1b}$ hereafter) with a strong signal in both (\autoref{fig:his_part}). We used an injection period of $20~ {\rm s}$ as we expect the gains to be stable on this timescale \citep{2020arXiv200701767L}.
Since the asynchronicity between the noise injection and the data sampling is constant throughout one observation, $f_{\rm diode}(t)$
can be expressed as a periodic function with a period of 20 seconds:
 \begin{gather}
\label{eq6}
f_{\rm diode}(t)=\left\{
\begin{aligned}
f_{\rm d}     , &~~~~~~~& {\rm if} ~t \in t_{\rm nd}^{1a} \, , \\
0.9-f_{\rm d} , &~~~~~~~& {\rm if} ~t \in t_{\rm nd}^{1b}\, , \\
0     , &~~~~~~~& {\rm if} ~t \in t_{\rm nd}^{0}\, ,
\end{aligned}
\right.
 \end{gather}
 where $t_{\rm nd}^{0}$ represents the time dumps without noise injection, and $f_{\rm d}$ is one of the parameters to be fitted in the calibration (see more details in \secref{sec:fitting}).

\subsection{Diffuse sky emission model}

The (large scale) diffuse celestial components include Galactic emission and the CMB, 
\begin{gather}
  T_{\rm diffuse}(t,\nu)= T_{\rm Gal}(t, \nu) +\overline T_{\rm CMB} ,
\end{gather}
where the Galactic emission $T_{\rm Gal} (t, \nu)$ is derived from {\tt PySM}
(\citealt{2017MNRAS.469.2821T}), and the CMB emission is set to be a uniform $\overline T_{\rm CMB} = 2.725$ K \citep{2009ApJ...707..916F}. We assume these models to be fixed during the calibration process, with no free fitting parameters, but will return to the impact of this assumption in \secref{sec:characteristics} and \ref{sec:comparison}. See \autoref{fig:models} for an example frequency-time (waterfall) plot of the Galactic emission.

Note that we use the {\tt s1} model from {\tt PySM}, which is based on the Haslam 408 MHz total intensity map (\citealt{2015MNRAS.451.4311R}; $56^\prime \pm 1^{\prime}$ resolution) on large scales, but which has added structure generated as a simple log-normal random field on small scales. This map is used at a resolution of {\tt nside=64}, corresponding to $\sim 55^\prime$ pixels. We do not apply a convolution with the MeerKAT beam to this map, instead relying on the limited map resolution to prevent the residuals from being contaminated with the synthetic small-scale structure. We have checked that the differences between the PySM model and Haslam map do not impact the calibration solutions (the effect on the gain is about 0.01\%).

\subsection{Elevation-dependent terrestrial emission}

Elevation-dependent terrestrial emission includes the atmospheric emission and spillover from the ground (i.e., ground pickup),
\begin{gather}
T_{\rm el}(t, \nu)=T_{\rm atm}(t,\nu)+T_{\rm spill} (t,\nu) \, .
\end{gather}
A spillover model for the MeerKAT telescope has been provided by EMSS Antennas\footnote{\url{https://www.emssantennas.com/}} at several frequencies and elevations. We apply a 2D interpolation of their results to obtain $T_{\rm spill}(t,\nu)$.
The contribution of atmospheric emission to the receiver system temperature can be modelled as
\begin{gather}
  T_{\rm atm}(t,\nu) = T^0_{\rm atm} (t) \left [ 1 - e^{-\tau\, {\rm cosec}(el)} \right ]\, ,
\end{gather}
where $T^0_{\rm atm} (t) = 1.12\, T_{\rm st} (t) - 50$ is the atmospheric temperature related to the temperature measured at the MeerKAT weather station $T_{\rm st}$ in K, ${\rm cosec}(el)$ is the relative airmass along the direction of the beam pointing as a function of the elevation angle $el$
and $\tau$ is the opacity of the atmosphere at zenith, which is related to the atmospheric temperature, 
air relative humidity, air pressure, telescope height above sea level, and frequency\footnote{See International Telecommunication Union Recommendation ITU-R P.676-9 for more details.}.
In contrast to the atmospheric emission which is applied to  $T^0_{\rm atm}\sim 270$ K, we ignore the atmospheric absorption since the transmission factor is very close to unity and will be multiplying temperatures of only a few K. 

\subsection{Calibration model fitting method}\label{sec:fitting}

\begin{figure*}
\centering
\includegraphics[width=2\columnwidth]{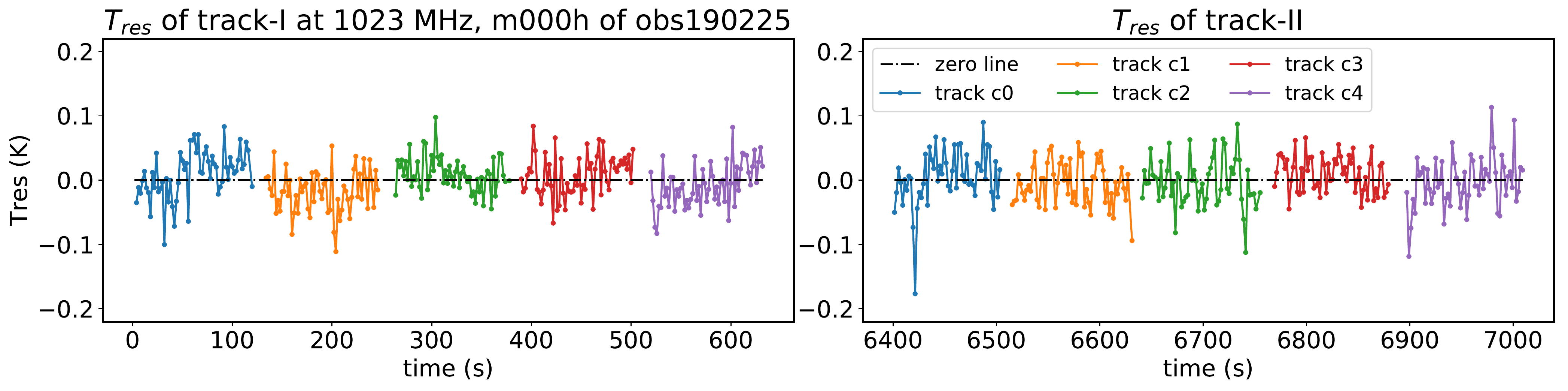}
\caption{The residual between the observed signal and best-fitting model (\autoref{eq:T_resi}) for the calibration source tracking data, in this case several pointings around 3C273. The colours denote the different pointings; either directly on source ({\tt c0}) or offset by $\sim 0.5$ deg ({\tt c1..4}) to probe different parts of the beam and surrounding diffuse emission; c.f. \autoref{fig:beam}.}
\label{fig:cali_track}
\end{figure*} 

We take a two-step approach to the calibration process.
The first step is to use the flux model of a strong point source during tracking observations to calibrate the bandpass and absolute flux, as well as the injection power of the noise diode. The second step then uses the noise diode injections to calibrate the time ordered data during the constant-elevation scanning phase. This is done per frequency, per dish, per polarisation and for each individual scan.

\subsubsection{Overall bandpass and absolute flux scale calibration}

We start by fitting the model in \autoref{equ:model} and \ref{equ:model2} to the raw auto-correlation signal from the tracking data around the calibrator point source. This is done for each polarisation (HH and VV) at each frequency.
Since we expect the gain to be smooth in time, we parameterise it as
\begin{gather}\label{eq:gain_par}
g(t)=\sum_{n=0}^{4} a_n P_n(x);~~~~~~ x=\frac{2(t-t_{\min})}{t_{\max}-t_{\min}}-1,
\end{gather}
where
$P_n(x)$ are the Legendre polynomials. We choose this model on the basis that the coefficients of an orthogonal polynomial basis are less likely to be strongly correlated than that of an ordinary polynomial, and thus are easier to handle with numerical fitting techniques \citep[c.f.][]{bradley1979correlation, shacham1997minimizing, tian1998comparison}.
We checked that the order of the polynomial was sufficient to represent the data; choosing higher orders produced noise-dominated coefficients.
The parameters to be fitted at each frequency are then $\{a_n\}$ for the gain function, plus $\overline T_{\rm diode}$, $T_{\rm rec}$, and $f_{\rm d}$.
We fit the parameters by maximizing \autoref{eq:fit} and setting the weights to $w(t_i)=1$.
For the calibration of the tracking parts of the TOD, we use the catalogue value of the flux of the calibrator point source, the beam model, and the difference of the raw signals between the centre and 4 outskirt pointings (e.g., see \autoref{fig:beam}) to get an approximate gain level, which we then use as a starting point for $a_0$. 
 
The hyperparameters of the Gaussian priors from \autoref{eq:fit}, $(\bar{p}_j, \sigma_j)$, are
  \begin{gather}
           \overline T_{\rm diode}(\nu) \sim \left( {T_{\rm diode}^{\rm ref} (v)},\ \sigma_{\rm diode} \right),\\ \notag
            T_{\rm rec} (\nu) \sim \left(T_{\rm rec}^{\rm ref} (\nu),\ 0.5T_{\rm rec}^{\rm ref}(\nu)\right),
      \end{gather}
where the reference values $T_{\rm rec}^{\rm ref}$ and $T_{\rm diode}^{\rm ref}$ have been previously measured for MeerKAT (see \secref{sec:diode}), $\sigma_{\rm diode}$ is the measured standard deviation of $T_{\rm diode}^{\rm ref}$ across the full frequency band and we use $0.5T_{\rm rec}^{\rm ref}$ for the standard deviation of the receiver temperature as a conservative prior.
\autoref{fig:cali_track} 
shows the fitting results for two tracking observations, before and after the scan.
In this example the fitted $\overline T_{\rm diode}$ are 20.08 K and 20.15 K at 1023 MHz for the two tracks, respectively. So later we use their mean value $\overline T_{\rm diode}=20.12 ~{\rm K}$ for the next step in the calibration. The rms of the residuals is below the $1\%$ level of the total temperature.

\subsubsection{Noise diode-based calibration of the scanning data}

During the constant elevation scans, there is no bright calibrator source in the field of view, and so \autoref{equ:model} becomes
\begin{gather}\label{equ:model_scan}
  T_{\rm model}(t, \nu)= T_{\rm diffuse} + T_{\rm el} 
    + T_{\rm diode}  + T_{\rm rec},
\end{gather}
where the terms on the right-hand side implicitly depend on $t$ and $\nu$. Given the longer duration of this data, in this case we allow the receiver temperature $T_{\rm rec}$ to depend on time, parameterising it using Legendre polynomials as we did previously for the gains.
The parameters to be fitted are then
\begin{gather} \label{eq:para2}
\begin{array}{ll}
\{a_n\}, &\text{for } g(t)=\sum_{n=0}^{4} a_n P_n(x),\\
\{b_n\}, &\text{for } T_{\rm rec}(t)=\sum_{n=0}^{3} b_n P_n(x),\\
f_d, &\text{constant number,}
\end{array}
\end{gather}
where $x$ was defined in \autoref{eq:gain_par}. 
We again checked that the order of the polynomials was sufficient to represent the data, with higher order choices returning noise-dominated coefficient values.

\begin{figure}
\centering
\includegraphics[width=.95\columnwidth]{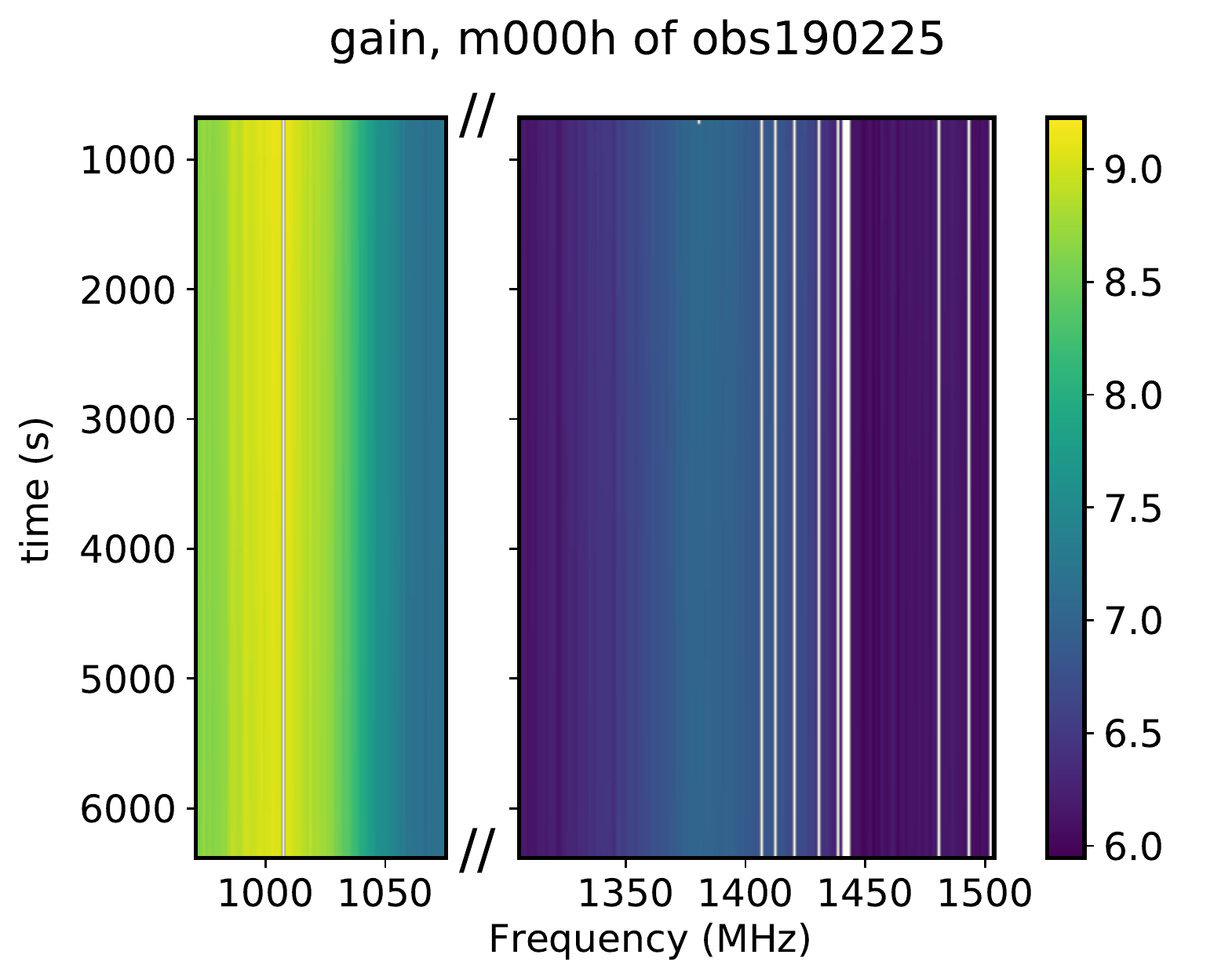}
\includegraphics[width=.95\columnwidth]{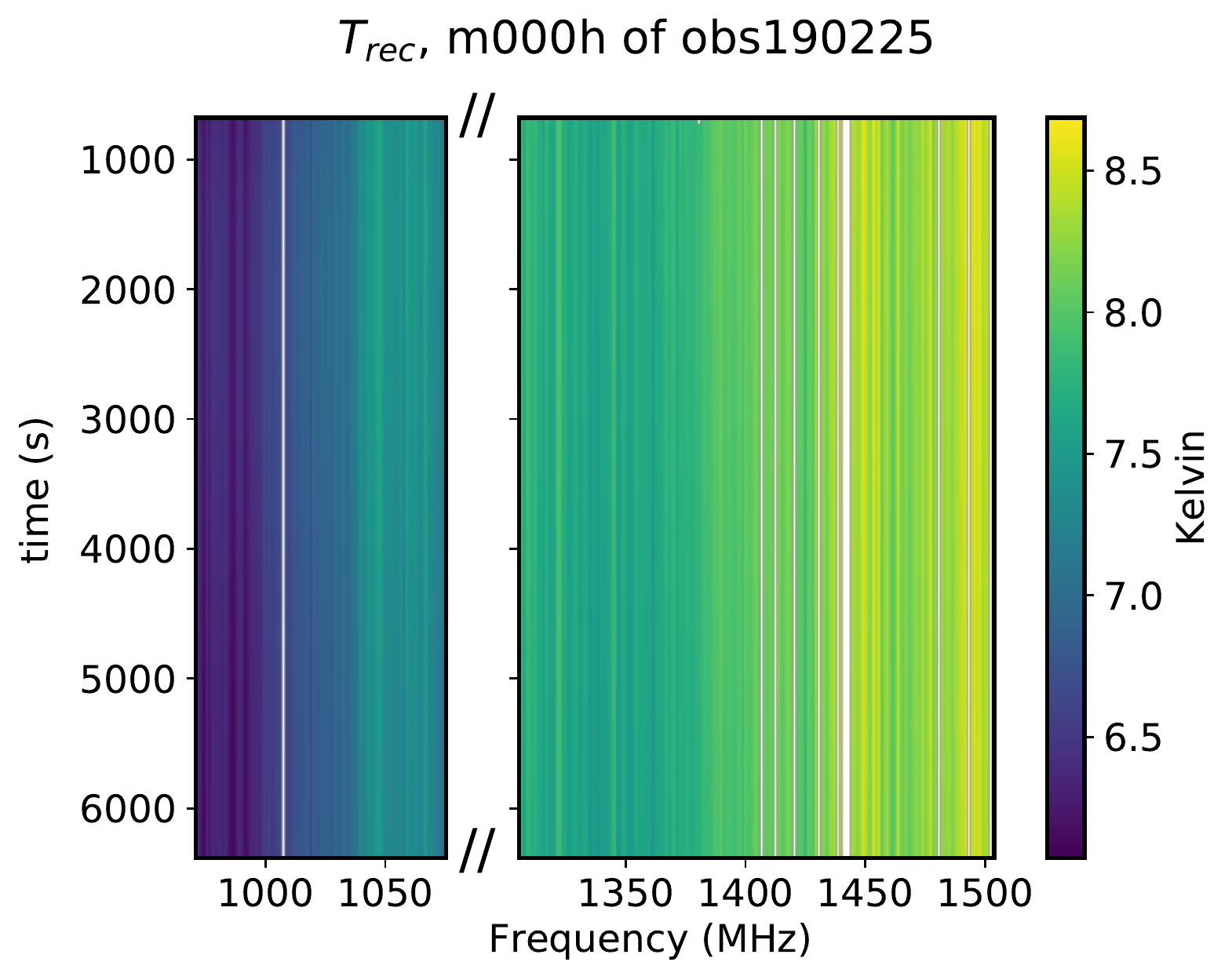}
\caption{Best-fit Legendre polynomial models for the time-dependent gain and $T_{\rm rec}$, using dish {\tt m000h} and {\tt obs190225} as an example. Note the smoothness of the solutions in time, with considerably more variation in the frequency direction (where smoothness was not enforced).}
\label{fig:gain_Trec}
\end{figure}

\begin{figure*}
\centering
\includegraphics[width=2\columnwidth]{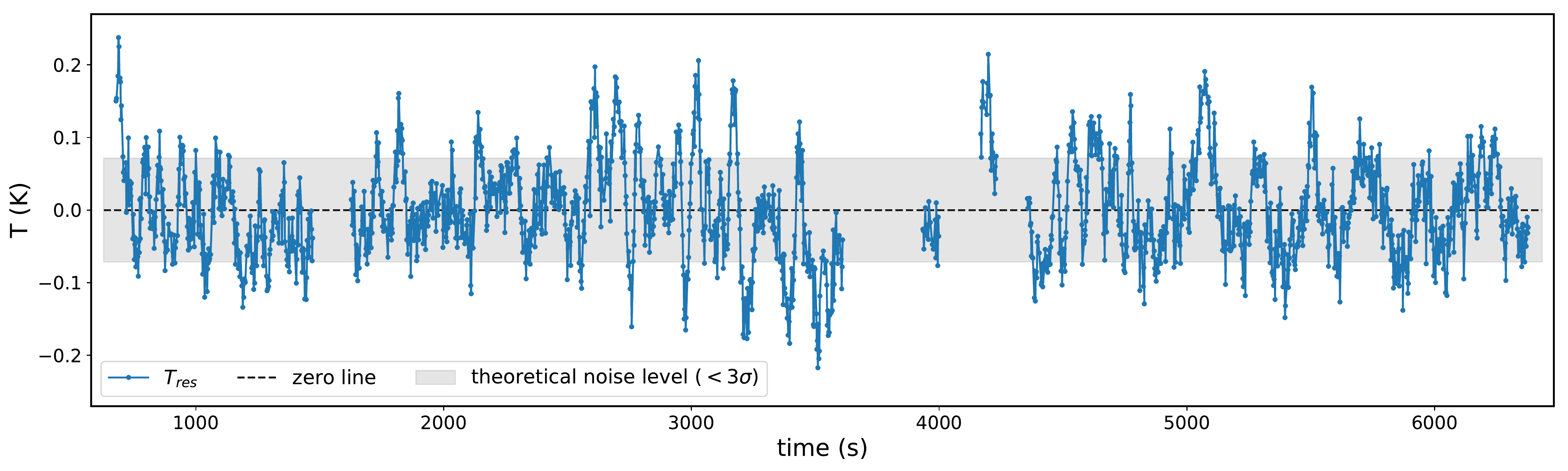}
\caption{The residual between the observed data and fitting model in a single channel (1023 MHz) as a function of time, using receiver {\tt m000h} and {\tt obs190225} as an example.}
\label{fig:cali_scan}
\end{figure*} 

\begin{figure*}
\centering
\includegraphics[width=2\columnwidth]{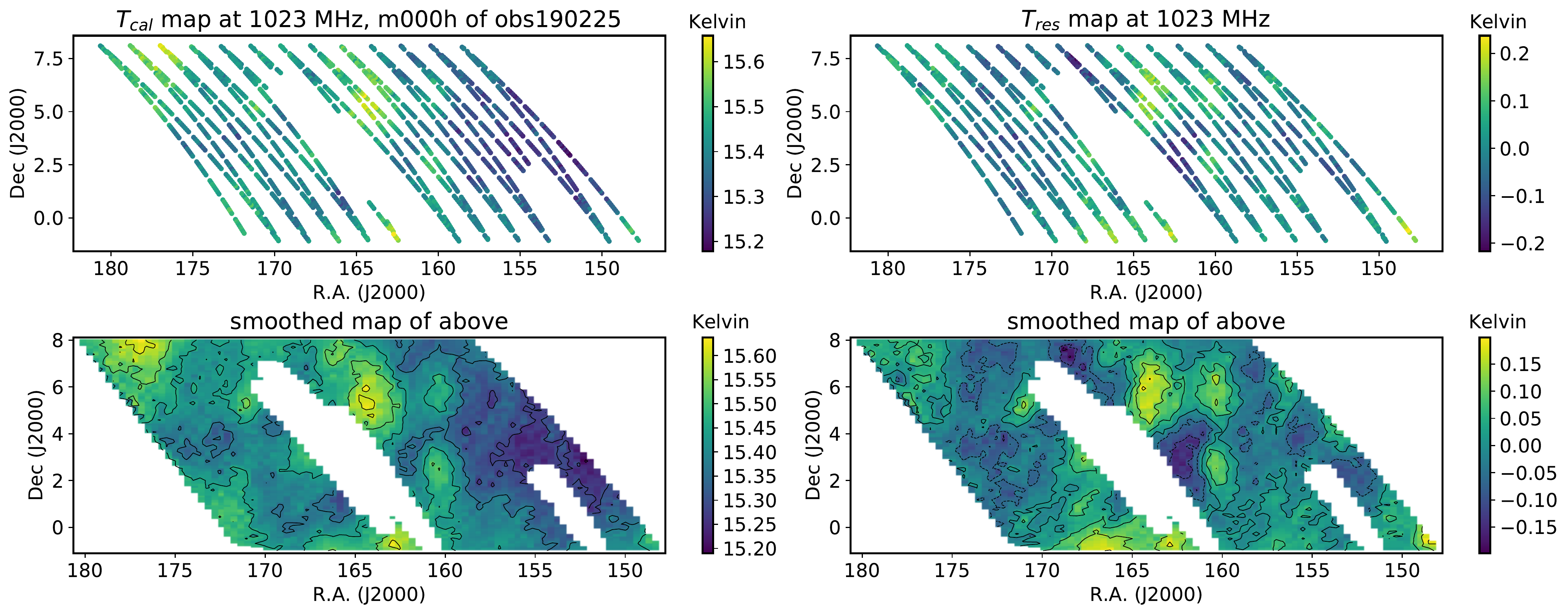}
\caption{The calibrated data (upper panel) and interpolation maps (lower panel) for the total temperature (left) and residual (right), for a single frequency channel at 1023 MHz, receiver {\tt m000h}, and observation {\tt obs190225}. }
\label{fig:cali_map}
\end{figure*} 

For the scan part, the gain is initially set to be $a_0=10$, which is close to the typical gain during the tracking phase. 
In this case, the hyperparameters $(\bar{p}_j, \sigma_j)$ for the prior are
\begin{gather}
     \overline T_{\rm rec} (\nu) \sim \left(T_{\rm rec}^{\rm ref} (\nu),\ 0.5 T_{\rm rec}^{\rm ref}(\nu)\right),
      \end{gather}
where $\overline T_{\rm rec} (\nu)$ is the average of the receiver temperature along time.
Again, we fit the parameters by solving \autoref{eq:fit}, except in this case we set the weights to $w(t_i)=1$ for the times when the noise diode is off and $w(t_i)=4$ when the diode is on (both $t_{\rm nd}^{1a}$ and $t_{\rm nd}^{1b}$ samples). This accounts for the fact that the noise diodes should be our main calibrators in this part of the timestream.

We repeated this approach for all frequency channels, receivers, and observation blocks independently. \autoref{fig:gain_Trec} shows an example of the fitted gain and $T_{\rm rec}$ solutions as a function of frequency and time for a single receiver and observation block. While the time variations are smooth, as guaranteed by our polynomial model, there is some variation in the solutions between neighbouring frequency channels, most noticeably for $T_{\rm rec}$. Smooth overall trends in the frequency direction are also observed, suggesting that a low-order polynomial parametrisation in frequency could also be used to reduce the number of degrees of freedom in the calibration model, at the expense of requiring the observations to be fit in frequency and time simultaneously.

An example of the residual, $T_{\rm res}$, from the TOD model fit for a single frequency channel is shown in \autoref{fig:cali_scan}. The calibrated signal and best-fit model agree well, typically to within $\sim 0.15$ K for the entirety of the scan, although some structure is visible in the residuals. This is most noticeable from $\sim 5000$ seconds onwards in this example, where oscillations on timescale of order 100 s can be seen. This is comparable to the duration of a single scan (around 200 sec).

We can study the structure of the residuals in more detail using \autoref{fig:cali_map}, which shows the calibrated data and its residual for the same frequency and receiver. Several of the brighter structures in the residuals are seen to be correlated with features in the calibrated map, suggesting a missing component in our sky model. This is indeed found to be the case: as we will discuss in subsequent sections, much of this structure can be explained by relatively bright ($> 1$ Jy) point sources that are known to exist in this field, but which were not included in our sky model. We investigate including point sources in \autoref{equ:model_scan} later, by using the final combined $T_{\rm sky}$ (see details  in \secref{sec:final_map}) to replace the $T_{\rm diffuse}$ model. It results in changes of about $0.01\%$ on the gain map and $0.05\%$ on the $T_{\rm sky}$ map. The results confirm that the unmodeled point sources have a very small impact on the calibration.

\section{Combined maps and data cubes} \label{sec:mapping}

In this section we present our methods for combining the calibrated data from all dishes, scans, and polarisations to form a final set of combined pixelised maps as a function of frequency. The increased depth of the data necessitates additional RFI flagging steps to remove outlying frequency channels and weak RFI features in individual stripes.

\subsection{Per-channel outlier removal for RFI (Round 2)} \label{sec:to_intensity}

The calibration process is done per frequency channel and per polarisation, ignoring any potential correlation between the calibration of different channels. However, the resulting solutions ($T_{\rm cal}(t, \nu)$, $T_{\rm res}(t, \nu)$, $T_{\rm rec} (t, \nu)$ and $g(t, \nu)$)) are expected to be relatively smooth in frequency. Despite this, contaminated data may still be present, even after the previous round of RFI flagging, and so we apply a per-channel filter to identify and mask such bad channels.  To do so, we use the sigma clipping method \citep{2013A&A...558A..33A}. For each dish and each observation, we apply spline fits to the time averages of $\overline T_{\rm cal}(\nu)$, $\overline T_{\rm res}(\nu)$, $\overline T_{\rm rec} (\nu)$, and $\bar g(\nu)$. We then flag any outlier frequency channels present in any of the quantities, for all times and both polarisations. Outliers are defined as channels in which any of the time-averaged quantities deviate from the spline fit by $>4\sigma$, where $\sigma$ is defined as  the sample standard deviations from the median. We perform the spline fitting and flagging process iteratively, running up to six iterations to make sure the final result is stable (although in most cases a stable result is attained after one or two iterations).

\subsection{From polarisation to intensity} \label{sec:pol2I}
So far all the analysis has been done per polarisation (HH and VV). We have calibrated and flagged TOD data for the total measured temperature $T_{\rm cal} (t,\nu)$ using \autoref{eq:T} and for the residual temperature $T_{\rm res}(t,\nu)$ using \autoref{eq:T_resi}. The calibrated sky temperature (per polarisation) is obtained by removing the telescope, ground and atmospheric contributions to the total signal:
\begin{gather} \label{eq:Tsky}
T_{\rm sky}(t,\nu)\equiv T_{\rm cal}(t, \nu) - T_{\rm el}(t, \nu) - T_{\rm rec} (t, \nu). 
\end{gather}
We can then construct equivalent TOD data arrays of total intensity (corresponding to Stokes I), as the mean value of the two calibrated polarisation temperatures, HH and VV, 
\begin{gather} 
T(t, \nu) = \frac {1} {2} \big ( T_{\rm HH}(t, \nu) + T_{\rm VV} (t, \nu)\big ). 
\end{gather}
We note that this is a pseudo-stokes I (or temperature $T$) as it does not correct for the primary beam shape and polarisation leakage. The polarisation contribution to stokes I is negligible at the current sensitivity level since the telescope leakage is below the 10\% level and the galactic polarisation is also around 10\% of the intensity. Correction for the beam asymmetries can be relevant in the presence of strong point sources (like we do for the calibrator). For diffuse emission, it should be enough to account for the beam smoothing using a symmetric beam model (see, e.g., \citealt{2020arXiv201110815M}).

\subsection{Map pixelisation} \label{sec:pixel}

The intensity maps from different blocks need to be added together to get a better sky coverage and higher signal-to-noise (S/N) ratio (this can be done for Stokes I, while Stokes Q, U and V, would require to compensate for the sky rotation before averaging). Since different scans have different pointings (ra, dec), the intensity maps need to be pixelised first and then joined together in the same coordinates system. We chose the Zenith Equal Area (ZEA) map projection method \citep{2013A&A...558A..33A} to get the coordinates of each TOD point on a pixel grid, denoted $({\rm ra_{pix}}, {\rm dec_{pix}})$, 
with a pixel size of $0.3$ deg chosen to be of about 1/3 of the beam size. We do the pixelisation per scan and per dish. The relation between time and pixel is not necessarily one to one given the pixel size.

\begin{figure}
\centering
\includegraphics[width=\columnwidth]{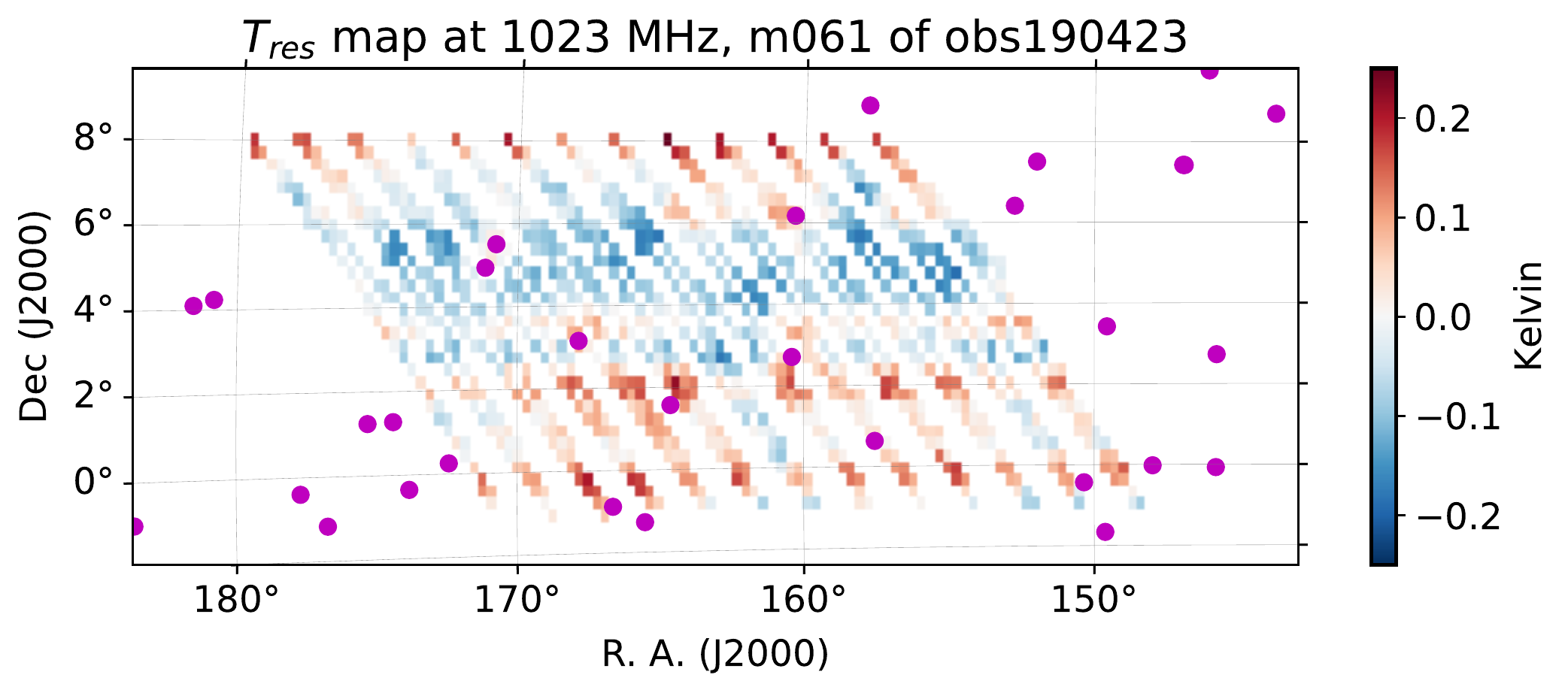}
\caption{Example of a pixelised intensity map with weak RFI contamination that would be removed by the Round 3 RFI filter. These data are for single scan at 1023 MHz, 
for dish {\tt m061} and {\tt obs190423}. Magenta dots mark the positions of point sources with flux $>1$ Jy at 1.4 GHz.}
\label{fig:check_weak_rfi_map}
\end{figure}

The final maps combining all scans and dishes, $\overline T_{\rm sky}({\rm ra_{pix}}, {\rm dec_{pix}},\nu)$, 
were created by performing a simple average of all unflagged data points that fell in each pixel,\footnote{Inverse $T^2_{\rm sys}$-weighted maps were also generated for testing purposes, but differed from the simple average maps by $\lesssim 5$ mK per pixel; a more detailed treatment of map making is deferred until later work \citep{2001A&A...374..358D,2001A&A...372..346N}.}  
\begin{gather} 
\overline T_{\rm sky}({\rm ra_{pix}}, {\rm dec_{pix}}, \nu) = \frac{1}{N_{\rm count}} \sum_{i=1}^{N_{\rm count}} T_i({\rm ra_{pix}}, {\rm dec_{pix}}, \nu),
\end{gather}
where $N_{\rm count}$ is the number of unflagged data points per pixel and per frequency channel, with $N_{\rm count, max}=366$ in the best case scenario. The final residual maps were created by applying the same averaging procedure to the individual residual maps. Note that in principle we would have up to $64\times7=448$ independent scans in the data, but this has not been the case due to malfunctioning dishes during the observation blocks. 

\subsection{Removal of low-level RFI features in the maps (Round 3)}\label{sec:weakRFI}

While for some astronomical observations one can safely ignore low levels of residual RFI, the faintness of the \textsc{Hi} IM clustering signal means further RFI removal is generally required. After the first rounds of RFI removal and calibration, some relatively faint line-shaped structures are detectable in maps across relatively broad frequency bands (see \autoref{fig:check_weak_rfi_map} for an example). We believe these features to be caused by geostationary satellites, e.g., due to heavily suppressed spillover of emission outside of their main transmitting band \citep{2018MNRAS.479.2024H}. 
The previous rounds of flagging likely missed most of this emission because it is faint in comparison to other RFI sources, and because the spatial coherence of the signal on the sky did not produce as obvious a pattern in the timestream domain.

\begin{figure}
\centering
\includegraphics[width=\columnwidth]{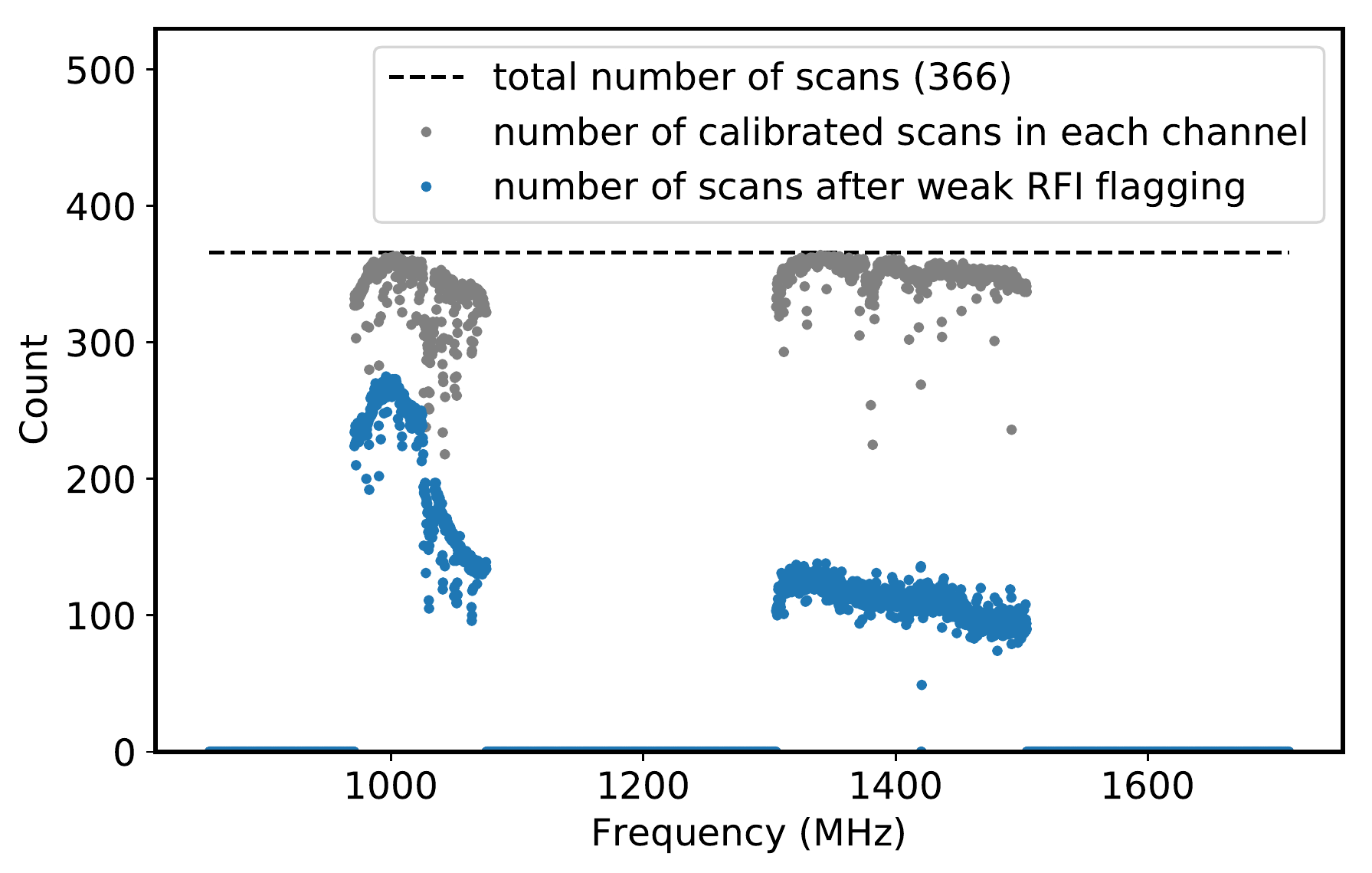}
\caption{The total number of scans available before (grey) and after (blue) the Round 3 weak RFI flagging. The higher frequency band is particularly badly affected by low-level broadband RFI.}
\label{fig:ch_counts}
\end{figure}

\begin{figure*}
\centering
\includegraphics[width=1.9\columnwidth]{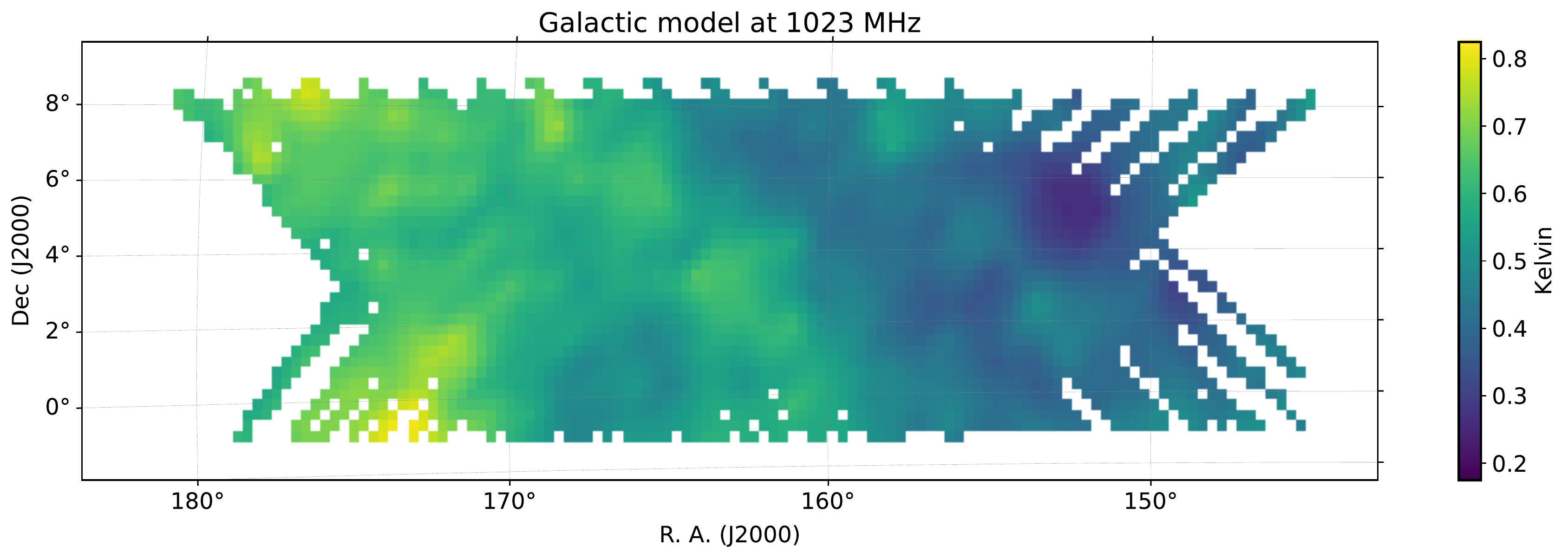}
\includegraphics[width=1.9\columnwidth]{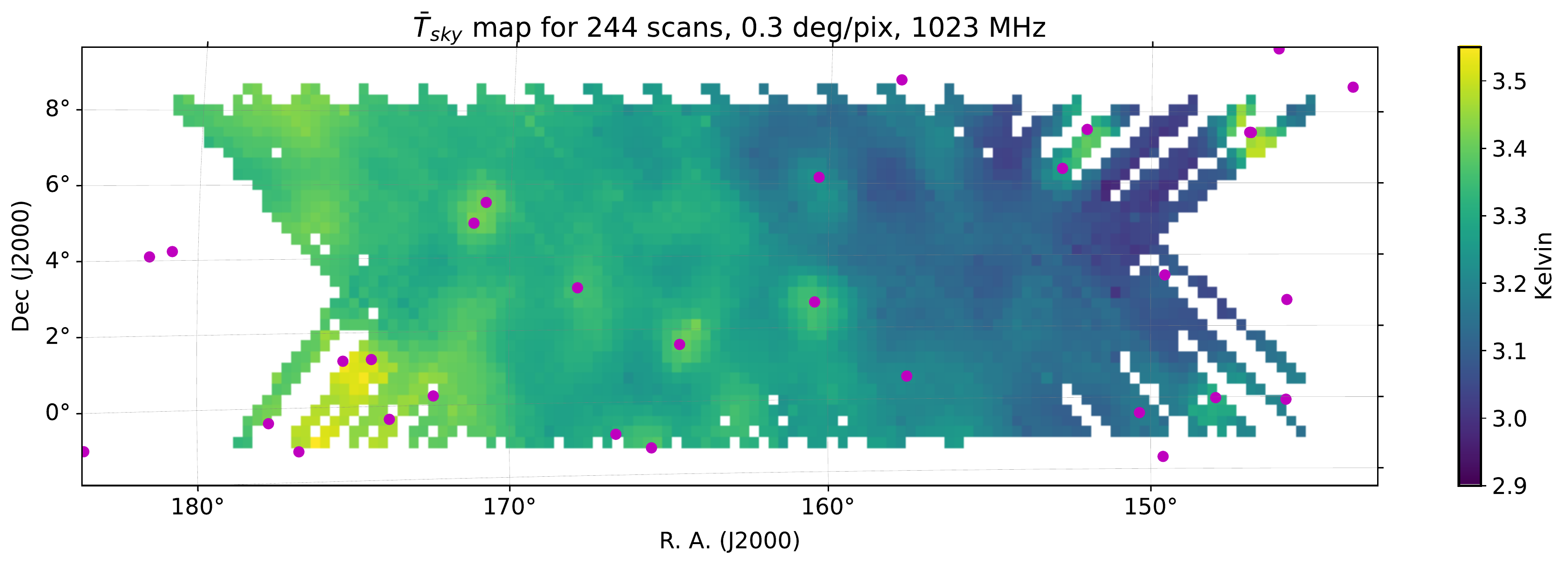}
\includegraphics[width=1.9\columnwidth]{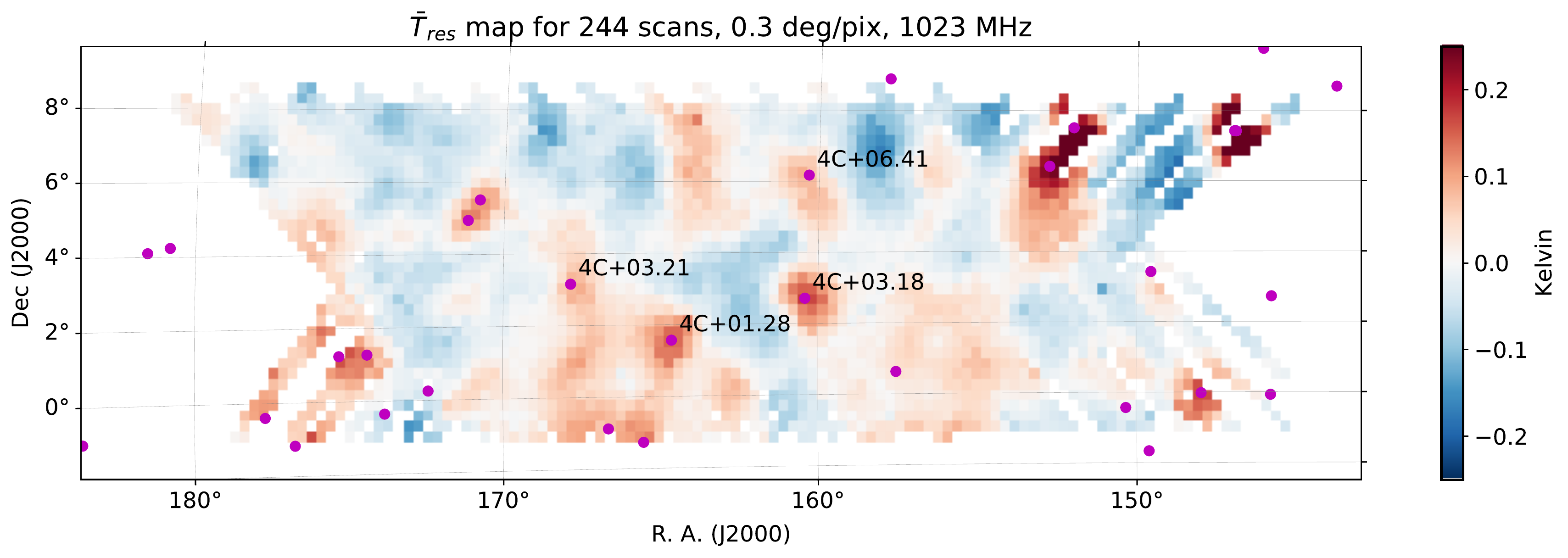}
\includegraphics[width=1.9\columnwidth]{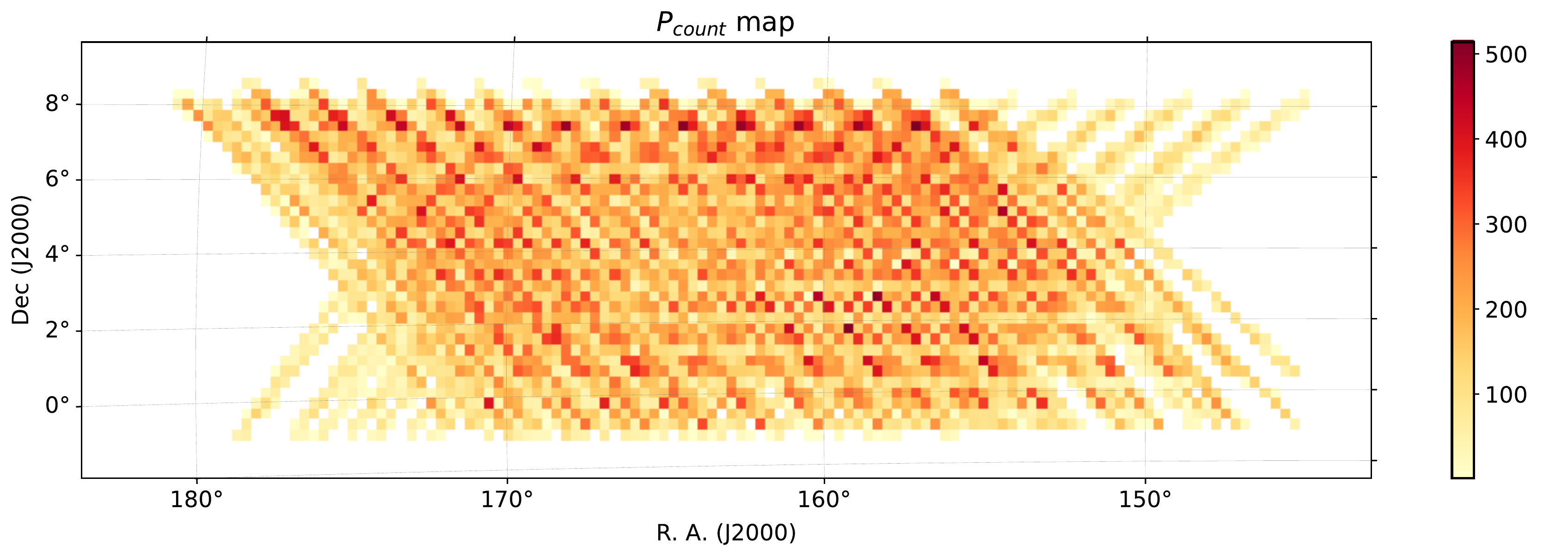}
\caption{The mean intensity maps at 1023 MHz from the combination of all scans after Round 3 RFI subtraction. The upper panel shows the model of diffuse Galactic emission derived from PySM with the same sky coverage as our data. The middle panels show the sky map and residuals that we obtained. The lower panel shows the total number of time samples (across all dishes and observations) that were combined in each pixel. Magenta dots mark the positions of point sources with flux $>1$ Jy at 1.4 GHz, which were not included in the sky model used for calibration.}
\label{fig:sky_map}
\end{figure*}

We therefore apply another round of RFI flagging to remove these features, taking advantage of the characteristic line structures left behind in the residual maps. As seen in  \autoref{fig:check_weak_rfi_map}, these structures occur at a fixed declination. 

Our flagging method proceeds as follows:
\begin{enumerate}
\item Point sources with flux $> 1$ Jy in the total combined residual maps, $\overline T_{\rm res}$, are temporarily masked (see the magenta points in \autoref{fig:check_weak_rfi_map}). Neighbouring pixels within 1 degree are also masked. 
\item For each frequency channel and declination, we calculate the mean and standard deviation of the total combined residual maps along the RA direction. 
\item Similar to the first step, we temporarily mask the point sources on the single scan maps (per dish), $T_{\rm res}$, and calculate the median value along the RA direction for each frequency and declination.
 \item Finally, we check whether the median at each scan, frequency, and declination is within 2.5 standard deviations of the mean of the total combined residual maps, and flag the whole scan in that frequency channel if any points exceed this threshold.
\end{enumerate}
We iterate this process three times. The $2.5\sigma$ threshold was chosen as it gave good results when comparing with affected scans by eye. The decision to flag the entire scan is aggressive, but reduces the chance that slightly weaker RFI at neighbouring declinations remains unflagged.
The result is a significant amount of additional flagging that particularly affects the higher-frequency sub-band, as shown in \autoref{fig:ch_counts}. 

\subsection{Final maps} \label{sec:final_map}

Our final output is the combined calibrated temperature maps for all scans and all dishes at all frequencies, after the three rounds of RFI flagging discussed previously. We show the final sky temperature map at 1023 MHz in \autoref{fig:sky_map}, along with the residual map for all observations combined, and the PySM diffuse Galactic emission model with the same sky coverage. Much of the diffuse Galactic structure has been recovered on large scales, and the contribution of the bright point sources that were not included in our calibration model is seen clearly in the residual in most instances. We use the recovered flux and frequency spectra of these point sources to evaluate the quality of our final calibrated maps in \secref{sec:ptr_spec}.

In comparing \autoref{fig:sky_map} with the map and residuals before Round 3 (weak RFI) flagging, we found little difference, which suggests that the weak RFI was not a large contribution in comparison with the dominant foreground signals. This is expected -- if it was brighter, this kind of RFI would likely have been flagged in previous rounds. This is encouraging if these data are to be used for, e.g., studying galactic foregrounds, as the effect of residual weak RFI does not seem to be significant. The efficacy of this round of flagging in helping to detect much fainter signals (such as the 21cm fluctuations) is less clear however.

Note that the residual map should in principle contain a series of other unmodelled components as well as the bright point sources, including noise, small-scale Galactic structures that are not included in PySM, the cosmological signal (below the noise level here), and other extragalactic emission, i.e., confusion noise. Calibration model errors and unmodelled instrumental artifacts will also contribute to the residuals. In subsequent sections, we will therefore use the properties of the residual maps as a way to check the quality of the data and our calibration solutions.
Further analysis should be applied to the full calibrated data, either $\overline T_{\rm sky}$ or $\overline T_{\rm cal}$, however, rather than $\overline T_{\rm res}$. This is because there is a risk of over-subtraction of sky components and therefore signal loss if the residuals are used directly.

\section{Characteristics of autocorrelation data} \label{sec:characteristics}

In this section we study the properties of the autocorrelation data, particularly with respect to the accuracy of the calibration and the effectiveness of the RFI flagging. Towards the end of this section, we also plot the per-dish sky maps to show the variations in data quality between dishes, and then calculate the principal components of the whole dataset as a way of demonstrating the correlation structure and dynamic range of the data.

\subsection{Gain solutions and bandpass shape}
It is instructive to study the structure of the gain solutions,  $g(t,\nu)$, that are obtained by the calibration pipeline. The ideal for an intensity mapping experiment is to find solutions that are smoothly varying in time and frequency, as in this case it is less likely that sky signal on the scales of interest will be erroneously absorbed into the gains, leading to a spurious modulation of the actual cosmological signal. Smooth variations in time are also indicative of the stability of the instrument. We also hope to find gain solutions that are independent of the sky; gains arise from purely instrumental effects, and so solutions that are appreciably correlated with the sky suggest the presence of errors in the sky model for example. 

In terms of time stability of the gains, first recall the gain solutions that were plotted in \autoref{fig:gain_Trec} 
for scan {\tt m000h} of {\tt obs190225}. The gain along time is rather flat, which means the gain does not change too much during the 1.5 hour observation. This is representative of the gain solutions we obtained for all observations and receivers. 

\begin{figure}
\centering
\includegraphics[width=\columnwidth]{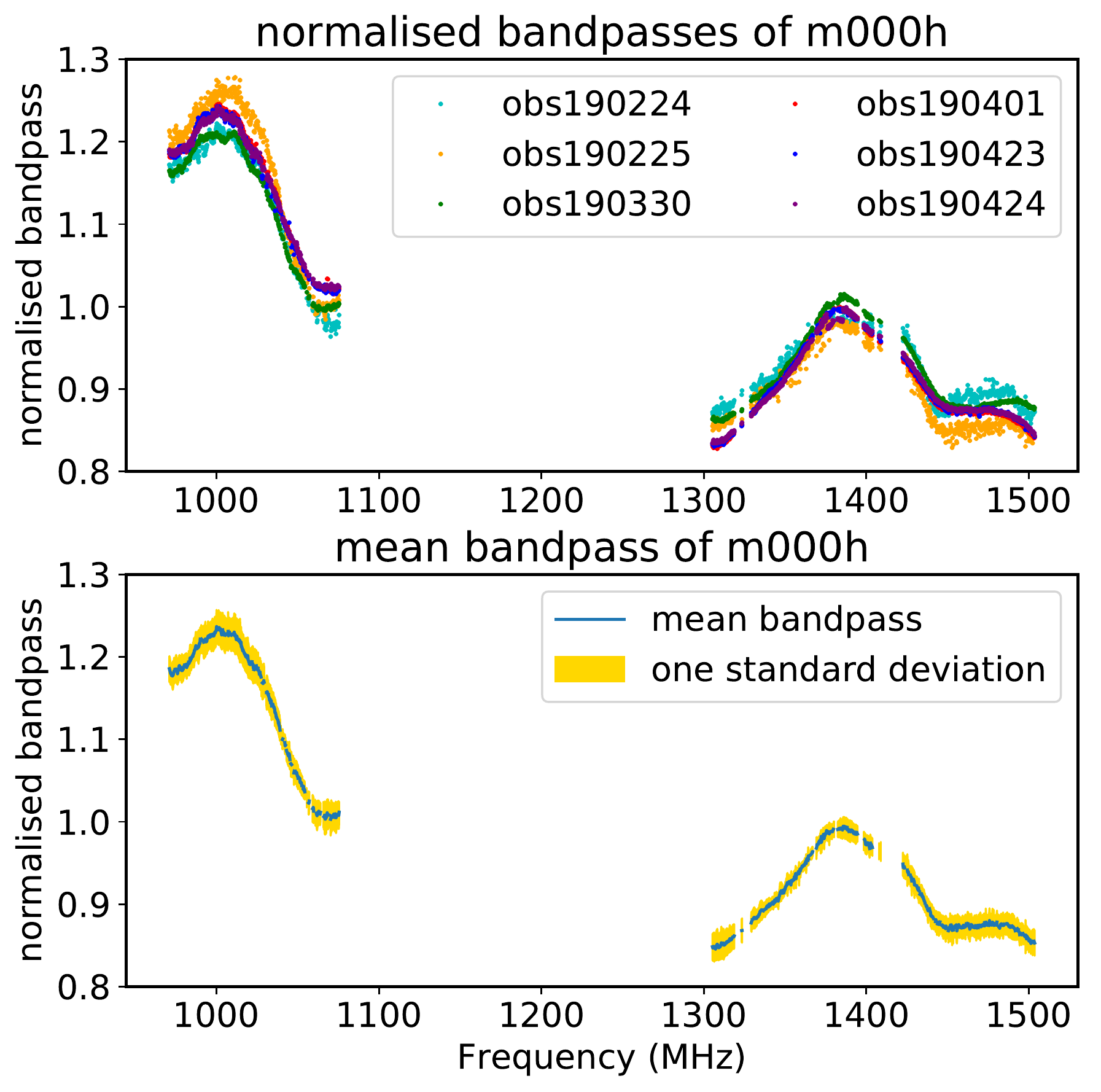}
\caption{Upper panel: The normalised bandpasses of the {\tt m000h} receiver for six observations, taken over a period of several months. Lower panel: The mean bandpass for all observations. The estimated standard deviation is a few percent or less across the entire band.}
\label{fig:gv}
\end{figure}

 Next, we calculated the time-averaged mean gain against frequency, $g(\nu) = \langle g(\nu, t)\rangle_t$, also known as the bandpass. The bandpass shape is expected to be stable for each receiver. To check this, we compared the bandpass from the same MeerKAT receiver but different observations, normalising each bandpass using its mean value (since the absolute bandpass values also incorporate the effects of gain drifts in time). \autoref{fig:gv} shows the resulting bandpass for receiver {\tt m000h} over several days of observations as an example. While there are variations from observation to observation, these are limited to $<10\%$, and there is good overall consistency in the bandpass shape across all observations. 

\subsection{RFI statistics}\label{sec:rfi_stat}

\begin{table}
\centering
 \caption{Percentage of discarded data, out of the initial ones, in each observation for both the lower and higher frequency band. The three RFI-flagging rounds - i.e., strong RFI flagging (R1), channel filtering (R2) and map based weak RFI identification (R3) - are reported separately to show the progressive identification of contamination in the data.   } 
 \label{tab:RFI}
 \begin{tabular}{c|c|c|c|c|c|c}
  \hline
  \hline
 & \multicolumn{3}{|c|}{{\rm Lower band [\%]}} &  \multicolumn{3}{|c|}{{\rm Higher band [\%]}} \\
 \hline
Block ID & R1  & R2 & R3 & R1 & R2  & R3 \\
\hline
{\tt obs190224} &  7.9 & 23.3 & 56.5 & 7.9  & 13.3  & 80.7 \\
{\tt obs190225} &  19.2 &24.5& 33.0& 19.8& 22.9& 56.9 \\
{\tt obs190330} &  18.1 & 19.0 & 30.2 & 18.0 & 18.7 & 99.9 \\
{\tt obs190401} & 4.6 & 8.6 & 81.7 & 4.9 & 5.2 & 60.4 \\
{\tt obs190423} &  6.5 & 29.4 & 71.5 & 6.8 & 31.8 & 91.6 \\
{\tt obs190424} &  10.3 & 13.1 & 72.5 & 10.6 & 10.9 & 94.5 \\
{\tt obs190711} &  26.8 & 30.8 & 32.1 & 23.4 & 28.8 & 39.4 \\
\hline
\end{tabular}
\end{table}

In our analysis, RFI flagging has been performed as a multi-step process starting from the masking of the strongest contamination, 
to a pre-channel outlier removal 
and finally a map-based cleaning of residual weak RFI sources.
In this section we discuss the fraction of data that have been discarded through these RFI flagging processes.

\begin{figure}
\centering
\includegraphics[width=\columnwidth]{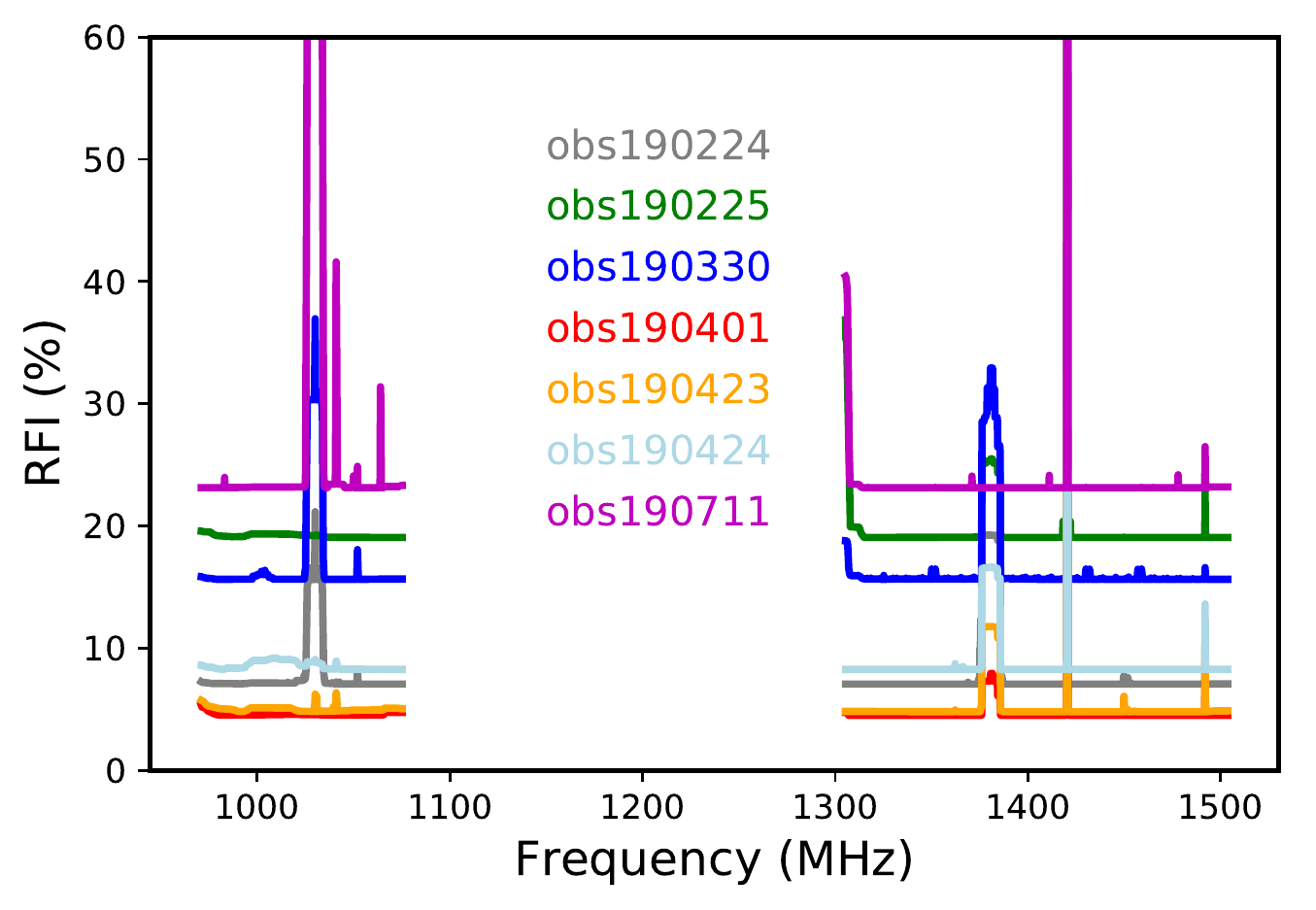}
\caption{The Round 1 RFI percentage flagged in the data as a function of frequency for the low and high band of interest. Different colors label the different observations.}
\label{fig:RFIstat_allscans}
\end{figure}

Globally, strong RFI flagging (i.e., Round 1 described in \secref{sec:rfi}) varies from a total of  $\sim 40$\% to $\sim 60$\% across observations. Focusing on the two target frequency bands, where most of RFI from satellite is avoided, flagging is below $\sim 25$\%, reaching values down to $\sim 5$\% for the best observations. 
We show in \autoref{fig:RFIstat_allscans} the fraction of RFI flagged data as a function of frequency for each observation. 

 We report in \autoref{tab:RFI} the fraction of flagged data with respect to the initial data, for Round 1 and for the subsequent RFI flagging rounds. 
 The flagging is computed over time and frequency, considering either the lower band (i.e., 971--1075 MHz, channels 550--1050) or the higher band (i.e., 1305--1504 MHz, channels 2150-3100).
 Channel filtering (i.e., Round 2 described in \secref{sec:to_intensity}) increases the flagged fraction from about $+ 1$\% to $+ 20$\% in the lower frequency band, with similar values for the higher frequency band. 
 Map based weak RFI removal (i.e., Round 3 described \secref{sec:weakRFI}) increases further the flagging up to very high percentage, in particular for the higher frequency band. 
 
\begin{figure}
\centering
\includegraphics[width=\columnwidth]{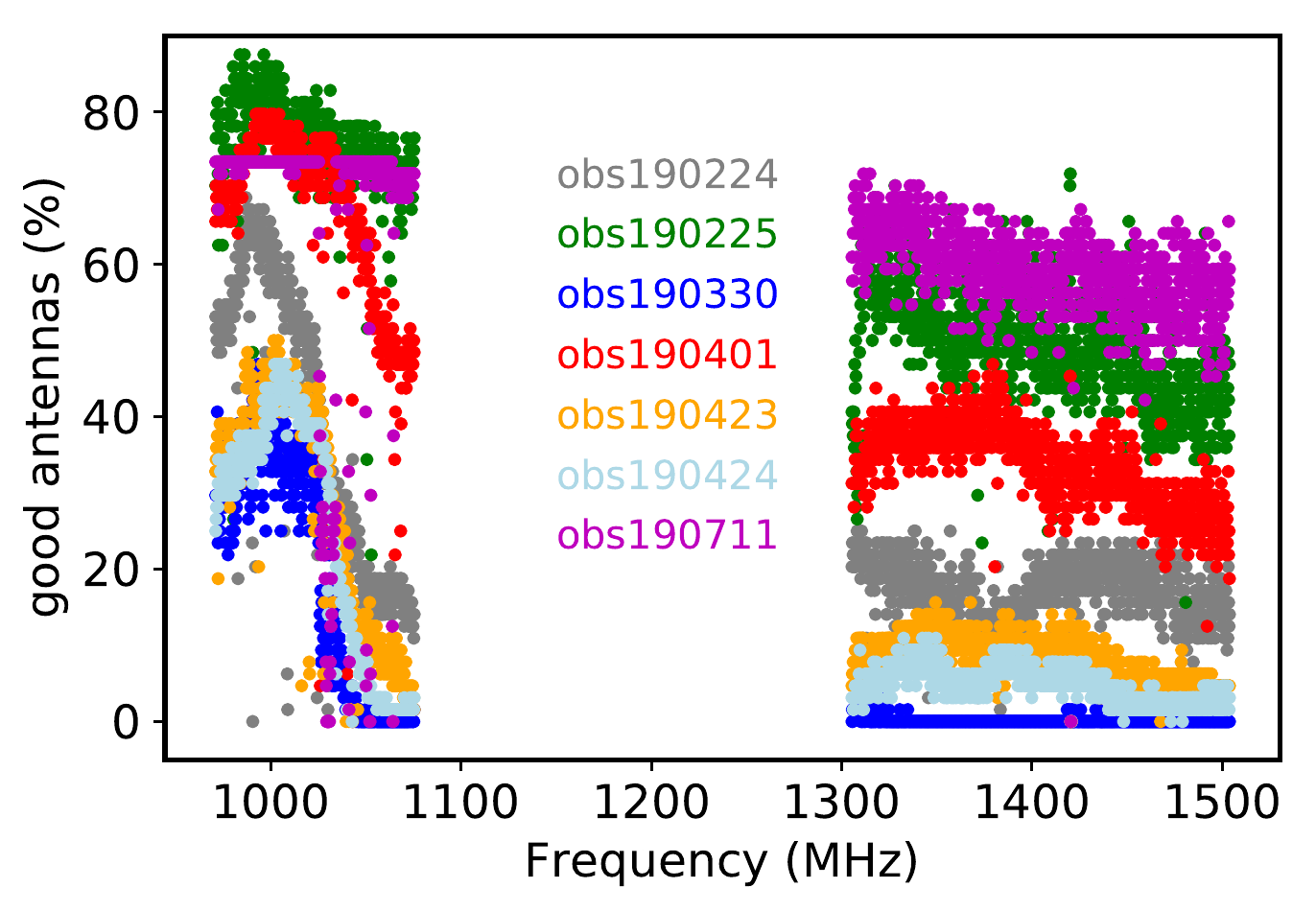}
\caption{The percentage of retained antennas (out of 64) at the end of Round 3 as a function of frequency for the low and high band of interest. As in ~\autoref{fig:RFIstat_allscans}, the different observations are labeled with different colors.}
\label{fig:RFIstat_goodants}
\end{figure}

 We show in \autoref{fig:RFIstat_goodants}, separately for each observation, the percentage of antennas, out the 64 MeerKAT dishes, for which we retain data at the end of the map making procedure, as a function of frequency. We recall that flagging is not only due to RFI but includes other issues such as possible bad functioning of a specific dish.

\subsection{Receiver temperature estimate}
For a radio receiver, the system temperature ($T_{\rm{sys}}$) is a combination of the internal temperature of the system as well as Galactic and atmospheric contributions:
\begin{equation}
T_{\rm{sys}}(t, \nu) = \overline T_{\rm CMB} + T_{\rm Gal}(t, \nu) + T_{\rm el}(t, \nu) + T_{\rm rec} (\nu)
\end{equation}
In reality the receiver temperature $T_{\rm rec}$ can also depend on time as discussed before in the calibration process. However, in this section we only consider a single observational block and data from a small enough region of the sky so as to assume that $T_{\rm rec}$ only varies across frequency.   

To provide an absolute calibration of the instrument, where each receiver measurement has no zero-level offset, the receiver and elevation-dependant temperatures would need to  be known to a high level of accuracy. This is not necessary for the purpose of \textsc{Hi} intensity mapping, since we are more concerned with fluctuations, but can be relevant for instance to make maps of the galactic synchrotron. 

The method employed by the MeerKAT single dish reduction pipeline to obtain $T_{\rm rec}$ uses a model for diffuse Galactic emission. This section provides a verification of the pipeline receiver temperatures using a method which does not require a Galactic emission model. Both the pipeline method and this "model-free" method require the subtraction of the ground spillover and atmospheric emission contributions ($T_{\rm el}(t, \nu)$) and the CMB monopole ($\overline T_{\rm CMB}$) from the full system temperature. Both methods also require RFI contaminated data to have been flagged and removed. 

The model-free method is based on the use of temperature-temperature (T-T) plots. T-T plots only require two different frequency observations of the same region of sky; if the dominant emission within said region is correlated across frequency then a linear regression across the region pixels will display a clear straight-line fit. The gradient of the fit will be dependant on the emission spectral index and the y-intercept will depend on any additive offsets present in each of the two observations \citep{wehus}. As $T_{\rm rec}$ is assumed not to change with location on the sky for this section, it is an additive offset which can be determined using the T-T plot method.

After the estimated ground spillover, atmospheric emission and CMB monopole are removed from the total calibrated temperature and RFI signals are masked out, the remaining contributions to the measured temperature are:
\begin{equation}
T_{\rm{cal}}^{\prime}(t, \nu) = T_{\rm{rec}}(\nu) + T_{\rm{sync}}(t, \nu) + T_{\rm{ueg}}(\nu), 
\end{equation}
where $T_{\rm{ueg}}(\nu)$ is the unresolved extragalactic point source offset level. We are assuming that the diffuse Galactic emission is purely synchrotron at these frequencies and Galactic latitudes. $T_{\rm{ueg}}(\nu)$ is assumed to be constant over all pixels but changes over frequency following a power law with a spectral index of around -2.7 \citep{2013MNRAS.434.1239B}. In reality, the unresolved point sources will also provide clustering and Poisson contributions to the total temperature which, unlike the mean temperature contribution, will vary across pixels. These contributions, however, will be minor with respect to $T_{\rm{sync}}$. 

If we consider two frequencies, $\nu_{1}$ and $\nu_{2}$, the diffuse Galactic synchrotron emission within those two frequency channels is related by a power law:
\begin{equation}
T_{\rm{sync}}(t, \nu_{2}) = T_{\rm{sync}}(t, \nu_{1})  \left(\frac{\nu_{2}}{\nu_{1}} \right )^{\beta_{sy}},
\end{equation}
meaning: 
\begin{equation}
\begin{split}
T_{\rm{cal}}^{\prime}(t, \nu_{2}) &= T_{\rm{rec}}(\nu_{2}) + T_{\rm{sync}}(t, \nu_{1})  \left(\frac{\nu_{2}}{\nu_{1}} \right )^{\beta_{sy}} + T_{\rm{ueg}}(\nu_{2}) \\
 &= T_{\rm{rec}}(\nu_{2}) + \left(T_{\rm{cal}}^{\prime}(t, \nu_{1}) - T_{\rm{rec}}(\nu_{1})  - T_{\rm{ueg}}(\nu_{1})  \right) \left( \frac{\nu_{2}}{\nu_{1}} \right )^{\beta_{sy}} \\ &+ T_{\rm{ueg}}(\nu_{2}). \nonumber
 \end{split}
\end{equation}
If the difference in frequency between $\nu_{2}$ and $\nu_{1}$ is less than a few MHz then we can assume that  $T_{\rm{rec}}(\nu_{2})  = T_{\rm{rec}}(\nu_{1}) $. Setting  $m = \left({\nu_{2}}/{\nu_{1}} \right )^{\,\beta_{sy}}$ and rephrasing the equation above as a linear regression between the total intensity maps yields: 
\begin{equation}
T_{\rm{cal}}^{\prime}(t, \nu_{2}) = m \, T_{\rm{cal}}^{\prime}(t, \nu_{1}) - m \, \left (T_{\rm{rec}}(\nu_{1})  + T_{\rm{ueg}}(\nu_{1}) \right)  + T_{\rm{rec}}(\nu_{1}) + T_{\rm{ueg}}(\nu_{2}).\nonumber
\end{equation}
Plotting $T_{\rm{cal}}^{\prime}(t, \nu_{2})$ against $T_{\rm{cal}}^{\prime}(t, \nu_{1})$ for several times, will produce a straight-line graph of gradient $m$ with a y-intercept ($c$) of: 
 \begin{equation}
c = -m \, \left ( T_{\rm{rec}}(\nu_{1})  + T_{\rm{ueg}}(\nu_{1}) \right )  + T_{\rm{rec}}(\nu_{1}) + T_{\rm{ueg}}(\nu_{2}).
\end{equation}
Therefore $T_{\rm rec}$ can be obtained from the fitted y-intercept and gradient as follows: 
 \begin{equation}
T_{\rm{rec}}(\nu_{1})  = \frac{c - T_{\rm{ueg}}(\nu_{2}) + m T_{\rm{ueg}}(\nu_{1})}{(1 - m)}.
\end{equation} 
To summarise, we assume that: 1) the receiver temperature is constant across the patch of the sky we are analysing (and the same for nearby frequencies); 2) we know the unresolved extragalactic point source offset level up to a certain accuracy (see below) and 3) the time changing quantity is related across frequencies by a spectral index (e.g. the Galactic synchrotron). Deviations from these assumptions will show up as errors in the linear fit. Moreover, any constant offset such as unmodeled ground pick up will also be absorbed into $T_{\rm{rec}}$ as before.

For the unresolved extragalactic point source offset level we use the 1.278\,GHz source counts under 12\,mJy from \citet{meerPS} and the recipe provided in \citet{2013MNRAS.434.1239B} to calculate the point source offset temperature at 1.278\,GHz. We then scale this value of 0.22\,K at 1.278\,GHz across our frequency range using a spectral index of -2.7. Allowing for 10 per cent errors on the offset temperature at 1.278\,GHz and a possible spectral index range of -2.5 to -2.9 leads to a 15 per cent error budget for the unresolved extragalactic point source contributions. As $T_{\rm rec}$ is $\sim$ 6 - 10\,K, our estimate of the unresolved point source level affects the $T_{\rm rec}$ measurements at the 1 per cent level. 

The MeerKAT channels are separated by roughly 0.2\,MHz. We calculate $T_{\rm rec}$ using the T-T plot method between channels in intervals of 5, thus providing a separation of 1\,MHz. To prevent anomalous measurements skewing the receiver temperature calculation, we consider several pairs of channels at one time. In practise this is achieved by selecting ten adjacent frequencies and performing linear regression between pairs of frequencies separated by 1\,MHz, e.g., frequency channel 1 and 6 out of the ten, frequency channel 2 and 7, frequency channel 3 and 8 and so on. This yields five estimates of $T_{\rm rec}$ for those ten adjacent frequencies, the median of which is used as the $T_{\rm rec}$ at middle frequency (channel 5).

\begin{figure}
\centering
\includegraphics[width=.85\columnwidth]{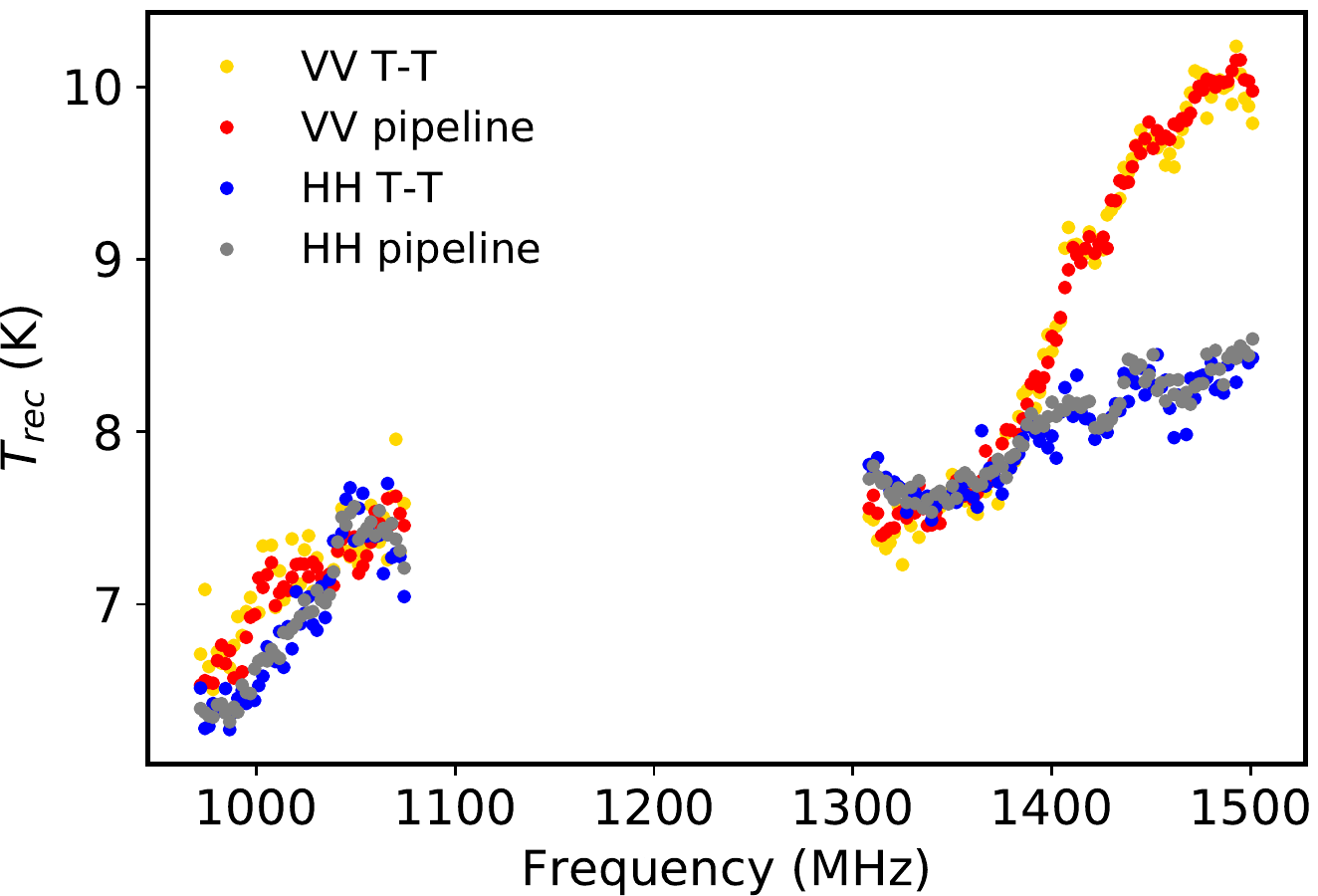}
\caption{$T_{\rm rec}$ as calculated by the T-T plot and pipeline method for {\tt m000h} and {\tt m000v} of {\tt obs190225}. Note that any unmodeled constant offset in time would be absorbed into $T_{\rm rec}$ (e.g., unmodeled ground pick up).}
\label{fig:trec}
\end{figure}

We choose to perform the linear regression across a small region of the sky within the observed WiggleZ field: $155 ^{\circ} < \rm{ra} < 157 ^{\circ}$ and $2 ^{\circ} < \rm{dec} < 4 ^{\circ}$. As the synchrotron spectral index is known to vary across the sky, the point of choosing a small region is to keep these variations as low as possible. The region selected is also free from notable point source emission, so as to specifically target diffuse synchrotron emission for the linear regression.  \autoref{fig:trec} shows $T_{\rm rec}$ for {\tt m000h} and {\tt m000v}, during a particular observation block ({\tt obs190225}) as calculated by both the T-T plot (model-free) method and the pipeline method. For the pipeline, since $T_{\rm{rec}}$ changes (slowly) with time, we choose the times corresponding to the same pixels on the sky we are using and average the values.
The two methods can be seen to be in close agreement with each other. Both methods also show large variations in $T_{\rm rec}$ over frequency.   

\begin{figure}
\centering
\includegraphics[width=.85\columnwidth]{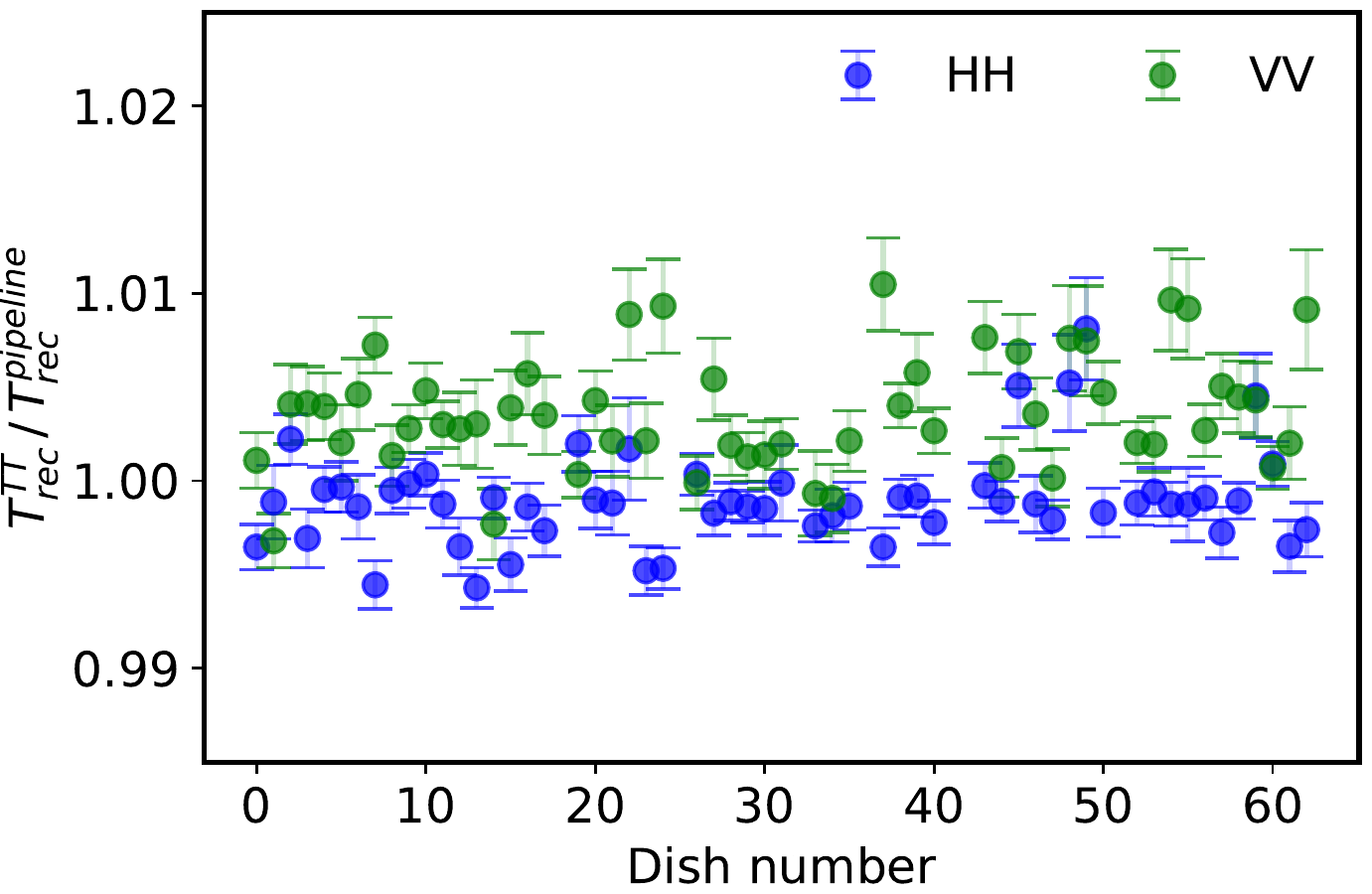}
\caption{Mean ratio of $T_{\rm rec}$ as calculated by the T-T plot and pipeline method for all the receivers observing during a particular scan ({\tt obs190225}). Error bars represent the standard deviation on the mean.}
\label{fig:alltrec}
\end{figure}

In \autoref{fig:alltrec} we plot the mean ratio of $T_{\rm rec}$ as calculated by the T-T plot (model-free) and pipeline method for all the receivers observing during {\tt obs190225}. The difference between these two methods is seen to be consistently below the one percent level. 

\subsection{Stability of the noise diode}
The noise diodes, as the TOD calibrator, need to be a stable source during the observation time scale. In the calibration, we calibrate the power of the noise diode twice, once in the beginning (track-I) and another at the end (track-II). Here we calculate the percentage variation of $\overline T_{\rm diode}$ as 
\begin{gather}\label{eq:Tnd_err}
    \delta_{\rm diode}(\nu)=\frac{\overline T_{\rm diode}^{\rm I}(\nu)-\overline T_{\rm diode}^{\rm II}(\nu)}{\frac{1}{2}\left (\overline T_{\rm diode}^{\rm I}(\nu)+\overline T_{\rm diode}^{\rm II}(\nu) \right )}.
\end{gather}

\begin{figure}
\centering
\includegraphics[width=\columnwidth]{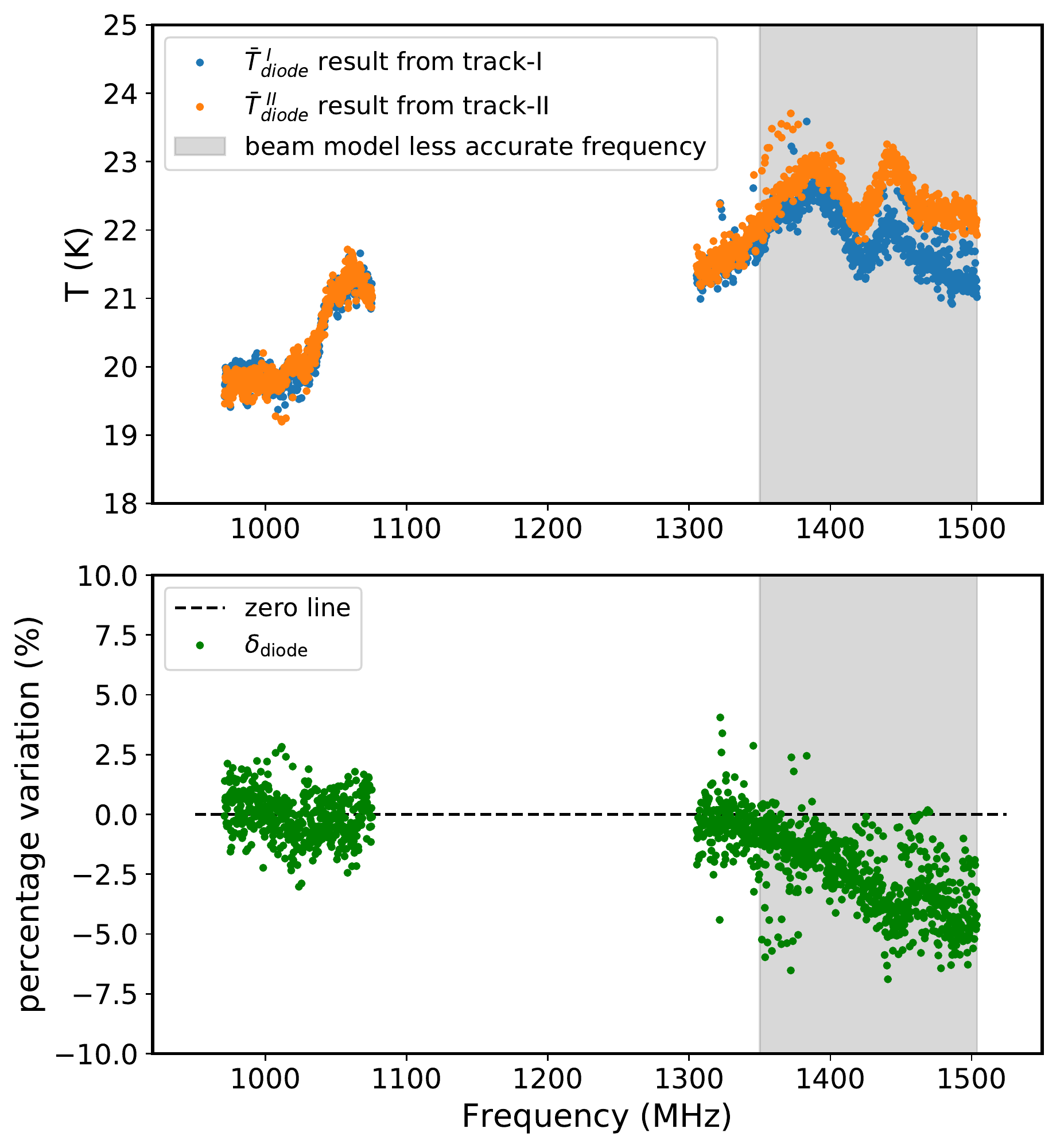}
\caption{The $\overline T_{\rm diode}$ results against frequency and the corresponding percentage variation (\autoref{eq:Tnd_err}), for dish {\tt m000h} of {\tt obs190225} as an example. The shaded area marks the frequency range where the beam model is reported to be less accurate.}
\label{fig:Tnd_comp}
\end{figure}

In \autoref{fig:Tnd_comp} we show the $\delta_{\rm diode}(\nu)$ result for {\tt m000h} of {\tt obs190225}. The lower-frequency band shows
$\delta_{\rm diode}\lesssim 2.5\%$ which will be reduced when we calculate the temperature map from HH and VV data and calculate the final map by averaging over 100 scans. The higher frequency band shows a clear bias above 1350 MHz, which may be caused by two factors: 1) as shown in \autoref{fig:ch_counts}, the RFI pollution is more significant at higher frequency, while the calibrator source is also close to the equator during the track observation;
2) as mentioned in \cite{2019arXiv190407155A}, the beam model is less accurate at higher frequencies (> 1350 MHz), which may cause errors when calibrating the power of the diode noise from the calibrator point source.
Further insight into this problem will be provided by an
upcoming observation that is far from the equator. In any case, our focus is mostly in the lower-frequency band given the cosmology goals.

\subsection{Residuals of the calibration}

\begin{figure}
\centering
\includegraphics[width=\columnwidth]{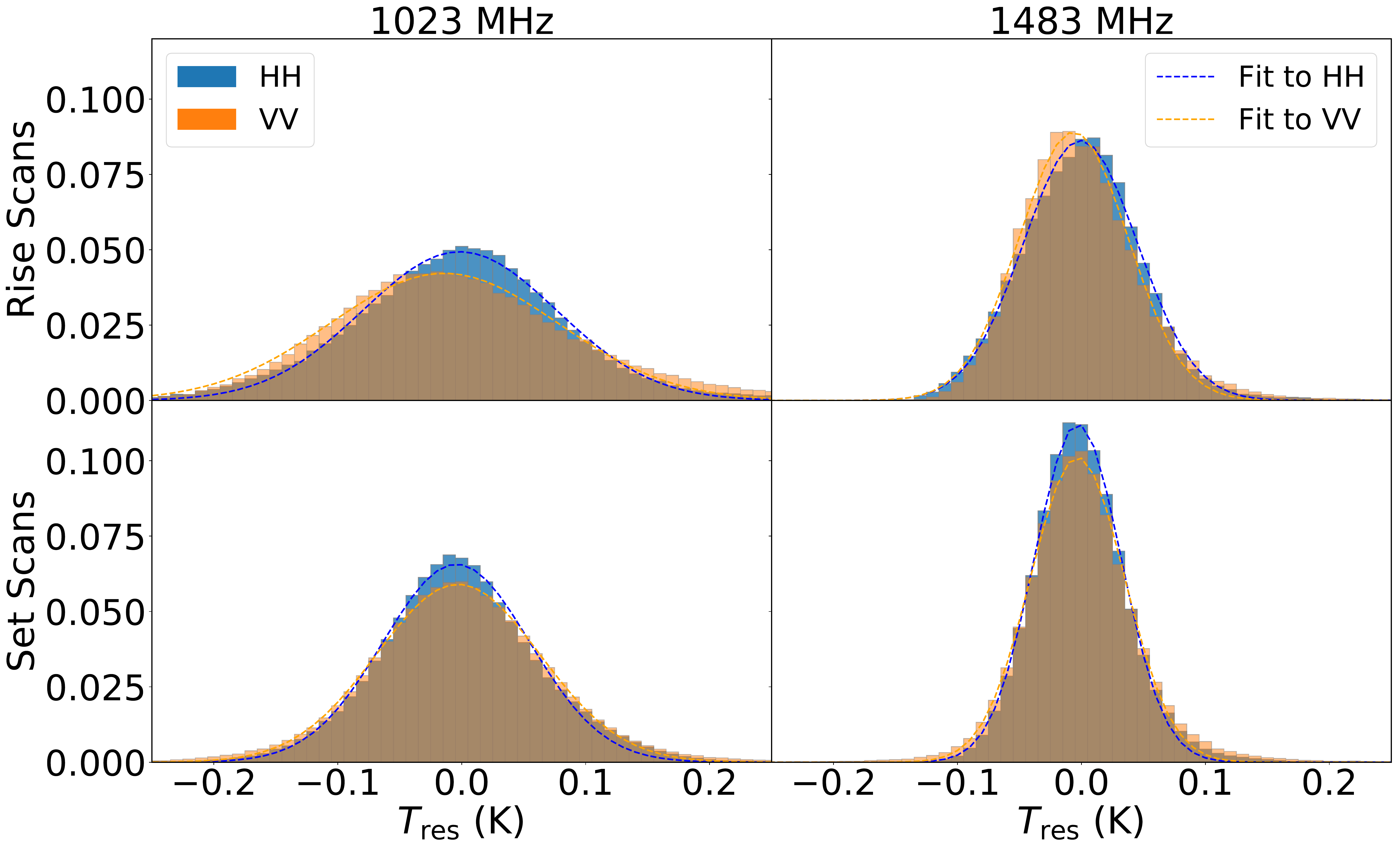}
\caption{Histograms of level-3 temperature residuals $T_{\rm res}(t,\nu)$ (\autoref{eq:T_resi}) for both polarisations. Each panel corresponds to a different combination of channel and scan type. The top row corresponds to rising scans while the bottom row to setting scans. Left column exemplifies a lower frequency while the right column a higher frequency of the total range.}
\label{fig:res_hists}
\end{figure}

In \secref{sec:pipeline} we presented a model (\autoref{equ:model}) to calibrate the sky. From this pipeline we are left with calibration residuals $T_{\rm res}(t,\nu_i)$ (\autoref{eq:T_resi}) for both HH and VV polarisations. If the model was a complete description of the observation, such residuals should behave as Gaussian random noise centered at zero. We already know that is not the case, as we did not include the point sources in the scan. However, note that any offset leftover in the scan will be absorbed into the receiver temperature, so that the residuals will still be centered at zero. Any skewness in the histogram would be an indication of the presence of contaminants, while the variance gives an estimate of the level of accuracy of our model.

In \autoref{fig:res_hists} we present normalised histograms of the residuals for the different polarisations using two example frequency channels and divided into ``rising'' and ``setting'' scans. The set scans look well behaved, i.e., both frequency channels have the distribution of the residuals centered near zero and reasonably described by Gaussian curves. The standard deviations of the setting scans are $\sigma^{\rm HH}_{1023}=0.06 \,{\rm K},\ \sigma^{\rm VV}_{1023}=0.07 \,{\rm K},\ \sigma^{\rm HH}_{1483}=0.04 \,{\rm K},\ \sigma^{\rm VV}_{1483}=0.05 \,{\rm K}$. This is still above the thermal noise, but corresponds to about a $0.3 \%$ error in our model, since our $T_{\rm sys}$ is about 16 K. In reality we have some outliers in the residuals that go up to 0.3 K (see \autoref{fig:sky_map}) which is still less than a $2\%$ correction to our model.

The rising scans show an interesting skewness in the distribution of the residuals for the VV polarisation. This might indicate the presence of residual RFI in the VV polarisation due to some residual omni-directional (vertical) transmitting antennas from terrestrial systems. Geographically speaking, the setting scans look northwest while the rising scans look northeast. As the effect was present in both rising observational nights, it is likely that a faint terrestrial vertically polarised source (GSM cell phone antennas, VHF radio, UHF TV signals) is present in the northeast direction from the MeerKAT site (there is indeed a communication tower 160 km away from the core of the MeerKAT in that direction). This possible RFI is however quite low and will get diluted in the final averaged maps. The main point is that corrections to our model which is used for the calibration have rms values below the $1\%$ level which should translate to gain fluctuations at the same level.

\subsection{Noise level estimation} \label{sec:Trms}

\begin{figure}
\centering
\includegraphics[width=.9\columnwidth]{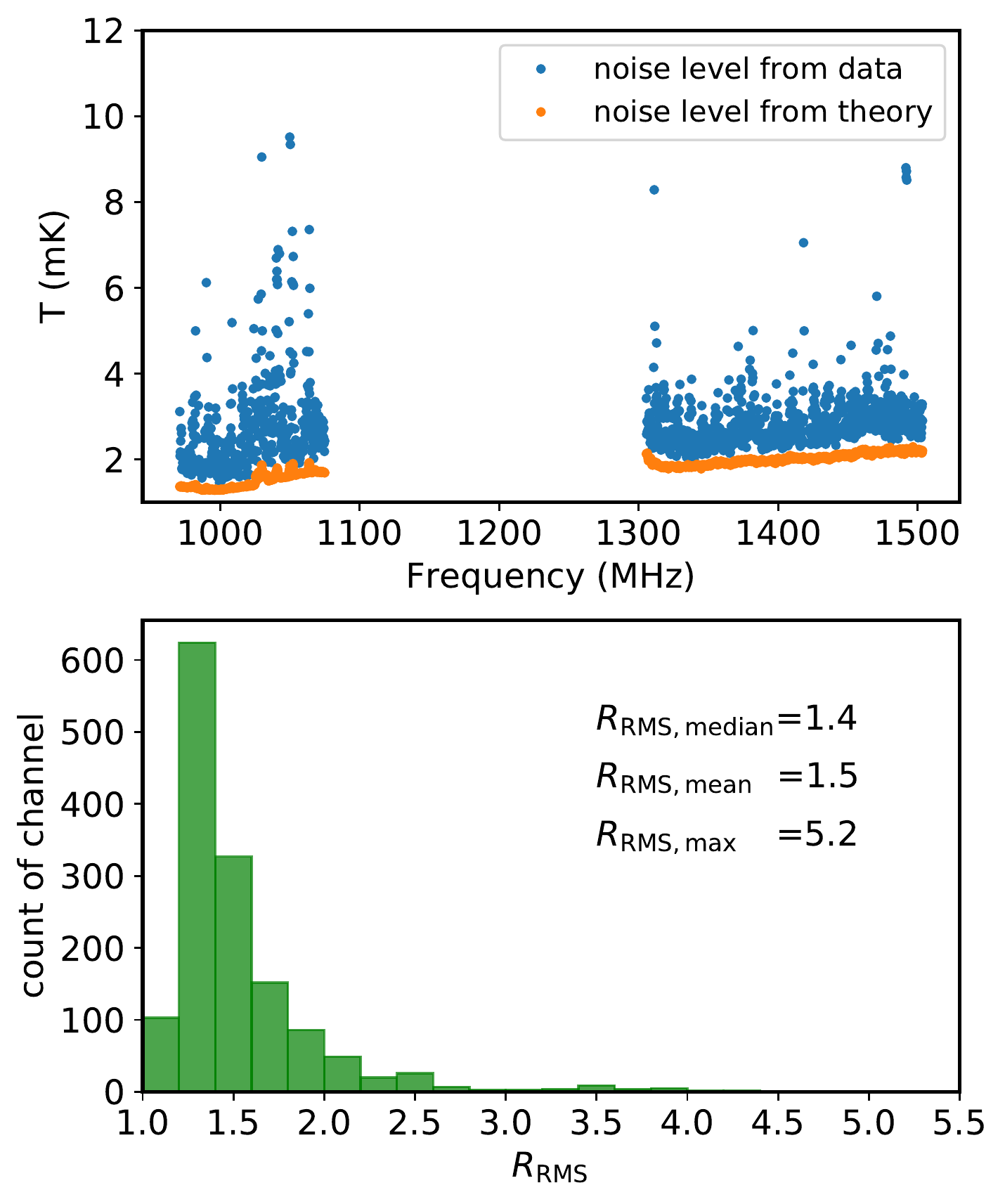}
\caption{The comparison between noise levels derived from theory equation and from data (upper panel) and the histogram of their ratio $R_{\rm RMS}=\Delta T_{\rm RMS} (\nu^{\star})/ \sigma_{\rm th} (\nu^{\star}) $ (lower panel).}
\label{fig:rms}
\end{figure} 

It is also useful to measure the noise level from the final maps and check if they are consistent with the theoretical expectation (and dropping as square root of time). Unfortunately, as we have seen, the residuals still have some contamination, in particular from point sources (\autoref{fig:sky_map}), so just calculating the variance of such maps would not work.
However, using the final data cube, $T_{\rm res}$, we can estimate the noise level using the difference between four neighboring channels (so-called ABBA), as
\begin{gather}
  \Delta T_i(\nu^{\star})= \frac{1}{2} \left( T^i_{\rm res}(\nu) + T^i_{\rm res}(\nu+\Delta \nu) \right ) \nonumber\\
  ~~~~~~~~~~~~~~~~~-\frac{1}{2}\left ( T^i_{\rm res}(\nu-\Delta \nu)+T^i_{\rm res}(\nu+2\Delta \nu)\right),
\end{gather}
where the index $i$ goes over all pixels in the map and $^{\star}$ means the combined result from the four channels $[\nu-\Delta \nu,\nu,\nu+\Delta \nu,\nu+2\Delta \nu]$. This will cancel any quantity that is linear in $\nu$ across these channels and should therefore cancel most, if not all, of the sky contributions.
If there was only white noise in the maps, the corresponding variance in each pixel would be,
\begin{gather}
     \sigma^2_{{\rm th},i}(\nu^{\star})=\frac{1}{4} \Big(\sigma^2_{{\rm th},i}(\nu)+\sigma^2_{{\rm th},i}(\nu+ \Delta \nu)+\sigma^2_{{\rm th},i}(\nu-\Delta \nu)\nonumber\\
     ~~~~~~~~~~~~~~~~~~~ + \sigma^2_{{\rm th},i}(\nu+ 2\Delta \nu) \Big),
\end{gather}
which would give the variance of a single map $\sigma^2_{{\rm th},i}(\nu)$ if they were the same for the four frequencies. We can calculate this theoretical variance by starting with the standard radiometer equation \citep{2009tra..book.....W} to get the noise in the initial time ordered data:
\begin{gather}
  \sigma^2_{\rm th}(t,\nu)=\frac {T^2_{\rm sys}(t,\nu)}{\Delta \nu \Delta t},
\end{gather}
where $\Delta \nu= 0.2$ MHz is the channel resolution and $\Delta t=2$ s. For the system temperature we can use the calibrated data itself for each polarisation, $T_{\rm sys}=T_{\rm cal,HH}(t,\nu)$ or $T_{\rm cal,VV}(t,\nu)$. In order to get the final data cube, $T^i_{\rm res}(\nu)$, the data goes through several stages of averaging. We propagate the variance taking into account this averaging in order to get to the final $\sigma^2_{{\rm th},i}(\nu)$.

From the data, our best estimate for the noise variance is the rms over the $\Delta T_i(\nu^{\star})$ map. Since the noise is not necessarily uniform across the map, we take a weighted rms,
\begin{gather} \label{eq:rms_obs}
     \Delta T_{\rm RMS}^2(\nu^{\star}) = 
     \frac {N_p} {N_p - 1}
     \left (\frac {\sum_i w_i\Delta T^2_i} {\sum_i w_i} -
     \left[\frac {\sum_i w_i\Delta T_i} {\sum_i{w_i}}\right]^2 \right),
\end{gather}
where the sums are over the number of pixels in the map, $N_p$, and we have suppressed the dependence of $w_i$ and $\Delta T_i$ on $\nu^\star$. 
For the weight, $w_i$, we used the theoretical expected variance itself, $w_i(\nu^{\star})=1/\sigma^2_{{\rm th},i}(\nu^{\star})$ (but note that, as explained above, such theoretical variance is calculated using $T_{\rm cal}$ as a proxy for the system temperature).
The theoretical rms which can be compared to $\Delta T_{\rm RMS} (\nu^{\star})$ in \autoref{eq:rms_obs}, is then simply
\begin{gather}
     \sigma^2_{\rm th} (\nu^{\star}) = N_p \left ( \sum_i \sigma^{-2}_{{\rm th},i}(\nu^{\star})  \right )^{-1}.
\end{gather}

In \autoref{fig:rms} we show the noise level against frequency and the histogram of ratio 
$R_{\rm RMS}=\Delta T_{\rm RMS} (\nu^{\star}) / \sigma_{\rm th} (\nu^{\star})$. The median value of $R_{\rm RMS}$ is $1.4$, while some channels are noisy and $R_{\rm RMS}$ can reach $5.2$. We see that we are reaching a noise level of about 2 mK which is within expectations and promising for our cosmology goals (averaging 100 pixels in the cube would reach an rms of 0.2 mK which is at the level of the \textsc{Hi} signal).

\begin{figure*}
\centering
\includegraphics[width=2\columnwidth]{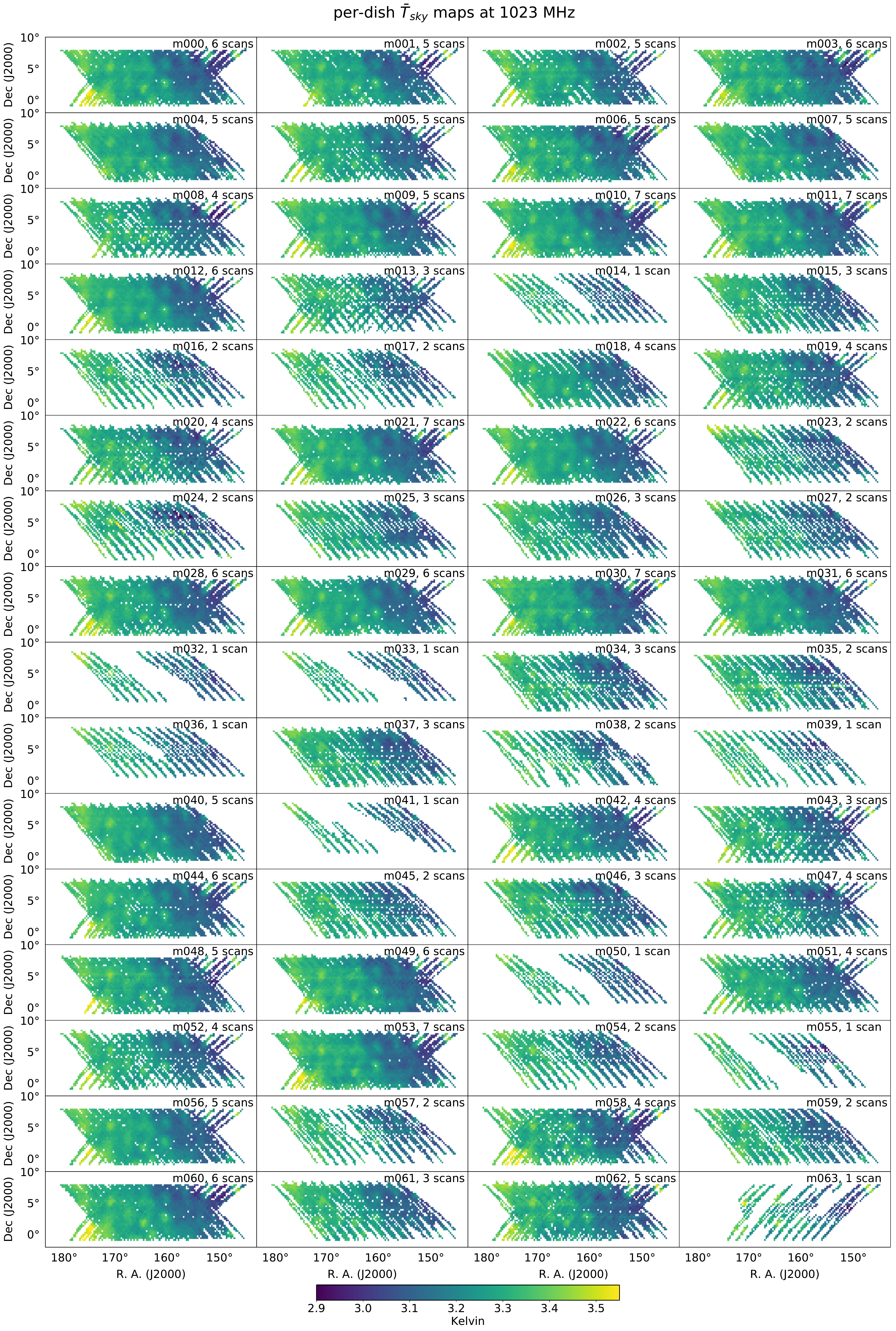}
\caption{Per-dish $\overline T_{\rm sky}$ maps at 1023 MHz.}
\label{fig:Tsky64}
\end{figure*} 

\begin{figure*}
\centering
\includegraphics[width=2\columnwidth]{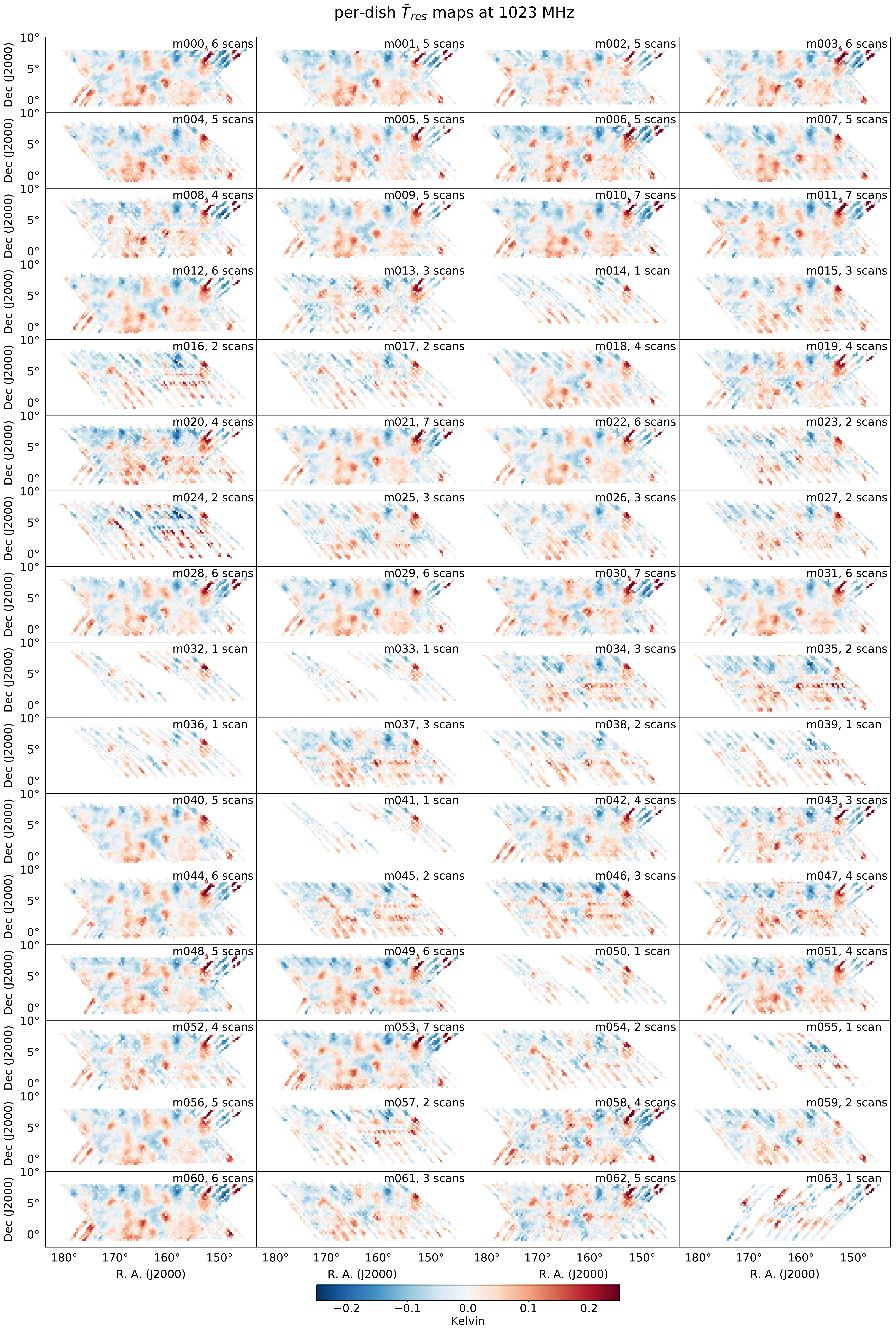}
\caption{Per-dish $\overline T_{\rm res}$ maps at 1023 MHz.}
\label{fig:Tresi64}
\end{figure*} 

\subsection{Per-dish maps}
In \autoref{fig:Tsky64} and \autoref{fig:Tresi64} we show the per-dish maps $T_{\rm sky}$ and $T_{\rm res}$, respectively, with all available data from the seven observations. These maps show a good consistency between each other and have no visible striping artifacts. We noticed that the different dishes have different "clean" scans left after the weak RFI flagging (\secref{sec:weakRFI}). It shows a slight relation to the position of the dish in the MeerKAT array, which denote that the weak RFI may come from the surrounding facilities or be affected by the terrain.

\begin{figure}
\centering
\includegraphics[width=0.9\columnwidth]{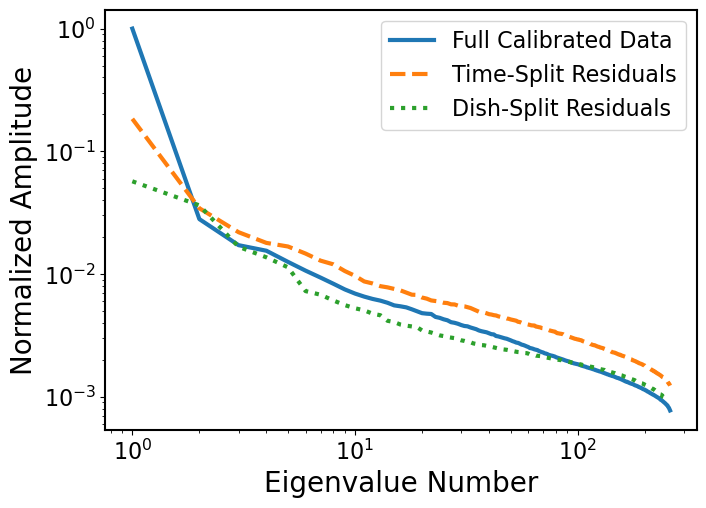}
\caption{Principal components from the frequency-frequency covariance matrix for the final $\overline T_{\rm sky}({\rm ra_{pix}}, {\rm dec_{pix}},\nu)$ calibrated maps, restricted to the $971<\nu<1025\,\text{MHz}$ frequency range least affected by RFI. Blue solid line shows the fully calibrated data from all dishes. Also shown are the principal components from the residuals of subtracting two subsections split by time blocks (orange dashed) and dishes (green dotted).}
\label{fig:Eigendecomp}
\end{figure} 

\begin{figure}
\centering
\includegraphics[width=0.9\columnwidth]{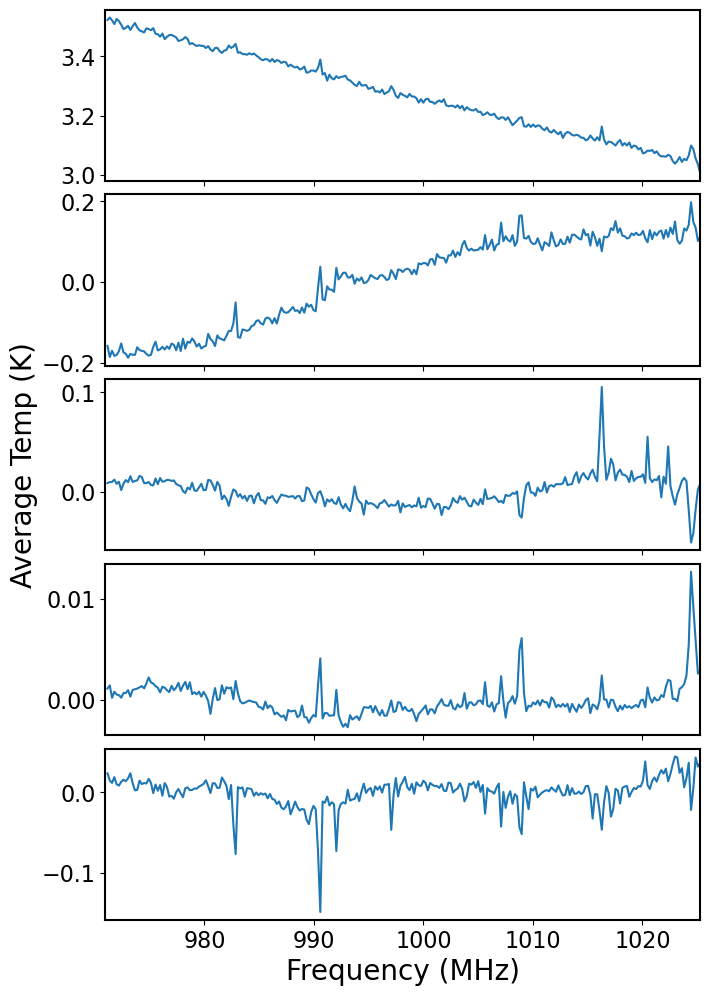}
\caption{First five eigenvectors (descending in order of magnitude from top to bottom) from the frequency-frequency covariance matrix for the final $\overline T_{\rm sky}({\rm ra_{pix}}, {\rm dec_{pix}},\nu)$ calibrated maps, restricted to the $971<\nu<1025\,\text{MHz}$ frequency range least affected by RFI.}
\label{fig:First5Eigenvec}
\end{figure} 

\subsection{Eigendecomposition of calibrated maps}

We expect the sky signal in the calibrated maps to be highly correlated across frequency. We can test this assumption by measuring the $\nu,\nu'$ covariance of the data given by $\mathbf{C} = \mathbf{X}^\text{T}\mathbf{X}/(N_p - 1)$. Here $\mathbf{X}$ represents the final calibrated maps $\overline T_{\rm sky}({\rm ra_{pix}}, {\rm dec_{pix}},\nu)$ unravelled into single vectors of length $N_p$ pixels at each frequency. The matrix $\mathbf{X}$ therefore has dimensions $N_\nu \times N_p$. We use data in the frequency range $971<\nu<1025\,\text{MHz}$ least affected by RFI, mask point sources in each map, then mean-centre the data for each channel. The eigendecomposition is then given by $\mathbf{C}\mathbf{V}=\mathbf{V}\mathbf{\Lambda}$, where $\mathbf{\Lambda}$ is the diagonal matrix of eigenvalues ordered by descending magnitude, and $\mathbf{V}$ the corresponding eigenvectors. \autoref{fig:Eigendecomp} shows the eigenvalues for the calibrated maps. One aim in calibrating the telescope for \textsc{Hi} intensity mapping is for the majority of foreground covariance to be contained in just a few dominant modes which can be removed to better isolate the underlying \textsc{Hi} signal, which should have a smooth, flat eigenvalue spectrum since it is approximately Gaussian. The fact that the full calibrated data (blue solid line) does not reach a flattened plateau after the first few modes (which we expect to contain foregrounds) suggests that there are frequency-correlated systematics in the data which dominate over the cosmological \textsc{Hi} signal.

Splitting the data into two subsections and subtracting one subsection map from the other, should leave residuals which are absent of constant astrophysical signal and any systematics which are constant throughout the observation. This is shown by the orange dashed and green dash-dotted lines where we split by time block and observing dishes. As expected, these residuals reduce the most dominant mode significantly which is likely the galactic synchrotron. However, there remains some reasonably correlated modes suggesting the data still contains some time-varying and dish-varying systematics which are correlated in frequency. Similar tests were also presented in \citet{Switzer:2015ria} (see their Figure 3).
These results indicate hints of systematics in the calibrated data which could be due to instrumental effects such as polarisation leakage, chromatic beam effects, calibration issues with the bandpass, $1/f$ noise or residual RFI. Further investigation into characterising these systematics will be the aim in future work. For instance, one way to mitigate the chromatic beam effects is to convolve the maps to a common resolution, something that was performed on GBT data. Also, this analysis was done on calibrated maps, $\overline T_{\rm sky}$, with models already subtracted, while for our future cosmological analysis we might work directly with the total calibrated map, $\overline T_{\rm cal}$ and apply prior filtering on the TOD in order to deal with the 1/f noise.

\autoref{fig:First5Eigenvec} shows the mean signal from the reconstructed first five eigenvectors of the $\nu,\nu'$ covariance matrix (this can also be compared to Figure 4 in \citealt{Switzer:2015ria}). The first (top) mode represents the synchrotron signal, which is expected to dominate, and the slope seen in this mode is consistent with a synchrotron spectral index. An ideal instrument would produce a perfectly smooth slope in this first mode and allow a simple polynomial model to be used to remove this contribution from the data. However, the slight oscillations through frequency motivates the requirement of a more sophisticated foreground removal technique for analysing cosmological \textsc{Hi}. The remaining modes are likely caused by instrumental response systematics as discussed in the previous paragraph and identified by \autoref{fig:Eigendecomp}. In order to probe the cosmological \textsc{Hi}, it is these modes, and further more dominant ones, that would need to be removed or mitigated before a detection is possible.

\section{Comparison with previous observations}\label{sec:comparison}

\subsection{Recovery of diffuse Galactic radio emission}

We aim to compare our calibrated sky data to realistic simulations of the diffuse Galactic radio emission, which are extrapolated from real data. Using the Global Sky Model (GSM) \citep{de_Oliveira_Costa_2008,Zheng_2016}, we generate realistic foreground maps at our frequencies of observation. The GSM is a model of diffuse Galactic radio emission, which uses 29 data sky maps to extrapolate this emission from 10 MHz to 5 THz. We select the frequency range of 971-1025 MHz due to containing the least RFI contamination, and generate accurate simulated foregrounds at each frequency channel of observation using the GSM. The output of the GSM extrapolation is a full sky map at a desired frequency, in \texttt{HEALPix}\footnote{\href{http://healpix.sourceforge.net}{https://healpix.sourceforge.io/}} \citep{Gorski2005,Zonca2019} format. Since we are only interested in the patch of sky targeted by our survey, we have to convert the GSM output into a flat-sky map that is in the same coordinates as our data. To do this, we first take the angular coordinate of each pixel in our MeerKAT map and then match it to the pixel in the \texttt{HEALPix} map with the closest angular coordinate. While this is an approximation, we find that it is sufficient in our case since the angular size of the map is not too large, and we recover similar structure in the foreground simulation as is present in the MeerKAT data. In addition, we also mean centre 
each frequency slice of the MeerKAT and foreground simulation maps, by calculating the mean temperature in each frequency slice and subtracting that mean from all the pixels in that particular slice. Finally, in order to make the simulated foreground maps match the MeerKAT data in resolution, we convolve the foreground simulation maps with a frequency dependent beam calculated according to the fitted 2D Gaussian beam model (\secref{sec:beam}). We also mask point sources in the MeerKAT data, since these are removed in the GSM simulations. The comparison is done using the MeerKAT sky temperature data, \autoref{eq:Tsky}, averaged over all scans and dishes, $\overline T_{\rm sky}({\rm ra_{pix}}, {\rm dec_{pix}},\nu)$.

\begin{figure}
    \centering
    \includegraphics[width=1\columnwidth]{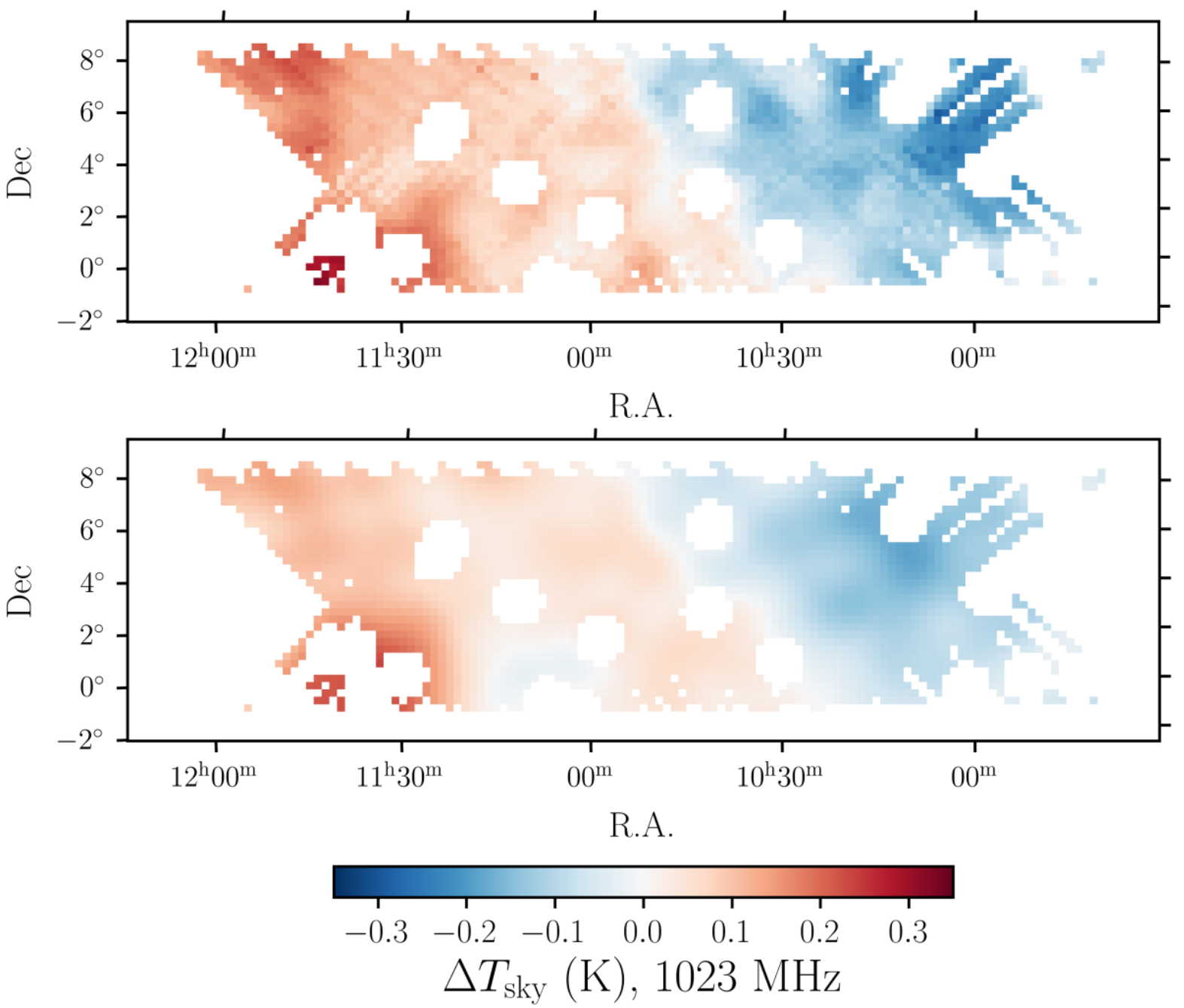}
    \caption{\textit{Top:} MeerKAT sky data at 1023 MHz, mean centered such that the colorbar shows $\Delta T_{\rm sky}$, the deviations from the mean map temperature in K. \textit{Bottom:} The same, but for the foreground simulation generated using GSM at the same frequency and angular coordinates. Point sources with flux $> 1$ Jy at 1.4 GHz and neighbouring pixels within 1 degree are masked on both maps.}
    \label{fig:MK_GSM_comparison}
\end{figure}

\begin{figure}
    \centering
    \includegraphics[width=0.9\columnwidth]{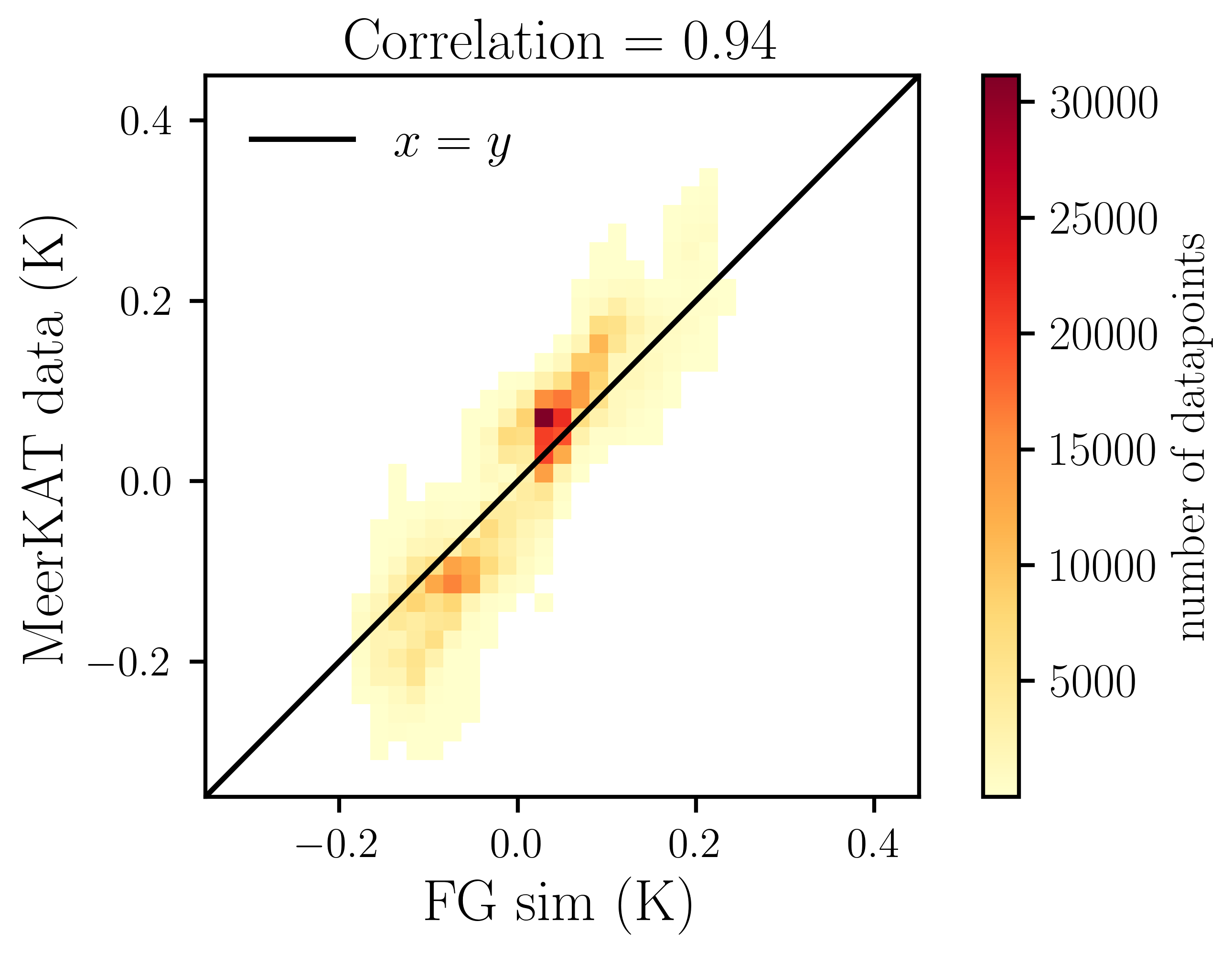}
    \caption{Correlation between the MeerKAT sky data (all dishes and scans) and the simulated foreground temperature fluctuations in the frequency range 971-1025 MHz. Darker colors indicate a higher concentration of datapoints, and the Pearson correlation coefficient between the two datasets is found to be 0.94. }
    \label{fig:MK_GSM_correlation}
\end{figure}

\autoref{fig:MK_GSM_comparison} shows a comparison between the MeerKAT calibrated sky map data at 1023 MHz, and the simulated GSM at the same frequency. It is clear from a visual inspection of these maps that similar structure is present in both. In order to check their agreement more quantitatively, we calculate the pixel correlation between the MeerKAT data cube and the GSM data cube in the frequency range 971-1025 MHz. Since this involved a large amount of data points, we show our results in the form of a 2D histogram in \autoref{fig:MK_GSM_correlation}, where the darkest colors indicate the highest concentration of data points. The Pearson correlation coefficient is found to be 0.94, indicating a strong correlation between our data and simulation. In addition, the cluster of points follows the $x=y$ line to a good degree. This demonstrates that the calibration pipeline yields a result that is highly correlated with a realistic simulation of the diffuse Galactic radio emission, both visually and quantitatively.

\subsection{Recovery of the Galactic \textsc{Hi} map}

\begin{figure}
\centering
\includegraphics[width=\columnwidth]{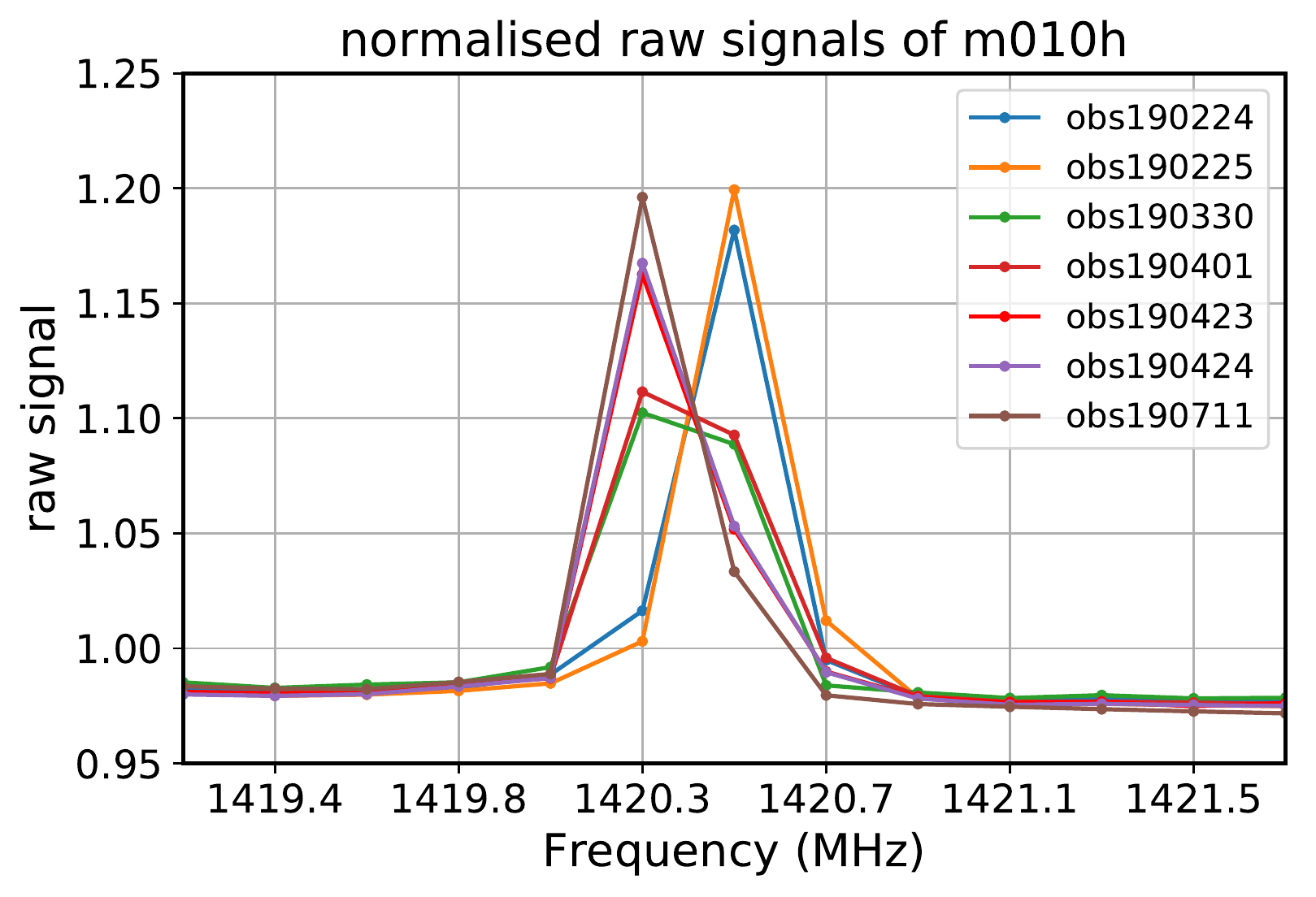}
\caption{The time averaged raw signals of seven observations, which are normalised by their mean values. The frequency is in the geocentric rest frame and the Galactic 21 cm lines are extended across several frequency channels due to the time and pointing related Doppler shifts.}
\label{fig:Galactic_HI_spec}
\end{figure}

\begin{figure*}
\centering
\includegraphics[width=2\columnwidth]{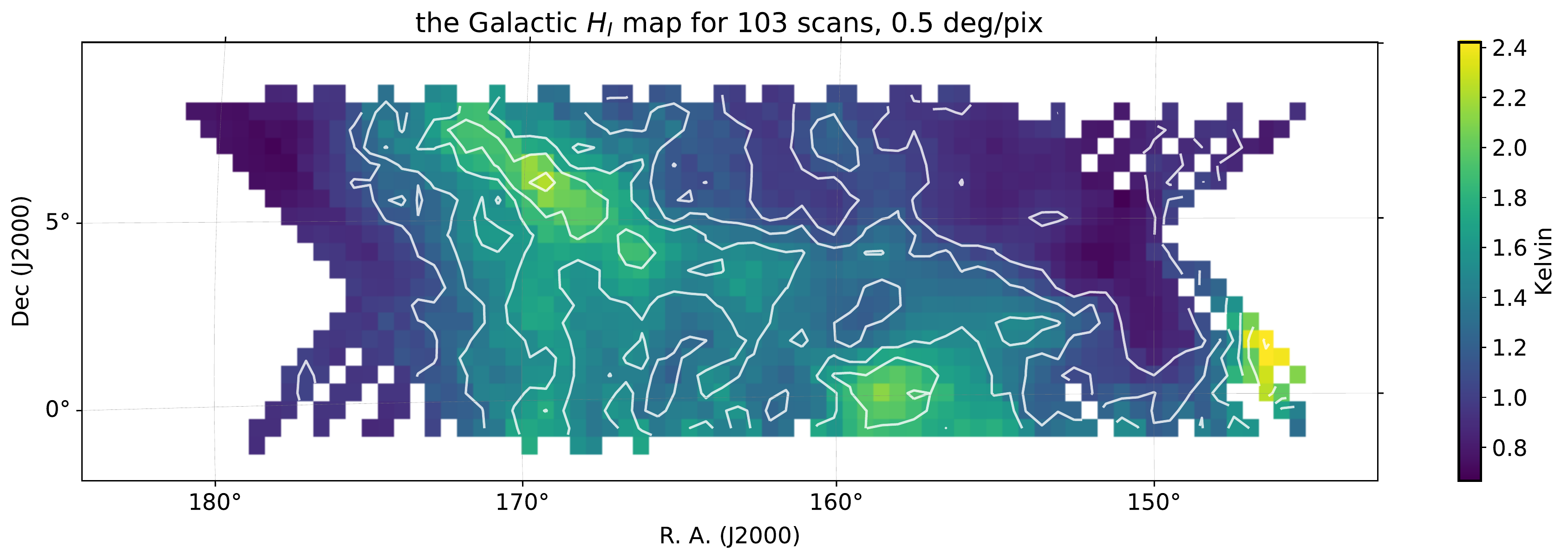}
\caption{The Galactic \textsc{Hi} intensity map which is extracted from frequency band $\nu_{\rm GHI} \in [1420.3, 1420.7]$ MHz.  The white contours are derived from the HI4PI \textsc{Hi} column density map \citep{2016A&A...594A.116H}, which is constructed from the Effelsberg-Bonn \textsc{Hi} Survey.}
\label{fig:Galactic_HI_map}
\end{figure*}

Another interesting test is to  compare our own \textsc{Hi} Galactic map to the literature.
The 21 cm emission of the Milky Way has different structures to the Galactic synchrotron emission due to their different radiation mechanisms. In \autoref{fig:Galactic_HI_spec} we show the normalised time averaged raw signals of seven observations across frequency. We see that the Galactic 21 cm lines, which correspond to a frequency of 1420.4 MHz, are extended across several frequency channels due to the the Doppler shifts. We start by selecting three frequency channels corresponding to $\nu_{\rm GHI} \in [1420.3, 1420.7]$ MHz, to extract the Galactic \textsc{Hi} emission. The channels $\rm ch_{GHI}$ are flagged before our calibration, so we interpolate the gain and $T_{\rm cal}$ from our previous results to these three channels so that,
\begin{gather}
    T_{\rm GHI}(t,\nu_{\rm GHI} )=\hat T_{\rm raw}(t, \nu_{\rm GHI})/ g(t, \nu_{\rm GHI}) - T_{\rm cal} (t, \nu_{\rm GHI}).
\end{gather}
This calculation is applied on the RFI flagged HH and VV data. The final Galactic \textsc{Hi} intensity map is shown in \autoref{fig:Galactic_HI_map}. We overlay the contours of the HI4PI \textsc{Hi} column density map \citep{2016A&A...594A.116H}, which show a good agreement to our map.

\subsection{Recovered point source fluxes vs catalogue values} \label{sec:ptr_spec}

\begin{figure}
\centering
\includegraphics[width=\columnwidth]{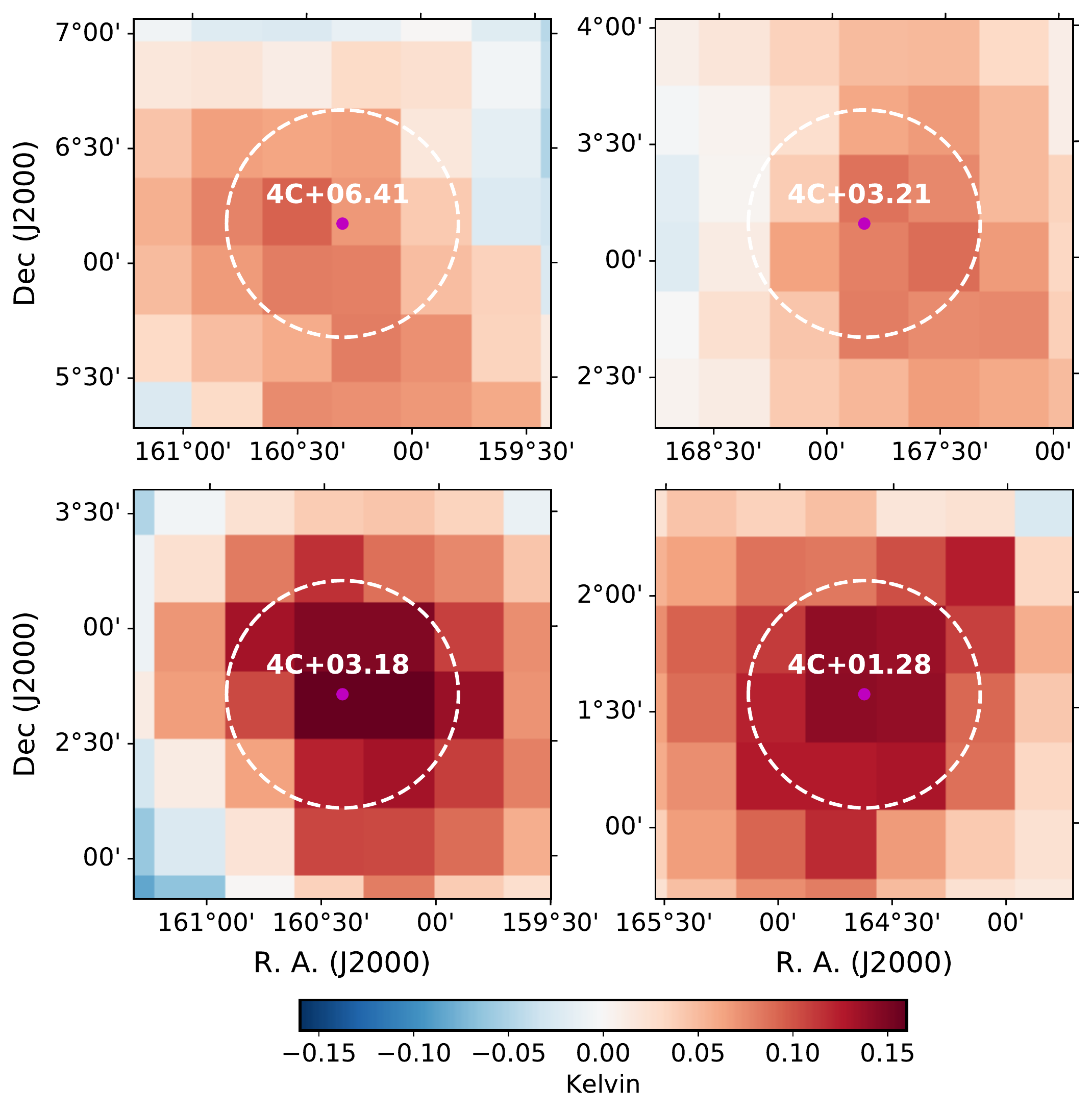}
\caption{$\overline T_{\rm res}(x,y, {\rm ch})$ maps for the four point sources in a binned channel with the center channel 800. The white dash circles label the positions of $r_{\rm cut}=0.5^{\circ}$.}
\label{fig:ptr_maps}
\end{figure} 

\begin{figure}
\centering
\includegraphics[width=.9\columnwidth]{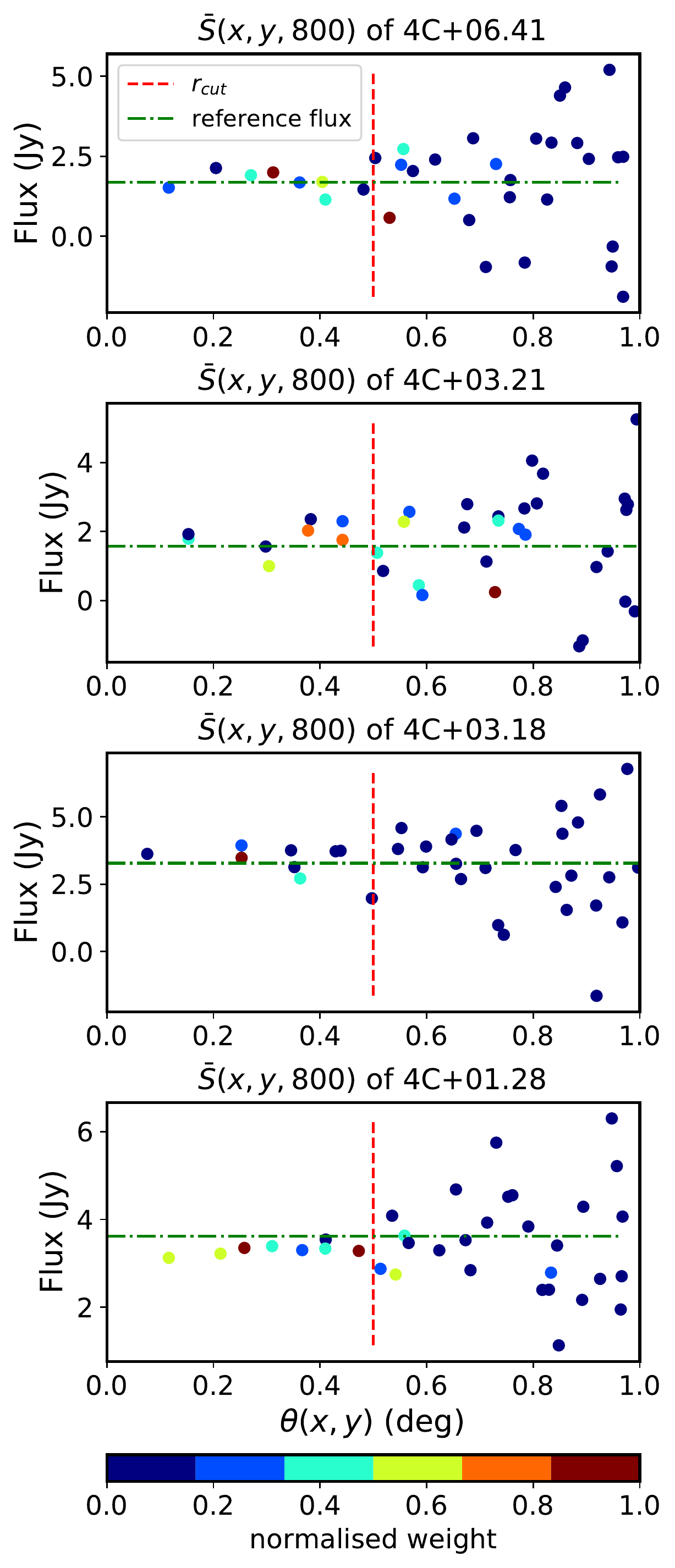}
\caption{$\bar S(x,y,{\rm ch})$ against $\theta(x,y)$ for the four point sources in a binned channel with the center channel 800 ($\nu=1023$ MHz). The colors on data show the normalisation of weight $w(x,y,{\rm ch})$ (\autoref{eq:w}).}
\label{fig:ptr_angle}
\end{figure} 

\begin{figure}
\centering
\includegraphics[width=.9\columnwidth]{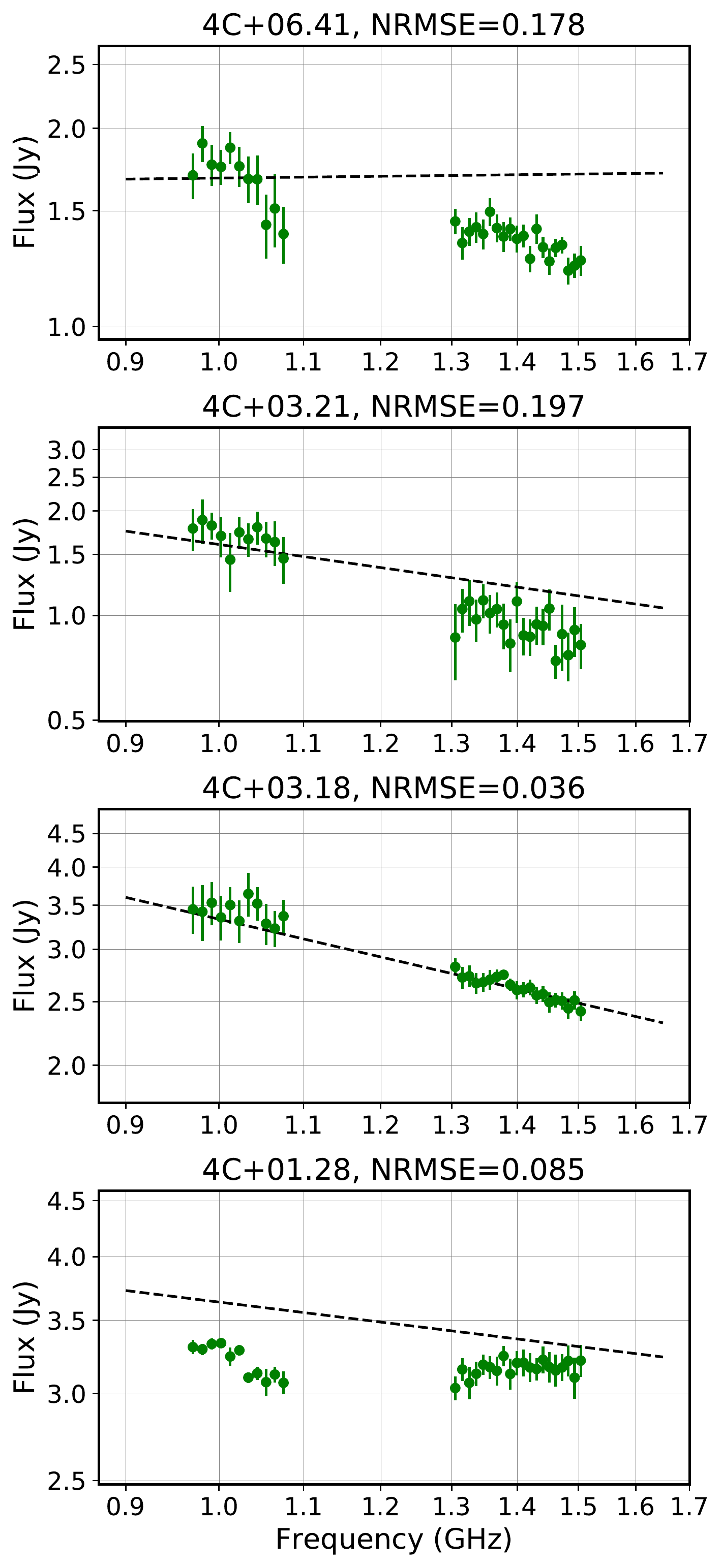}
\caption{The extracted fluxes of point sources compared with the spectra calculated from the Parkes data (black dashed line).}
\label{fig:spec}
\end{figure} 

We select four radio point sources in the WiggleZ 11hr field, the position of which are marked on the $\overline T_{\rm res}$ map in  \autoref{fig:sky_map}, 
and extract their fluxes from the final $\overline T_{\rm res}$ data cube in order to compare to the literature. One issue is that we are scanning over a large area and therefore the signal to noise on each of the point sources is relatively low. To compensate for that, we average the flux over 11 channels ($[{\rm ch}-5, {\rm ch}+5]$) and combine the nearby pixels around the point sources ($r_{\rm cut}=0.5^{\circ}$) with appropriate weights.

The frequency averaged flux on a given pixel $(x,y)$ is
\begin{gather}
\bar S(x,y,{\rm ch}) =2 k_B \frac {\overline T_{\rm res}(x,y, {\rm ch})} {{A_{\rm eff}({\rm ch)}} P_{\rm B}(x,y,{\rm ch})} \times 10 ^{26} ~ {\rm Jy}, 
\end{gather}
where $\overline T_{\rm res}(x,y, {\rm ch})$ is the weighted channel mean temperature (see \appref{sec:channel_weight}), 
$A_{\rm eff}$ is the effective area,  
and $P_{\rm B}$ is the normalized power pattern amplitude at pixel (x,y) which is displaced from the source being studied with an angle $\theta(x,y)$. Here we use the fitted 2D Gaussian beam model (\secref{sec:beam})
\begin{gather}
    P_{\rm B}(x,y, {\rm ch})=\exp\left(-\frac {\theta^2(x,y)} {2\sigma_{\rm Beam}^2}\right).
\end{gather}
In \autoref{fig:ptr_maps} we show the $\overline T_{\rm res}(x,y, {\rm ch})$  map at a given frequency for the four point sources with the $r_{\rm cut}=0.5^{\circ}$ circle indicated as a white dashed line.
In \autoref{fig:ptr_angle} we show $\bar S(x,y,{\rm ch})$ against $\theta(x,y)$ for the four point sources. The plot shows the $\theta(x,y)$ range $0-1^{\circ}$, while in the calculation we only use data with $\theta< 0.5^{\circ}$. Including more pixels increases the chance of adding contamination from other sources.

We then average $\bar S(x,y,{\rm ch})$ over the selected pixels with appropriate weights to obtain the mean spectral fluxes and standard errors. See \appref{sec:pix_weight} for more details. The extracted fluxes and the spectrum calculated from Parkes data are compared in \autoref{fig:spec}. In order to access the quality of the fit, we calculate the normalized root mean square error (NRMSE) for the full band as the percentage error: 
\begin{gather}
    NRMSE=\sqrt{ \frac{1}{N_{\rm ch}} {\sum\limits_{{\rm ch}} \left(\frac {\bar S({\rm ch})- S_{\rm ref} ({\rm ch})} {S_{\rm ref} ({\rm ch})}
    \right)^2}},
\end{gather}
where $S_{\rm ref}$ is the reference flux derived from Parkes observation data between 408 MHz and 1.4 GHz (or 2.7 GHz if 1.4 GHz data is absent), ch $\in [550, 600, ..., 1050]$ and $[2150, 2200, ..., 3100]$, and $N_{\rm ch}=31$ is the total number of fluxes measured across frequency. 

The NRMSE are listed in  \autoref{fig:spec}, which shows strong correlations to the source types and also to the brightness, pixel count and declination:
\begin{itemize}
\item 
4C +06.41 and 4C +01.28 are BL Lac objects (\citealt{2017A&A...602A..86L}; \citealt{2008AJ....135.2453P}). Monitoring by the OVRO-40m telescope shows that these sources are significantly variable at 15\,GHz over decade timescales \citep{2011ApJS..194...29R} and we interpret variability to be the cause of the discrepancy between the measured flux spectra and the Parkes predictions.
\item
4C +03.21 is the only one that has no redshift record and is labelled as `blank field' in Parkes Catalog (1990, version 1.01) in these four point sources. 
The Parkes data at 408~MHz, 1.41~GHz and 2.7~GHz are not consistent with a single power law spectrum, so it is not surprising that the MeerKAT data have a relatively poor fit to the 408~MHz to 1.41 GHz spectral prediction (NRMSE=0.197).
\item
4C+03.18 is identified as a radio galaxy and its flux might be expected to be stable at low radio frequency. We find an excellent agreement with the Parkes prediction with NRMSE=0.036.
\end{itemize}
Our flux estimation also has some other potential risks, such as the observations being done in scanning mode rather than tracking and the $\overline T_{\rm res}$ maps possibly containing Galactic structures that are unmodelled in PySM. 

\section{Conclusions}\label{sec:conclusions}

We presented a full data reduction pipeline for single dish data obtained from a pilot survey through the MeerKAT open time call. The observation was done in the L-band (856 -- 1712 MHz, 4096 channels)
and targeted a single patch of about 200 deg$^2$ with the aim of testing the single dish \textsc{Hi} intensity mapping technique. The scan patch covered the 11hr field of the WiggleZ Dark Energy Survey, which is a large-scale spectroscopic survey of emission-line galaxies selected from UV and optical imaging. The total (usable) observation time was about 10.5 hours, which, with about 60 dishes on average, corresponds to 630 hours effectively (before any flagging). 

The data reduction pipeline for the time-ordered data includes steps for flagging of human-made radio frequency interference, bandpass and absolute calibration using known point sources, and calibration of receiver gain fluctuations based on interleaved signal injection from a noise diode. \autoref{fig:pipe} shows the flowchart of the pipeline.

We construct a model for all the components that will contribute to the total signal, which includes the calibrator point sources, celestial diffuse components, elevation related emission (atmospheric emission and ground pickup), noise diode signals and  the receiver temperature.
By comparing the model to the time ordered data using a prescribed likelihood with priors, we fit the free parameters and use the gain solution to obtain the calibrated temperature. 

We show that the pipeline is sufficiently accurate to recover maps of diffuse celestial emission and point sources, showing a good correlation with Galactic synchrotron maps in the literature. 
We also extracted the Galactic \textsc{Hi} emission, showing that the reconstructed map has an excellent agreement to the HI4PI \textsc{Hi} column density map \citep{2016A&A...594A.116H}, which is constructed from the Effelsberg-Bonn \textsc{Hi} Survey.

The reconstructed maps  have no visible striping artifacts and a good level of consistency between per-dish maps and external datasets. 
We noticed that the different dishes have different "clean" scans left after the three rounds of RFI flagging. This shows a slight relation to the position of the dish in the MeerKAT array, which denotes that the weak RFIs may come from the surrounding artificial facilities or be affected by the terrain.

The residual maps have rms amplitudes below 0.1 K, corresponding to $<1\%$ of the model temperature. This shows that the emission missing in our sky model, such as point sources, has a negligible impact on the calibration. 
We also estimate the noise level in the final data cube after averaging over all dishes and scans using the difference between four neighboring channels. The rms is about 2 mK, with a median value of $1.4~\times$ the expected theoretical noise level, although there are deviations in some of the channels.

Finally, we select four radio point sources in the WiggleZ 11hr field and extract their fluxes from the total $\overline T_{\rm res}$ data cube in order to compare to the Parkes results.
The flux of the radio galaxy (4C+03.18) is recovered within $3.6\%$, while other point sources have reasonably larger errors: $17.8\%$ and $8.5\%$ for two BL Lac objects (4C +06.41 and 4C +01.28), and $19.7\%$ for the `blank field' source (4C +03.21).
This also demonstrates that the autocorrelation can be successfully calibrated to give the zero-spacing flux and potentially help in the imaging of MeerKAT interferometric data \citep{2019AJ....158....3R}.

In conclusion, we have shown that it is possible to calibrate the single dish data from the MeerKAT interferometer with high accuracy using sky calibrators and noise diodes. No obvious show stoppers were found although the RFI contamination remains challenging. This work opens the door to use MeerKAT and the future SKA to measure the \textsc{Hi} intensity mapping signal and probe Cosmology on degree scales and above. In a follow up paper we will be using this data to constrain the \textsc{Hi} power spectrum and its cross-correlation with galaxy surveys.

\appendix

\section{Weighted flux for Point sources} \label{sec:flux_eqs}
\subsection{Channel-weighted mean map} \label{sec:channel_weight}
On combining $2n+1$ channels, the channel mean temperature is
\begin{gather}
   \overline T_{\rm res}(x,y,{\rm ch})=\frac {1}{N (x,y,{\rm ch})}\sum\limits_{i=ch-n}^{ch+n}  T_{\rm res}(x,y,i)N_{\rm count}(x,y,i), 
\end{gather}
where the number of scans and channels contributing to the channel-weighted mean map is
\begin{gather}\label{eq:N}
N(x,y,{\rm ch})=\sum\limits_{i=ch-n}^{ch+n} N_{\rm count}(x,y,i),
\end{gather}
and $N_{\rm count}$ is the number of (unflagged) scans in the pixel and frequency channel. The weight assigned to pixel $(x,y)$ is
\begin{gather} \label{eq:w}
    w(x,y,{\rm ch})=\frac{N(x,y,{\rm ch})}{\sigma^2(x,y,{\rm ch})} P_{\rm B
    }^2(x,y,{\rm ch}),
\end{gather}
where 
\begin{gather}
    \sigma^2(x,y,{\rm ch})=\frac {N(x,y,{\rm ch})} {N(x,y,{\rm ch})-1} V (x,y,{\rm ch}),
\end{gather}
and the variance is
\begin{gather}
V(x,y,{\rm ch})\\\notag
    =\frac{1}{N(x,y,{\rm ch})} \left( \sum\limits_{i=ch-n}^{ch+n} N_{\rm count}(x,y,i)T_{\rm res}^2(x,y,i) \right)- {\overline T_{\rm res}^2(x,y,{\rm ch})}.
\end{gather}
Note here $N > 2n+1$ should be calculated from \autoref{eq:N}.

\subsection{Pixel weighted flux} \label{sec:pix_weight}
The weighted flux of the point source is
\begin{gather}
    \bar S({\rm ch}) = \frac {\sum \limits_{(x,y)} w(x,y,{\rm ch}) \bar S(x,y,{\rm ch})} {\sum \limits_{(x,y)} w(x,y,{\rm ch})}.
\end{gather}
Its standard deviation is
\begin{gather}
    \sigma_S({\rm ch}) =\sqrt{ \frac {\sum \limits_{(x,y)} w(x,y,{\rm ch}) \left(\bar S(x,y,{\rm ch}) -\bar S({\rm ch})\right)^2 } {\sum \limits_{(x,y)} w(x,y,{\rm ch})}},
\end{gather}
and the flux error is
\begin{gather}
    \epsilon_S({\rm ch}) = \sqrt{\frac {\sum \limits_{(x,y)} w^2(x,y,{\rm ch})} {\left(\sum \limits_{(x,y)} w(x,y,{\rm ch})\right)^2}} \sigma_S({\rm ch}). 
\end{gather}

\balance
\section*{Acknowledgements}
We would like to thank Stuart Harper for several useful discussions during the development of this work.
JW, MGS, MI, YL and BE acknowledge support from the South African Radio Astronomy Observatory 
and National Research Foundation (Grant No. 84156).
PB acknowledges support from Science and Technology Facilities Council grant ST/T000341/1. SC is supported by STFC grant ST/S000437/1. PS is supported by the STFC [grant number ST/P006760/1] through the DISCnet Centre for Doctoral Training. AP is a UK Research and Innovation Future Leaders Fellow, grant MR/S016066/1, and also acknowledges support by STFC grant ST/S000437/1. JF was supported by the University of Padova under the STARS Grants programme {\em CoGITO: Cosmology beyond Gaussianity, Inference, Theory and Observations} and by UK Science \& Technology Facilities Council (STFC) Consolidated Grant ST/P000592/1. MS acknowledge funding from the INAF PRIN-SKA 2017 project 1.05.01.88.04 (FORECaST). GB acknowledges support from the Ministero degli Affari Esteri della Cooperazione Internazionale—Direzione Generale per la Promozione del Sistema Paese Progetto diGrande Rilevanza ZA18GR02 and the National Research Foundation of South Africa (grant No. 113121) as part of the ISARP RADIOSKY2020 Joint Research Scheme.
We acknowledge the use of the Ilifu cloud computing facility, through the Inter-University Institute for Data Intensive Astronomy (IDIA). 
The MeerKAT telescope is operated by the South African Radio Astronomy Observatory, 
which is a facility of the National Research Foundation, an agency of the Department 
of Science and Innovation.

\section*{Data availability}
The analysis pipeline software is available on request. Data products are also available on request but are subject to an access policy within a 12-month proprietary period.

\bibliographystyle{mnras}
\bibliography{bibtex} 

\bsp	
\label{lastpage}
\end{document}